\documentclass[11pt]{article}
\pdfoutput=1
\usepackage{epsfig}
\usepackage{amsfonts}
\usepackage{amsmath}
\usepackage{amssymb}
\usepackage{dsfont}
\usepackage{hyperref}
\usepackage{color}
\usepackage{cite}

\newcommand{\al}{\alpha}
\newcommand{\bt}{\beta}

\newcommand{\MT}{M_{\slashTsub}}

\newcommand{\slashT}{\slash\hspace{-0.5em}T}
\newcommand{\slashTsub}{\slash\hspace{-0.4em}T}

\newcommand{\Nb}{\bar N}

\newcommand{\tb}{\bar \theta}

\newcommand{\mpi}{m_{\pi}}
\newcommand{\MQCD}{\Lambda_\chi}

\newcommand{\Or}{\mathcal O}

\newcommand{\vL}{\ensuremath{\mathcal{L}}}
\newcommand{\vp}{\varphi}
\newcommand{\sq}{^{2}}
\newcommand{\ga}{\gamma}

\newcommand{\dslash}[1]{#1 \llap{/\kern-0.5pt}}
\newcommand{\Dslash}[1]{#1 \llap{/\kern+1.5pt}}
\newcommand{\DDslash}[1]{#1 \llap{/\kern+2.3pt}}
\newcommand{\dslashh}[1]{#1 \llap{/\kern+1pt}}

\newcommand{\bea}{\begin{eqnarray}}
\newcommand{\eea}{\end{eqnarray}}
\newcommand{\bma}{\begin{pmatrix}}
\newcommand{\ema}{\end{pmatrix}}
\newcommand{\nn}{\nonumber}

\newcommand{\dHg}{d_{\mathrm{Hg}}}
\newcommand{\dRa}{d_{\mathrm{Ra}}}
\newcommand{\dXe}{d_{\mathrm{Xe}}}
\newcommand{\Exp}[1]{\cdot 10^{-#1}}

\begin{document}
\begin{titlepage}

\begin{flushright}
LA-UR-15-27548
\end{flushright}

\vspace{2.0cm}

\begin{center}
{\LARGE  \bf 
Direct and indirect  constraints 
on CP-violating
\\[0.3em] 
Higgs-quark and Higgs-gluon interactions}
\vspace{2.4cm}

{\large \bf Y.-T. Chien$^a$, V. Cirigliano$^a$, W. Dekens$^b$, J. de Vries$^c$, and E. Mereghetti$^a$ } 
\vspace{0.5cm}

{\large 
$^a$ 
{\it Theoretical Division, Los Alamos National Laboratory,
Los Alamos, NM 87545, USA}}

\vspace{0.25cm}
{\large 
$^b$ 
{\it Van Swinderen Institute, Faculty of Mathematics and Natural Sciences, University of Groningen, Nijenborgh 4,
NL-9747 AG Groningen, The Netherlands}}

\vspace{0.25cm}
{\large
$^c$ {\it Institute for Advanced Simulation, Institut f\"ur Kernphysik, 
and J\"ulich Center for Hadron Physics, Forschungszentrum J\"ulich, 
D-52425 J\"ulich, Germany}}

\end{center}

\vspace{1.5cm}

\begin{abstract}

We investigate direct and indirect constraints on the complete set of anomalous CP-violating Higgs couplings to quarks and gluons originating from dimension-6 operators,  by studying their signatures  at the LHC and in electric dipole moments (EDMs). 
We show that  existing uncertainties in hadronic and nuclear matrix elements have a significant impact on the interpretation of EDM experiments, 
and we  quantify  the improvements needed to fully exploit the power of EDM searches. 
Currently, the best bounds on the anomalous  CP-violating Higgs interactions come from a combination of EDM measurements and the data from LHC Run 1.
We  argue that Higgs production cross section and branching ratios measurements at  the LHC Run 2 will not improve the constraints significantly. 
On the other hand, the bounds on the couplings scale roughly linearly with EDM limits, so that future theoretical and 
experimental EDM developments can have a major impact in pinning down interactions of the Higgs.

\end{abstract}

\vfill
\end{titlepage}

\section{Introduction}

The discovery of a 125 GeV  boson at the Large Hadron Collider (LHC) 
is a breakthrough towards a deeper understanding of 
the mechanism of electroweak symmetry breaking~\cite{Aad:2012tfa,Chatrchyan:2012ufa}. 
Current data are consistent  with the spin-parity assignment $J^P = 0^+$ 
and  indicate that the couplings of this boson to the gauge vector bosons ($\gamma$, $g$, $W$, $Z$)  
and the third family of fermions ($t$, $b$,  $\tau$) are
consistent with those of the standard model (SM) Higgs boson~\cite{Agashe:2014kda}.

The current level of accuracy, however,  leaves  room for possible deviations from the SM picture. 
In fact, the Higgs  couplings  to gauge  bosons and $t,b$ quarks are known with an uncertainty of $O(20-30\%)$~\cite{Agashe:2014kda}, 
while the couplings of the Higgs to first and second generation fermions are much less constrained~\cite{Kagan:2014ila,Perez:2015aoa}.
Clearly,  better knowledge of the Higgs couplings will  shed light on the nature of EWSB mechanism and  will also have  
non-trivial implications  for other aspects of Higgs phenomenology (such as Higgs portal Dark Matter~\cite{Bishara:2015cha}). 
Improving the sensitivity and constraints to the Higgs couplings is a major goal of  Run 2 at the LHC 
and  is  becoming an increasingly important target for low-energy indirect probes. 

The analysis of non-standard Higgs couplings can be conveniently performed within an effective field theory (EFT)   framework.
There are at least two scenarios that  can be used to describe current data:   
(i)   Linear realization, in which the observed Higgs forms an electroweak (EW)  doublet with the would-be Goldstone 
modes associated with spontaneous breaking of the EW group (that manifest themselves as longitudinal degrees of 
freedom of the massive gauge bosons  $W^\pm$ and $Z$).  In this framework the leading  dimension-6  operators 
describing new physics and in particular new Higgs and EW dynamics have been classified in Refs. \cite{Buchmuller:1985jz,Grzadkowski:2010es}. 
(ii) The other option is that the boson discovered at the LHC is actually a light composite state associated to new strong dynamics.  
Explicit models of composite Higgs have been put forward starting with the pioneering work of Refs.~\cite{Kaplan:1983fs,Dugan:1984hq}.  
This class of models can be best analyzed within the framework of the electroweak chiral Lagrangian 
with a light singlet Higgs state~\cite{Giudice:2007fh,Contino:2013kra,Buchalla:2013rka,Buchalla:2013eza}. 

In both scenarios (i) and (ii) outlined above, there already exist  EFT analyses of non-standard Higgs couplings.  Most global 
analyses (see \cite{Agashe:2014kda} and references therein)  
make assumptions about the flavor and CP structure of the Higgs couplings, such as minimal flavor violation~\cite{D'Ambrosio:2002ex,Cirigliano:2005ck}, 
that reduce the number of operators considered.  
While systematic studies of flavor-violating Higgs couplings exist in the literature~\cite{Blankenburg:2012ex,Harnik:2012pb},  
analyses of  CP-violating (CPV)  Higgs couplings  have typically focused on subsets of 
operators~\cite{Huber:2006ri,Brod:2013cka,Dekens:2013zca,Kamenik:2011dk,Aguilar-Saavedra:2014iga}. 
Here we wish to  initiate a systematic study  of the flavor-diagonal  CPV couplings of the Higgs, starting with 
its couplings  to quarks and gluons and  leaving the discussion of  couplings to weak gauge bosons and fermions to future work. 
Our study is primarily motivated by the need to learn as much as possible in a model-independent way  about the  recently discovered Higgs,
 including its  CP properties (for recent discussions of CP violation in the Higgs sector  in the context of the Two-Higgs Doublet Model  
see Refs.~\cite{Inoue:2014nva,Fontes:2015xva,Chen:2015gaa,Fontes:2015mea,Cheung:2014oaa}). 
Moreover,   CPV in the Higgs sector might have implications for weak scale baryogenesis  in a number of scenarios beyond the SM (BSM).
And finally,  we expect  strong bounds on non-standard  CPV Higgs  couplings from permanent electric dipole moments, 
somewhat in contrast to the  CP-conserving couplings,  which are harder to constrain. 

In this work we focus on the linear  EFT realization for the Higgs sector  and leave 
the discussion of  strongly interacting light Higgs to a future study. 
Within this setup,  our analysis involves both indirect and direct constraints, along the lines described below: 

\begin{itemize}

\item We identify the  dimension-6  CPV Higgs couplings to quarks and gluons and 
discuss their renormalization group evolution from the scale of new physics down to the hadronic scale, 
including all the relevant SM heavy particle thresholds (Section \ref{Sec2}). 

\item  In Section~\ref{Sec3} we study in detail the indirect constraints coming from  electric dipole moments (EDMs). 
All bounds are derived assuming that  the Peccei-Quinn mechanism \cite{Peccei:1977ur} is at work. 
We pay special attention to the role of hadronic and nuclear uncertainties. 
We present bounds corresponding to current and prospective experimental  sensitivities and we assess the impact of improving the 
theoretical uncertainties on hadronic and nuclear  matrix elements. 

\item In Section~\ref{Sec4} we study the direct constraints from LHC Higgs production and decay as well as 
$t \bar t$ and $t \bar t h$ production, presenting bounds from current data and 
prospective sensitivities at LHC Run 2.  We focus here on CP-conserving observables that depend on the square of the CP-violating couplings as these observables
currently give the strongest constraints.

\item In our analysis we first obtain bounds on the effective couplings  by ``turning on" one  coupling  at the time at the high scale. 
We subsequently study the case in which two operators are switched on simultaneously (Section \ref{Sec-interplay}).  

\item In our concluding discussion (Section \ref{Sec-discussion}) we compare  the strength of the indirect and direct bounds 
for the various couplings.  We summarize the current status and describe the 
impact of prospective sensitivities in both  planned EDM searches and  Run 2 at the LHC. 

\end{itemize}

\section{The set of operators and its renormalization-group evolution}
\label{Sec2}
Our analysis assumes the existence of new physics involving heavy degrees of freedom, that  
modify the low-energy dynamics via a number of $SU(3)_C\times SU(2)_W \times U(1)_Y$-invariant 
local operators of dimension 5 and higher~\cite{Buchmuller:1985jz,Grzadkowski:2010es}.
Here we are interested in CPV operators involving  the Higgs doublet, quarks, and gluons, 
so   at  some  scale $\MT \gg v$     ($v\simeq 246$ GeV is the Higgs vacuum expectation value (vev))
we consider the following  effective Lagrangian, 
\bea
{\cal L}_{\rm eff} &=& {\cal L}_{\rm SM} + {\cal L}_6 \nn \\
\vL_6&=&
 -\, \theta^\prime \frac{\alpha_s}{32\pi} \,\varepsilon^{\mu\nu\alpha\beta}G^a_{\mu\nu}G^a_{\alpha\beta}(\vp^\dagger \vp)
\ + \  \sqrt{2}\vp^\dagger \vp\left(\bar q_L Y_u^{\prime} \, u_R  \,  \tilde \vp  \ +\  \bar q_L Y_d^{\prime} \, d_R \, \vp \right)\nn\\ 
&&
- \frac{1}{\sqrt{2}} \, \bar{q}_L  \, \sigma \cdot G  \,  \tilde{\Gamma}_u  \, u_R \ \frac{\tilde \vp}{v} 
\ - \ \frac{1}{\sqrt{2}} \, \bar{q}_L  \, \sigma \cdot G  \,  \tilde{\Gamma}_d  \, d_R \ \frac{\vp}{v} 
+ \text{h.c.} \,\,\,, 
\label{Dim6def}
\eea
where ${\cal L}_{\rm SM}$ denotes the SM Lagrangian. The operators in $\mathcal L_6$ are written in terms of the Higgs doublet $\vp$, the left-handed quark doublet $q_L$, the right-handed quark singlets $u_R$ and $d_R$, and the gluon field strength $G_{\mu\nu}^a$. We have introduced the  notation $\tilde \vp = i \sigma_2 \vp^*$ and  $\sigma \cdot G \equiv \sigma^{\mu \nu} G_{\mu \nu}^a t^a$, and we 
are suppressing generation indices. 
The  $3\times 3$ matrices $Y_{u,d}^{\prime}$ and $\tilde \Gamma_{u,d}$ induce anomalous Yukawa interactions and  quark color dipole moments, respectively. The $\theta'$ term represents a CPV interaction between the Higgs field and gluons. Note that the couplings $Y_{u,d}^\prime$ and $\theta^\prime$ have mass-dimension $-2$, while $\tilde{\Gamma}_{u,d}$ has mass dimension $-1$, due to the explicit factor of $1/v$  associated with $\tilde \Gamma_{u,d}$. Finally, $\varepsilon^{\mu\nu\alpha\beta}$ denotes the completely antisymmetric tensor with $\varepsilon^{0123}=+1$.

In the unitary gauge,  we can write the Higgs doublet as $\vp  =(0,\, v+h)^T/\sqrt{2}$. 
To $O(h^0)$ the couplings $Y_{u,d}^\prime$ then contribute to  the quark mass matrices,
while $\theta^\prime$ produces a shift in the SM QCD $\tb$ term.~\footnote{We assume in this work that  
the Peccei-Quinn mechanism \cite{Peccei:1977ur} is at work, so that  $\tb$  (including the shift $\delta \tb = (1/2) v^2 \theta'$)
 relaxes to  $\tb_{\rm ind} \neq0$ due to the 
distortion of the axion potential induced by the higher dimensional  operators.} 
The remaining  $O(h)$  terms in Eq.\ \eqref{Dim6def} give rise to effects that are not described by the SM, 
and in particular  induce anomalous $\bar{q} q h$  and CPV Higgs-gluon  interactions. 

Working  in the  basis in which the full quark mass matrices (including SM and BSM effect from $Y_{u,d}^\prime$) are diagonal, 
the operators of Eq. \eqref{Dim6def} in combination with the SM Yukawa interactions (${\cal L}_Y$), give the following contributions to ${\cal L}_{\rm eff}$,~\footnote{
Denoting the Standard Model Yukawa couplings by  ${\cal L}_Y = - \sqrt{2} \bar{q}_L Y_d d_R \varphi   - \sqrt{2} \bar{q}_L Y_u u_R \tilde{\varphi}$, 
the quark mass matrices are given by  $m_{u,d} = v \left(Y_{u,d} - \frac{v^2}{2}  Y_{u,d}^\prime \right)$.  
Upon expanding  ${\cal L}_Y + {\cal L}_6$ to first order in $h$ and expressing the couplings in terms of $m_{u,d}$ and $Y_{u,d}^\prime$,  
we obtain  Eq.~(\ref{Dim6v0}).}
\bea 
\label{Dim6v0}
\vL_{Y}+\vL_{6}&=&
- \bar{d} m_d d  - \bar{d} \left[ \frac{m_d}{v} - v^2  {\rm Re}\, Y_d^\prime \right] d \, h    
+ v^2  \bar{d}\,  i \gamma_5  \left[ {\rm Im}\, Y_d^\prime \right]  d \, h  
\nn \\
&-&  \bar{u} m_u u  - \bar{u} \left[ \frac{m_u}{v} - v^2  {\rm Re} \,Y_u^\prime \right] u \, h    
+ v^2  \bar{u}\,  i \gamma_5  \left[ {\rm Im}\, Y_u^\prime \right]  u \, h  
\nn \\
&-& \frac{1}{2}  \bar{u}  \, \sigma \cdot G \, \left[ {\rm Re}\, \tilde{\Gamma}_u  + i \gamma_5  \, {\rm Im}\, \tilde{\Gamma}_u \right]  u \ \left(1 + \frac{h}{v} \right)
\nn \\
&-& \frac{1}{2}  \bar{d}  \, \sigma \cdot G \, \left[ {\rm Re}\, \tilde{\Gamma}_d  + i \gamma_5  \, {\rm Im}\, \tilde{\Gamma}_d \right]  d \ \left(1 + \frac{h}{v} \right)
-\theta^\prime \frac{\alpha_s}{8\pi}h v \frac{1}{2} \,\varepsilon^{\mu\nu\alpha\beta}G^a_{\mu\nu}G^a_{\alpha\beta}~, 
\eea
where we use the  compact matrix notation ${\rm Re}\, A \equiv 1/2 (A + A^\dagger)$ 
and ${\rm Im }\, A \equiv 1/(2 i) (A - A^\dagger)$.
$m_{u,d}$ are the real and diagonal quark mass matrices,  while ${\rm Re}\, Y_{u,d}^\prime$, ${\rm Im}\, Y_{u,d}^\prime$, 
and $\tilde \Gamma_{u,d}$ are not necessarily diagonal.

Studies of the flavor-violating couplings induced by $Y_{u,d}^\prime$ have appeared in the literature~\cite{Blankenburg:2012ex,Harnik:2012pb}, 
while the CPV third generation couplings have been studied in \cite{Huber:2006ri,Brod:2013cka}.   
As for the gluon dipole operators,  the EDM constraints on
light quark diagonal couplings have been studied in Ref.~\cite{Dekens:2013zca,Sala:2013osa}, and the top chromo-EDM (CEDM) 
has been  studied in several papers, see for example Refs.~\cite{Kamenik:2011dk,Hayreter:2013kba,Aguilar-Saavedra:2014iga}.
In this work we focus on  the CPV flavor-diagonal couplings 
arising from Eq.~(\ref{Dim6v0}) and we ignore the real and the flavor-violating parts of $Y_{u,d}^\prime$.
In  the dipole operator sector,  we focus 
on  the top CEDM as it strongly mixes with $\mathrm{Im}\,Y'_t$ and $\theta'$ 
(in Appendix~\ref{sect-appA} we summarize the bounds on the light quark CEDMs $\tilde{d}_q$, $q \neq t$). 
That is, we take at the high scale  $\tilde \Gamma_{d}= 0$ 
and $\tilde{\Gamma}_u^{ij}  = \delta_{i3} \delta_{j3} \tilde{\Gamma}_t$, and extend the results of   
Ref.~\cite{Kamenik:2011dk}  by taking into account  hadronic uncertainties, including additional mixing effects, 
and considering additional collider constraints from Higgs production.

In summary, our starting point high-scale CPV and flavor-diagonal  effective Lagrangian reads:
\begin{equation}
\label{Dim6}
\vL_{6}^{CPV}   =
-\theta^\prime \frac{\alpha_s}{8\pi}h v \frac{1}{2} \,\varepsilon^{\mu\nu\alpha\beta}G^a_{\mu\nu}G^a_{\alpha\beta}
\  + \ 
\sum_{q=u,d,c,s,t,b}  \  v\sq {\rm Im}\, Y'_q   \, \bar  q i \ga_5 q    \  h 
\ - \ \frac{i}{2} \tilde d_t\, g_s    \   \bar t     \sigma \cdot G   \gamma_5  t \   \left(1+\frac{h}{v}\right)+\dots
\end{equation}
where now $\tilde d_t\equiv {\rm Im}\,\tilde \Gamma_t/g_s$, $\textrm{Im}\, Y^{\prime}_q$ denotes the diagonal entries of the matrices in Eq. \eqref{Dim6v0}, 
and the dots stand for interactions involving two or more Higgs fields.

\begin{figure}[t]
\centering
\includegraphics[scale=0.9]{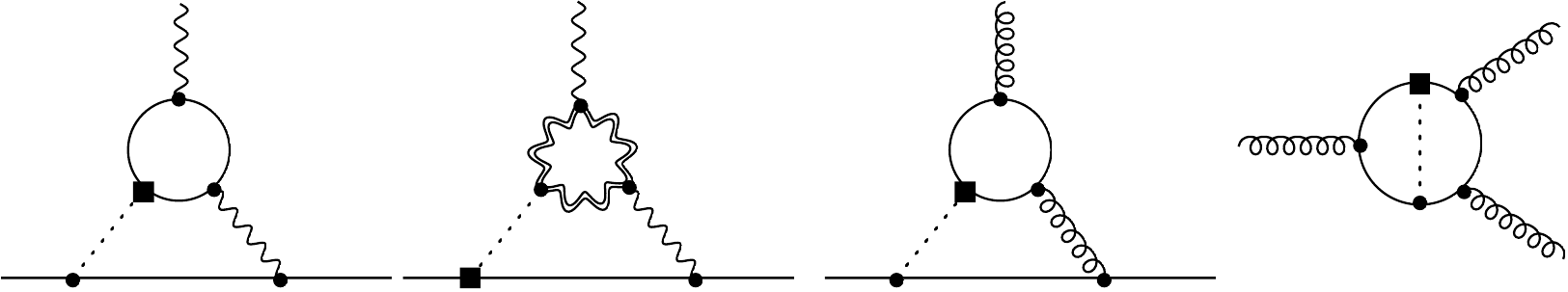} \\
\caption{\small Examples of two-loop threshold contributions to the quark (C)EDMs and Weinberg operator. A solid (dashed) line denotes
   a quark (Higgs bosons). Wavy (curly) lines denote a photon (gluon) and a double wavy line a $W^{\pm}$ boson. Circles denote SM vertices and squares CPV Yukawa interactions. Not all possible diagrams are shown.} 
\label{Barrzee}
\end{figure}
\subsection{Renormalization-group equations}
In order to connect the above operators to measurements taking place at energies below $\MT$, the renormalization-group equations (RGEs) governing the  scale  dependence of these operators are required. Relating these interactions to EDM  experiments necessitates evolving them down to the QCD scale, $\Lambda_\chi\simeq 1\, {\rm GeV}$, below which QCD becomes strongly coupled and non-perturbative techniques are required, see Sect. \ref{Sec:interpretation}.
As the operators in Eq.\ \eqref{Dim6} do not contribute to EDMs directly, but only through their 
mixing contributions to other operators, we require an extended basis of operators 
that includes the light  fermion (C)EDMs and the Weinberg operator.
Accordingly,  we extend the effective Lagrangian as follows: 
\bea \label{ExtendedO}
\vL_6^{CPV}  \to  \ \  \vL_6^{CPV}  
&-&   \frac{i}{2}\sum_{f=e,u,d,s,c,b} d_f \,  
\bar f \sigma \cdot    F  \ga_5  f
-\frac{i}{2}\sum_{q=u,d,s,c,b}\tilde d_q \, g_s \ \bar q   \sigma \cdot G   \gamma_5  q   \nn\\
&+&d_W \frac{1}{6}f_{abc}\varepsilon^{\mu\nu\al\bt}G^a_{\al\bt}G^b_{\mu\rho}G_{\nu}^{c \, \rho}.
\eea
Taking the basis, $\vec C_q = (d_q/e Q_q m_q,\, \tilde d_q/m_q,\, d_W/g_s,\, {\rm Im}\, Y'_q,\, \theta')^T$, the one-loop QCD RGEs can be written as  \cite{Weinberg:1989dx, Wilczek:1976ry,  BraatenPRL, Degrassi:2005zd,Kaplan:1988ku, Grojean:2013kd, Bhattacharya:2015rsa},
\bea
\frac{d\vec{C_q}(\mu)}{d\ln\mu}=\frac{\al_s}{4\pi}
\bma 
8C_F & -8C_F &0&0&0\\
0& 16C_F-4N & 2N& 0& -1/4\pi\sq\\
0&0&N+2n_f+\bt_0&0&0\\
0&-18C_F\big(\frac{m_q}{v}\big)^3&0&-6C_F&12 C_F\frac{\al_s}{4\pi}\frac{m_q}{v}\\
0&-8\frac{4\pi}{\al_s}\big(\frac{m_q}{v}\big)\sq &0&0&0
\ema \cdot \vec C_q(\mu) ,
\label{RGEs}\eea
where $C_F = \frac{N\sq-1}{2N}$, $\bt_0 = (11 N- 2n_f)/3$, and $N$ ($n_f$) is 
the number of colors (flavors). As we are only interested in the operators of Eq.\ \eqref{Dim6}, we have $d_f(\MT)=\tilde d_{u,d,s,c,b}(\MT)=d_W(\MT)=0$ as boundary conditions.~\footnote{Note  the RGE evolution strictly speaking does not preserve the form of Eq.~(\ref{Dim6v0}), 
as the chromo-EDM operators induce at one-loop level a pseudoscalar quark mass term,  not present in (\ref{Dim6v0}). 
The pseudoscalar masses  can be eliminated through an axial  transformation of the quark fields, 
which has the net effect of changing the 4-2 entry of the anomalous dimension in Eq. \eqref{RGEs}
from $- 30 C_F (m_q/v)^3$ to $- 18 C_F (m_q/v)^3$. Due to the Yukawa suppression, this effect is only relevant for the top quark.}
Note that the electron EDM introduced in Eq.\ \eqref{ExtendedO}, $d_e$, does not appear in the RGEs since it is not affected by one-loop QCD renormalization. However, it is generated by threshold corrections that are discussed below.

\subsubsection{Evolution to $\mu=m_t$}\label{sec:mt}
When considering the contribution of the dimension-6 operators in Eq.\ \eqref{Dim6} to collider observables, it is mainly the mixing among the operators, $\tilde d_t,\,   Y'_q$, and  $\theta'$, themselves that is relevant. The above RGEs can be used to first run the couplings down to $\mu=m_t$, where the top CEDM and Yukawa coupling are integrated out. At this scale, the top Yukawa induces a contribution to $\theta'$ through a top loop, 
\bea
\theta'(m_t^-) = \theta'(m_t^+) - \frac{v}{m_t}{\rm Im}\, Y'_t(m_t). 
\eea
We present the numerical results of this procedure in Table \ref{Table:mt}, where we employ the following values \cite{Agashe:2014kda},
\bea
\al_s(M_Z) &=& 0.118, \quad M_Z=91.2,\nn\\
m_u(\MQCD) &=& 3.1\, \text{MeV},\quad m_d(\MQCD) = 6.5\, \text{MeV}, \quad m_s(\MQCD) = 128\, \text{MeV},\nn\\
m_c(m_c) &=& 1.28\, \text{GeV},\quad m_b(m_b) = 4.18\, \text{GeV},\quad m_t(m_t) = 160\,\text{GeV}.
\eea
All quark masses are given in the $\overline{\textrm{MS}}$ scheme.
 A fixed-order perturbation-theory solution of Eq.\ \eqref{RGEs} approximates the exact solution to $20\%\, (45\%)$ at $\MT =1\,(10)$ TeV\footnote{This implies neglecting the $\mu$ dependence of $\al_s$ and $m_q$ in Eq.\ \eqref{RGEs}, as these would constitute higher-order effects. The approximate results are obtained when taking $\al_s=\al_s(\MT)$ and $m_q=m_q(m_t)$.}.
\begin{table}\footnotesize
\centering
$\begin{array}{c||cccccccc}
\MT = 1\, {\rm TeV}& \tilde{d}_t(\MT)/m_t(\MT) &{\rm Im\,} Y_q'(\MT) & \theta'(\MT)\\\hline\hline
{d}_t(m_t^+)/m_t(m_t^+) & 0.089\,  e\, & - & 2.5 \cdot 10^{-5}e \,Q_t\\
\tilde d_{t}(m_t^+)/m_t(m_t^+) & 0.87 \, & - & 3.4 \cdot 10^{-4} \\
{d}_{q\neq t}(m_t^+)/m_{q\neq t}(m_t^+) &4.6 \cdot 10^{-5}e\, Q_q  & - & 2.5 \cdot 10^{-5}e \,Q_q\\
\tilde {d}_{q\neq t}(m_t^+)/m_{q\neq t}(m_t^+) & 9.1 \cdot 10^{-4}  & - & 3.4 \cdot 10^{-4} \\
{\rm Im\,} Y_q'(m_t^+) &0.076\, \delta_{q t}{}^{\star}  & 1.12 & -1.2\cdot 10^{-3}\, \delta_{q t}{}^{\star}\\ 
\theta'(m_t^+) & 5.2  & - & 1 \\ \end{array}$ 
\caption{\small The contributions of the operators in Eq.\ \eqref{Dim6} at $\MT = 1$ TeV, to the operators at $\mu = m_t$. A dash, $``-"$, indicates no, or a negligible, contribution. The ${}^\star$ denotes that we neglected tiny contributions to the CPV Yukawa couplings of lighter quarks.
\label{Table:mt} }
\end{table}

\subsubsection{Evolution to $\mu=\MQCD$}
\begin{figure}[t]
\centering
\includegraphics[scale=0.6]{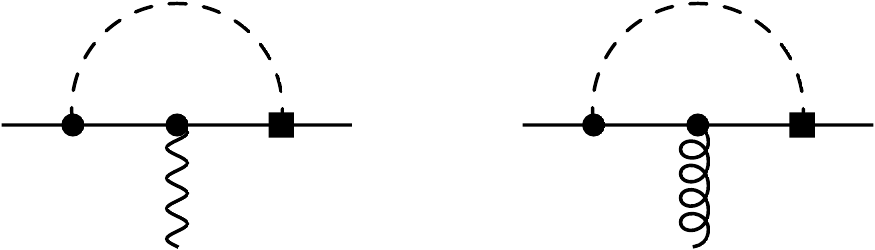} \\
\caption{\small One-loop threshold contributions to the quark (C)EDMs involving the CPV Yukawa interactions. Notation is as in Fig.\ \ref{Barrzee}.} 
\label{Fig:OneLoop}
\end{figure}
Evaluating the contributions to EDMs is somewhat more involved. At low energies, around $\Lambda_\chi$,
 the light-quark (C)EDMs, $d_{u,d,s}$ and $\tilde d_{u,d,s}$, and the Weinberg operator $d_W$, contribute to EDMs, while the charm- and bottom-quark CEDMs facilitate indirect contributions. As a result, the mixing with the additional operators in Eq.\ \eqref{ExtendedO} determines the contribution to EDMs. Apart from the mixing, the matching corrections at the different thresholds are relevant as well. 

First the RGEs of Eq.\ \eqref{RGEs} are used to run the operators from $\mu=\MT$ to $\mu=m_t$, where we integrate out the top quark and the Higgs boson. This implies that  the  couplings $\theta'$, $Y_t'$, and  $\tilde d_t$  and their corresponding operators are removed from the EFT  below  
$\mu=m_t$.  Eliminating these operators gives rise to several threshold corrections to the operators in Eq.\ \eqref{ExtendedO}. The Yukawa interactions contribute to the (C)EDMs \cite{Barr:1990vd,Gunion:1990iv,Abe:2013qla,Jung:2013hka} and the Weinberg operator \cite{Weinberg:1989dx,Dicus:1989va} through Barr-Zee diagrams, shown in Fig.\ \ref{Barrzee}. The quark (C)EDMs receive additional contributions from the one-loop diagrams shown in Fig.\ \ref{Fig:OneLoop} \cite{Hisano3}. The top CEDM gives rise to a one-loop threshold contribution to the Weinberg operator \cite{BraatenPRL,Boyd:1990bx}. In total we have the following matching conditions,
\bea\label{mtThreshold}
d_f(m_t^-) &=& d_f(m_t^+)- 24e v\frac{\alpha}{(4\pi)^3}Q_f \sum_{q'}Q_{q'}\sq\bigg[ f(x_{q'})\frac{m_f}{m_{q'}} {\rm Im}\, Y_{q'}'(m_t^+)+g(x_{q'}){\rm Im}\, Y_f'(m_t^+)\bigg]\nn\\
&&+4eQ_f\frac{\al}{(4\pi)^3}v\bigg[3f(x_W)+5g(x_W)\bigg]{\rm Im}\, Y_f'(m_t^+)+\frac{eQ_q}{2\pi\sq}\frac{m_f\sq}{m_h\sq}v\bigg(\frac{3}{4}+\ln\frac{m_f}{m_h}\bigg) {\rm Im}\, Y_f'(m_t^+)\nn\\
\tilde d_q(m_t^-) &=& \tilde d_q(m_t^+)+4 v\frac{\alpha_s}{(4\pi)^3}\sum_{q'}\bigg[ f(x_{q'})\frac{m_q}{m_{q'}} {\rm Im}\, Y_{q'}'(m_t^+)+g(x_{q'}){\rm Im}\, Y_q'(m_t^+)\bigg]\nn\\
&&-\frac{1}{2\pi\sq}\frac{m_q\sq}{m_h\sq}v\bigg(\frac{3}{4}+\ln\frac{m_q}{m_h}\bigg) {\rm Im}\, Y_q'(m_t^+)
,\nn\\
d_W(m_t^-)&=&d_W(m_t^+)-\frac{g_s^3}{32\pi\sq m_t} \tilde d_t(m_t^+)+4 \frac{g_s^3}{(4\pi)^4}\frac{v}{m_t}h(m_t,m_h){\rm Im}\, Y_t'(m_t^+) ,
\eea
where $m_t^+$ ($m_t^-$) indicates a scale just above (below) $m_t$, $x_i\equiv \frac{m_i\sq}{m_h\sq}$, and the functions $f$, $g$, and $h$ are given by,
\bea
&f(z) \equiv \frac{z}{2}\int_0^1 dx \frac{1-2x(1-x)}{x(1-x)-z}\ln \frac{x(1-x)}{z}\,\,, \qquad g(z) \equiv \frac{z}{2}\int_0^1 dx \frac{1}{x(1-x)-z}\ln \frac{x(1-x)}{z},&\nn\\
&h(m,M) = \frac{m^4}{4}\int_0^1 dx\int_0^1 du \frac{u^3x^3(1-x)}{[m\sq x(1-ux)+M\sq(1-u)(1-x)]\sq}\,\,\,.&
\eea
\begin{table}
\scriptsize
$\begin{array}{c||cccccccc}
\MT = 1\, {\rm TeV} & {\rm Im\,} Y_u' & {\rm Im\,} Y_d' &
  {\rm Im\,} Y_c' &{\rm Im\,} Y_s' &
  {\rm Im\,} Y_t'& {\rm Im\,} Y_b' & \theta'& \tilde{d}_t/m_t \\\hline\hline
 d_u/m_u & 15\,e &-
   & 2.8\cdot 10^{-5} \, e & - & 7.3 \cdot 10^{-5}\, e & 7.1\cdot 10^{-5} \,e & 9.3\cdot 10^{-5}\, e
   & 4.2 \cdot 10^{-4}\,e \\
 \tilde{d}_u/m_u & 26 & - & 9.8 \cdot 10^{-5}
   & - & 1.9\cdot10^{-4} & 1.7 \cdot 10^{-4}& 1.7 \cdot 10^{-4}& 1.0 \cdot 10^{-3}\\
 d_d/m_d & - & -3.5\,
   e & -1.4 \cdot 10^{-5}\,e & - & -3.7 \cdot 10^{-5}\,e & -3.5 \cdot 10^{-5}\,e &
   -4.7 \cdot 10^{-5}\,e &-2.1 \cdot 10^{-4}\,e \\
 \tilde{d}_d/m_d &- & 12 & 9.8 \cdot 10^{-5}
   & - & 1.9\cdot 10^{-4} & 1.7\cdot 10^{-4} & 1.7 \cdot 10^{-4} & 1.0 \cdot 10^{-3} \\
 d_s/m_s&- &
  - & -1.4\cdot 10^{-5}\, e & -0.18\, e & -3.7 \cdot 10^{-5}\,e & -3.5 \cdot 10^{-5}\, e
   & -4.7\cdot 10^{-5}\, e & -2.1 \cdot 10^{-4}\,e \\
 \tilde{d}_s/m_s &-&-& 9.8 \cdot 10^{-4} & 0.62 & 1.9 \cdot 10^{-4}& 1.7 \cdot 10^{-4}& 1.7 \cdot 10^{-4} &1.0 \cdot 10^{-3}\\
 d_e/m_e & - &
  - & 2.5\cdot 10^{-5}\, e & 1.3 \cdot 10^{-6}\, e&
   7.0 \cdot 10^{-5}\, e & 1.3 \cdot 10^{-5}\, e & -7.2 \cdot 10^{-8}\,e & 4.7 \cdot 10^{-6}\,e \\
 d_W &- &-& -1.5 \cdot 10^{-3} &
- & 2.7\cdot  10^{-6} & -2.3 \cdot 10^{-4}& -7.3\cdot 10^{-6} & -1.9\cdot 10^{-3} \\
\end{array}$
\caption{\small The contributions of the operators in Eq.\ \eqref{Dim6} to the operators which contribute to EDMs (Eq.\ \eqref{ExtendedO}) at low energies, $\MQCD\simeq 1\, {\rm GeV}$. Here we assumed the scale of new physics to be $\MT = 1$ TeV. A dash, $``-"$, indicates no, or a negligible, contribution.  }
\label{Table:1GeV}
\end{table}

The first loop terms, contributing to the quark (C)EDMs in Eq.\ \eqref{mtThreshold}, are due to the Barr-Zee diagrams involving quark loops.~\footnote{For the Barr-Zee diagrams involving quarks other than the top in the loop, one should in principle apply the procedure of Ref.\ \cite{Brod:2013cka} in order to correctly handle the appearance of large logarithms, e.g.\ $\ln m_q/m_h$. However, for all Yukawa couplings we find larger contributions from the diagrams involving the top quark, we therefore approximate the diagrams involving lighter quarks by the expressions in Eq. \ref{mtThreshold}.} The second term, contributing to quark EDMs, originates in Barr-Zee diagrams involving an internal $W^\pm$ loop. The remaining terms are from the one-loop graphs in Fig.\ \ref{Fig:OneLoop}. The loop terms contributing to $d_W$ arise from, respectively, the top qCEDM threshold correction and the fourth Barr-Zee diagram in Fig.~\ref{Barrzee}. The contribution of $\tilde d_t$ to the Weinberg operator was also considered in Ref.\ \cite{Kamenik:2011dk}, however, due to its mixing with $\theta'$, we obtain somewhat larger contributions of $\tilde d_t$ to the operators at $1$ GeV (shown in Table \ref{Table:1GeV}).

Below $\mu=m_t$, our basis consists of the operators explicitly listed in Eq.\ \eqref{ExtendedO}. The RGEs can then be used to run down to $\mu=m_b$ and subsequently to $\mu=m_c$. At these thresholds the bottom and charm quarks and their (C)EDMs are integrated out, which results in additional threshold corrections to the Weinberg operator,
\bea
d_W(m_{c,b}^-)&=&d_W(m_{c,b}^+)-\frac{g_s^3}{32\pi\sq m_{c,b}} \tilde d_{c,b}(m_{c,b}^+) .
\label{eq:dWthreshold}
\eea
After the charm threshold the remaining operators can be evolved to $\MQCD$ using Eq.\ \eqref{RGEs}. The numerical result of this analysis is presented in Table \ref{Table:1GeV} for $\MT=1$ TeV.
A fixed-order perturbation approximation, as the one mentioned in Sec.\ \ref{sec:mt}, is less accurate below $m_t$ as $\al_s$ runs faster in this regime. However, the solution to the RGE below $m_t$ is simpler, as only the quark (C)EDMs and the Weinberg operator are involved, and is explicitly given in Ref.\ \cite{Degrassi:2005zd}.

\section{Constraints from electric dipole moments}
\label{Sec3}

Strong constraints on the CPV higher-dimensional operators can be derived from experimental upper bounds on EDMs. The strongest constraints arise from measurements on the neutron \cite{Baker:2006ts}, the ${}^{199}$Hg atom \cite{Griffith:2009zz}, and the ThO molecule \cite{Baron:2013eja}. To interpret the experimental upper bounds, it is necessary to express the  observables in terms of the Wilson coefficients of the dimension-6 operators 
and the corresponding hadronic and nuclear matrix elements.  In particular, for the operators involving quarks and gluons this is problematic due to the non-perturbative nature  of QCD at low energies. Nevertheless, various techniques have been applied to calculate EDMs directly in terms of CPV quark-gluon operators.  Depending on the operator under investigation, the techniques vary in their sophistication and accuracy. Typically in the literature, the uncertainty of the calculations is not taken into account and only the central values of the results are considered. In this work we take into account the theoretical uncertainty  and show that in some cases this drastically weakens the constraints  on possible CPV from BSM physics.

\subsection{Experimental status and prospects}
We briefly summarize the current experimental status and the outlook on future possibilities. At the moment, the strongest constraints have been set on the EDMs of the neutron \cite{Baker:2006ts}, $d_n$, the ${}^{199}$Hg atom \cite{Griffith:2009zz}, $\dHg$, the ${}^{129}$Xe atom \cite{PhysRevLett.86.22}, $\dXe$, and on the energy shift indicating $T$ violation in the ThO molecule \cite{Baron:2013eja}. As discussed below, for our purposes the latter can be interpreted as a constraint on the EDM of the electron, $d_e$. Recently, a first measurement of the EDM of the atom ${}^{225}$Ra, $\dRa$, has been reported \cite{Parker:2015yka}, but the experiment is not precise  enough to impact the constraints discussed below. 

\begin{table}[t]
\begin{center}\small
\begin{tabular}{||c|cccccc||}
\hline
&$d_e$ & $d_n$ &$d_{p,D}$ & $\dHg$ & $\dXe$ & $\dRa$\\
\hline
\rule{0pt}{3ex}
current limit &$8.7 \cdot 10^{-29} $  &$ 2.9 \cdot 10^{-26}$  & x  & $2.6 \cdot 10^{-29}$  & $5.5 \cdot 10^{-27}$   & $4.2\cdot 10^{-22}$ \\
expected limit &$5.0 \cdot 10^{-30}  $  &$ 1.0 \cdot 10^{-28} $ &$ 1.0 \cdot 10^{-29}$  & $1.0 \cdot 10^{-29}$ & $5.0 \cdot 10^{-29}$ & $1.0 \cdot 10^{-27}$\\
\hline
\end{tabular}
\end{center}
\caption{\small Current and expected EDM constraints ($90\%$ confidence level) in units of e cm.}
\label{tableEDMconstraints}
\end{table}

The outlook of EDM experiments is very positive. Measurements on $d_n$ and $d_e$ are expected to improve by one to two orders of magnitude, while the limits on $\dXe$ and $\dRa$ will be improved by several orders of magnitude. On the longer time-scale, experiments are being developed to measure the EDMs of light nuclei (proton and deuteron and perhaps helion) in electromagnetic storage rings \cite{Pretz:2013us, Eversmann:2015jnk}. These experiments have a projected sensitivity of  $10^{-29}$ e cm.   
In Table~\ref{tableEDMconstraints}  we summarize the current limits  and expected sensitivities for a variety of EDMs. 
The future sensitivities are meant to be only indicative at this stage  (see \cite{Kumar:2013qya,Chupp:2014gka} and references therein).

\subsection{Theoretical interpretation} 

\label{Sec:interpretation}

\subsubsection{Nucleon EDMs}

\begin{table}[t]
\begin{center}
\scriptsize
\begin{tabular}{||c|ccc|ccc| c||}
\hline
& $d_u(1\,\mathrm{GeV})$  & $d_d(1\,\mathrm{GeV})$  & $d_s(1\,\mathrm{GeV})$ & $e\,\tilde d_u(1\,\mathrm{GeV})$ & $e\,\tilde d_d(1\,\mathrm{GeV})$ & $e\,\tilde d_s(1\,\mathrm{GeV})$ & $e\,d_W(1\,\mathrm{GeV})$  \\
\hline
\rule{0pt}{3ex}
$d_n$ & $-0.22\pm0.03 $ & $ 0.74\pm0.07 $  &$0.0077\pm0.01$ & $-0.55\pm0.28$&$-1.1\pm0.55\,$ & xxx & $\pm(50\pm40)$ \,MeV \\
$d_p$ & $0.74\pm0.07 $ & $ -0.22\pm0.03 $  &$0.0077\pm0.01$ & $1.30\pm0.65$&$0.60\pm0.30\,$ & xxx & $\,\mp(50\pm40)$ \, MeV \\
\hline
\end{tabular}
\end{center}
\caption{\small Central values and ranges of nucleon-EDM matrix elements.}
\label{tableNucleon}
\end{table}

For the dimension-6 operators under investigation the low-energy operators relevant for hadronic and nuclear EDMs are the light quark (C)EDMs and the Weinberg operator. Theoretically, by far the best understood operators are the quark EDMs whose contributions to the nucleon EDMs have been recently calculated with lattice-QCD techniques \cite{Bhattacharya:2015esa, Bhattacharya:2015wna}, see Table~\ref{tableNucleon}. The uncertainties on the up and down qEDM contributions are $10\%-15\%$, whereas the strange qEDM contribution is consistent with zero and thus highly uncertain. Although the matrix element is smaller than for the up and down qEDMs, the Wilson coefficients typically scale with the quark mass which means that the largest uncertainty arises from the strange EDM contribution.

Unfortunately no  lattice-QCD calculations exist for the qCEDM contributions (see Ref.~\cite{Bhattacharya:2015rsa} for preliminary steps towards such a calculation). 
Instead, the most-used results are obtained with QCD sum rules \cite{Pospelov_qCEDM, Pospelov_deuteron, Pospelov_review, Hisano1} which are consistent with a chiral perturbation theory ($\chi$PT) calculation combined with naive dimensional analysis (NDA)\cite{deVries2010a}. The matrix elements  are shown in Table~\ref{tableNucleon}, where it must be stressed that these results apply only if a Peccei-Quinn (PQ) mechanism is invoked to remove the QCD $\bar\theta$ term \cite{Peccei:1977hh}.~\footnote{Note that  in the presence of BSM sources of CP violation, the PQ mechanism relaxes $\bar \theta$ to a finite value $\bar{\theta}_{\rm ind}$ (induced theta-term),  proportional  to the coefficient of the new physics operator. Currently the effect of $\bar{\theta}_{\rm ind}$ is taken into account within the QCD sum rule approach.  Progress  in lattice-QCD evaluations of $d_{n} (\bar \theta)$ 
(for recent results see~\cite{Shintani:2008nt,Guo:2015tla,Shindler:2015aqa})  
 will also improve the contribution to the nucleon EDM proportional to $\bar{\theta}_{\rm  ind}$ .}
The uncertainty is estimated to be significant, $\mathcal O(50\%)$, for the light qCEDM contributions. More problematic is the dependence of the nucleon EDMs on the strange CEDM. Typically, in  the PQ scenario, the contribution from the strange CEDM is taken to vanish \cite{Pospelov_qCEDM, Hisano1}. However, a recent calculation based on $SU(3)$ $\chi$PT found a much larger dependence~\cite{Hisano2}. Here we assume no dependence on the strange qCEDM, but stress that this issue has not been resolved.~\footnote{If the PQ mechanism is not invoked, but the strong CP problem is solved via other ways, for example via extreme fine-tuning, then the matrix elements of the up and down CEDMs shift by $O(1)$ factors \cite{Pospelov_qCEDM}, while the strange CEDM matrix elements are not expected to vanish. 
In this work we do not pursue this scenario, and assume that  the PQ mechanism is at work.}

The least is known about the Weinberg operator. No systematic calculation exists and we must rely on estimates. An estimate based on QCD sum rules \cite{Pospelov_Weinberg} gives a somewhat smaller estimate than NDA \cite{Weinberg:1989dx}. Here we take a range, see Table \ref{tableNucleon}, which covers both estimates and also vary the sign of the matrix element. In principle, the matrix elements of $d_n$ and $d_p$ have an independent sign and magnitude. However, because the coupling to photons goes via the electromagnetic quark current, 
assuming that the (larger) isovector component dominates 
we take as the benchmark case that $d_p$ has a relative sign with respect to $d_n$, but we vary their magnitude independently. 
We comment later on the importance of fixing the relative sign.

\subsubsection{EDMs of light nuclei}

EDM of light nuclei receive two main contributions. The one-body component is determined by the EDM of the constituent nucleons, $d_n$ and $d_p$.
A second contribution  is due to modifications to the nuclear wavefunction induced by the CPV nucleon-nucleon potential.
For the operators under consideration, an analysis based on chiral EFT indicates that the CPV potential is dominated by  two CPV pion-nucleon  interactions\footnote{For the Weinberg operator, important contributions can arise from CPV nucleon-nucleon interactions, but these vanish for $d_D$ \cite{deVries2011a,Jan_2013} which is the main focus here.} \cite{deVries:2012ab,Bsaisou:2014oka}
\begin{equation}
\mathcal L = \bar g_0\,\Nb \vec\pi \cdot \vec\tau N + \bar g_1\,\Nb \pi_3 N\,\,\,,
\end{equation}
where $N = (p\,\,n)^T$ is the nucleon doublet, $\vec \pi$ the pion triplet, $\vec \tau$ the Pauli matrices, and $\bar g_{0,1}$ two low-energy constants (LECs). Because the quark EDMs contain an explicit photon, their contribution to $\bar g_{0,1}$ is suppressed by $\alpha_{\mathrm{em}}/\pi$ and can therefore be neglected. The Weinberg operator is chiral invariant  and therefore its contribution to $\bar g_{0,1}$ is suppressed by $\mpi^2/\Lambda_\chi^2$ where $\Lambda_\chi \sim 1$ GeV \cite{Pospelov_deuteron,deVries2011a}. Nevertheless, power counting indicates that nuclear EDMs can still significantly depend on $\bar g_{0,1}$ induced by the Weinberg operator \cite{deVries2011a,Bsaisou:2014zwa}, but explicit calculations show that the largest contributions arise from the constituent nucleon EDMs \cite{deVries2011b,Dmitriev:2003sc}. We will therefore neglect $\bar g_{0,1}$ from the Weinberg operator. That leaves us with the quark CEDMs, that do induce large values of $\bar g_{0,1}$ as indicated by QCD sum rules \cite{Pospelov_piN}
\begin{eqnarray}
\bar g_0 = (5\pm 10)(\tilde d_u + \tilde d_d)\, \mathrm{fm}^{-1}\,\,\,,\qquad
\bar g_1 = (20^{+40}_{-10})(\tilde d_u - \tilde d_d)\,\, \mathrm{fm}^{-1}
\,\,\,.
\end{eqnarray}
These values are consistent with an $SU(2)$ $\chi$PT analysis \cite{deVries2010a}. 

So far, no EDM measurements have been performed on charged particles. As we show in this work, measurements on different systems are crucial to isolate or constrain possible new physics, and we therefore investigate the potential impact of these measurements. The EDM of the deuteron is given by \cite{LiuTimmermans,deVries2011b,Jan_2013}
\begin{eqnarray}
d_{D} &=&
(0.94\pm0.01)(d_n + d_p) + \bigl [ (0.18 \pm 0.02) \,\bar g_1\bigr] \,e \,{\rm fm} \,\,,
 \label{eq:h2edm} 
\end{eqnarray}
where the small uncertainties are taken from Ref.~\cite{Bsaisou:2014zwa}. The ${}^3$He EDM has been analyzed within the same framework \cite{deVries2011b,Bsaisou:2014zwa} and depends on $\bar g_0$ as well. In addition, it depends on CPV nucleon-nucleon interactions induced by the Weinberg operator. We do not consider a ${}^3$He EDM measurement in what follows.

\subsubsection{Atomic EDMs}

Next we focus on diamagnetic atoms. Schiff's theorem \cite{Schiff:1963zz} tells us that the EDM of a point-like nucleus is screened by the electron-cloud, ensuring that the total atomic EDM vanishes. However, in heavy diamagnetic atoms the conditions for Schiff's theorem are violated by the finite size of the nucleus. For the operators we consider, the dominant contributions\footnote{In principle, important contributions could arise from CPV electron-quark interactions, but these are highly suppressed for the dimension-6 operators under investigation. } to diamagnetic atomic EDMs then arises from the nuclear Schiff moment $S_A$ \cite{Dzuba:2012bh}. The atomic EDM can then be written in term of an atomic screening factor $\mathcal A_A$ times $S_A$. The latter can be expressed as a function of $\bar g_{0,1}$ and the nucleon EDMs $d_{n,p}$:
\begin{eqnarray}
d_A &=& \mathcal A_A\, S_A\, \\
S_A &=&   \left( a_0\, \bar g_0 +  a_1\, \bar g_1  \right) e\, \mathrm{fm}^3  \ + \   \left( \alpha_n  \, d_n  + \alpha_p  \, d_p \right) \mathrm{fm}^2 \,\,\, . 
\end{eqnarray}
Whereas the atomic uncertainties are rather minor ($\mathcal O(20\%)$, see Refs.~\cite{Dzuba:2009kn,Singh:2014jca,Singh:2015aba}), the dependence of $S_A$ on $\bar g_{0,1}$ is far more uncertain due to the complicated nuclear many-body problem \cite{deJesus:2005nb,Ban:2010ea} (for a detailed discussion, see Ref.~\cite{Engel:2013lsa}). We give the best value and range for $\mathcal A$ and $a_{0,1}$ in Table~\ref{tableEngel}. In addition, the nuclear Schiff moments depend on the constituent nucleon EDMs. As far as we are aware, this has only been calculated for $\dHg$ \cite{Dmitriev:2003sc}, with the result
$\alpha_n = 1.9 (1)$ and $ \alpha_p = 0.20 (6)$. 
We neglect possible contributions from CPV short-range nucleon-nucleon interactions but stress that this assumption is untested.

\begin{table}[t]

\begin{center}\small
\begin{tabular}{||c|c|cc|cc||}
\hline
& Atomic screening& Best values &\hspace{-7mm} of $a_{0,1}$ & Estimated ranges &\hspace{-22mm} of $a_{0,1}$  \\
&$\mathcal A (\mathrm{fm}^{-2})$&  $a_0$ & $a_1$  &  $a_0$ & $a_1$  \\
\hline
 ${}^{129}\mathrm{Xe}$ &$(0.33\pm0.05)\cdot 10^{-4} $  &   $-0.10$   &$-0.076$ & $\{-0.063,\, -0.63\}$ & $\{-0.038,\, -0.63\}$ \\
${}^{199}\mathrm{Hg}$ &$-(2.8\pm0.6)\cdot 10^{-4}$ & $0.13$  & $\pm0.25$& $\{0.063,\, 0.63\} $&  $\{-0.38,\, 1.14\} $\\
${}^{225}\mathrm{Ra}$ & $-(7.7\pm0.8)\cdot 10^{-4}$&  $-19$   &$76$  & $\{-12.6,\, -76\}$ & $\{51,\,303 \}$\\
\hline
\end{tabular}
\end{center}
\caption{\small Central values and ranges of atomic and nuclear matrix elements according to Refs.~\cite{Dzuba:2009kn,Engel:2013lsa}.}
\label{tableEngel}
\end{table}

Finally, we discuss the constraint on the electron EDM. For the operator set discussed in this paper, there appear no significant contributions to CPV electron-quark interactions\footnote{A tree-level Higgs exchange involving $\mathrm{Im}\,Y_{u,d}'$ induces a contribution to $\bar q\,i\gamma^5 q\,\bar e e$ that is suppressed by the electron Yukawa coupling.}, such that paramagnetic EDMs are dominated by $d_e$.  The strongest constraint then arises from the ThO measurement which gives \cite{Baron:2013eja}
\begin{eqnarray}
d_e &\leq&8.7\cdot 10^{-29}\,e \,\mathrm{cm} \,\,\,,
\end{eqnarray}
at $90\%$ confidence level (c.l.).
The conversion of the ThO measurement into a bound on $d_e$ entails a theoretical uncertainty from atomic and molecular dynamics, 
estimated at the $15\%$ level \cite{Skripnikov,Fleig:2014uaa}. Since this is  substantially below the hadronic and nuclear uncertainties, 
we neglect it in our analysis.

\subsection{Analysis strategy: central, conservative, and minimized bounds}
\label{sec:strategies}

In most of the existing literature, when discussing EDM constraints on BSM physics, the theoretical uncertainty of the hadronic and nuclear matrix elements is not taken into account. Bounds are obtained by considering the central values given in the previous section, leading to strong constraints  on many BSM models. In this work, we investigate how the constraints are softened if we do consider the range of the matrix elements. To do so, we present bounds obtained by three different choices of matrix elements:
\begin{enumerate}
\item \textbf{Central:} Here we take the central value of the hadronic and nuclear matrix elements. This is the usual method of deriving EDM constraints on BSM physics.
\item \textbf{Conservative:} In this case we minimize the absolute value of each hadronic and nuclear matrix elements within their given range. For example, in case of the qCEDMs we take $d_n = -0.27\,e\,\tilde d_u -0.55\,e\,\tilde d_d$. For ranges which include zero, such as the dependence of $d_n$ on the strange qEDM or the dependence of $d_{\mathrm{Hg}}$ on $\bar g_1$, we set, somewhat arbitrarily,
the matrix elements to one tenth of the central value. For example, $d_n=0.0008\,d_s$. 
\item \textbf{Minimized:} Here we vary the matrix elements within  their allowed range 
assuming a flat distribution,  and minimize 
the total $\chi^2$ of the set of EDM experiments. This method corresponds to the Range-fit (Rfit) procedure defined in Ref.~\cite{Charles:2004jd}. It always gives the weakest constraint of the three methods discussed in this work as it allows for cancellations between different contributions. 
This approach gives the most conservative (perhaps over-conservative, but realistic) constraints.
\end{enumerate}
For matrix elements with an uncertain sign, such as the dependence of $d_n$ and $d_p$ on $d_W$, we calculate the bounds for all permutations of the signs and present the most conservative one. 

\begin{table}[t]
\centering
\footnotesize
\begin{tabular}{||c|cccccccc||}
\hline
  &$v^2 \mathrm{Im}\, Y'_u$ & $v^2 \mathrm{Im}\, Y_d' $  &$v^2 \mathrm{Im}\, Y'_c$ & $v^2 \mathrm{Im}\,Y'_s$ & $v^2 \mathrm{Im}\, Y'_t$& $v^2 \mathrm{Im}\, Y'_b$ & $v^2\, \theta^\prime$& $v^2 \tilde d_t/m_t$\\
\hline
 $d_e$ &x & x & $0.022$& $0.42$& $7.8\cdot 10^{-3}$& $0.041$ & x & $0.12$ \\
\hline
\rule{0pt}{3ex}
$d_n$ Cen. & $1.6\cdot 10^{-6}$ & $8.1\cdot 10^{-7}$  &$1.2\cdot 10^{-3}$ & $5.1\cdot 10^{-4}$ &$0.047$&$9.3\cdot 10^{-3}$&$0.056$&$1.1\cdot 10^{-3}$ \\
$d_n$ Con. & $2.8\cdot 10^{-6}$ & $1.4\cdot 10^{-6}$  &$6.1\cdot 10^{-3}$ & $5.1\cdot 10^{-3}$ &$0.084$&$0.068$&$0.089$&$7.1\cdot 10^{-3}$ \\
$d_n$ Min. &  $2.8\cdot 10^{-6}$ & $1.5\cdot 10^{-6}$  &$6.6\cdot 10^{-3}$ & x &$0.11$&x&$0.23$&$4.7\cdot 10^{-2}$ \\
\hline
$d_{\mathrm{Hg}}$ Cen. & $4.0\cdot 10^{-7}$ & $3.2\cdot 10^{-7}$  &$2.3\cdot 10^{-3}$ & $7.7\cdot 10^{-4}$ &$0.036$&$0.023$&$0.041$&$2.5\cdot 10^{-3}$ \\
$d_{\mathrm{Hg}}$ Con. & $1.6\cdot 10^{-5}$ & $2.9\cdot 10^{-6}$  &$0.015$ & $0.011$ &$0.19$&$0.18$&$0.21$&$0.018$ \\
$d_{\mathrm{Hg}}$ Min. & x & x  &x & x&x&x & x & x\\
\hline
Comb. Cen. & $3.9\cdot 10^{-7}$ & $3.0\cdot 10^{-7}$  &$1.1\cdot 10^{-3}$ & $4.3\cdot 10^{-4}$ &$7.6\cdot 10^{-3}$&$8.4\cdot 10^{-3}$&$0.033$&$1.0\cdot 10^{-3}$ \\
Comb. Con. & $2.7\cdot 10^{-6}$ & $1.3\cdot 10^{-6}$  &$5.5\cdot 10^{-3}$ & $4.6\cdot 10^{-3}$ &$7.8\cdot 10^{-3}$&$0.035$&$0.082$&$6.6\cdot 10^{-3}$ \\
Comb. Min. &  $2.8\cdot 10^{-6}$ & $1.5\cdot 10^{-6}$  &$6.3\cdot 10^{-3}$ & $0.42$ &$7.8\cdot 10^{-3}$&$0.041$&$0.23$&$4.3\cdot 10^{-2}$ \\
\hline
Future Min.&  $1.9\cdot 10^{-6}$ & $0.97\cdot 10^{-6}$  &$2.3\cdot 10^{-3}$ & $8.7\cdot 10^{-4}$ &$7.8\cdot 10^{-3}$&$0.011$&$0.052$&$1.6\cdot 10^{-3}$\\
\hline
\end{tabular}
\caption{\small $90\%$ upper bounds on the CPV couplings (at the scale $\MT=1$ TeV) due to current EDM constraints, assuming that a single operator dominates at the high scale. 
Row $1$ is the bound from $d_e$, Rows $2-4$ are bounds from the $d_n$ with the three strategies explained in the text, Rows $5-7$ are the same but using $d_{\mathrm{Hg}}$. Rows $8-10$ are bounds due to the combined EDM limits. Row $11$ shows the combined minimized bounds in case of improved matrix elements, see Sect.~\ref{future_matrix} for more details. An `x' indicates that the bound is larger than $1$.
\label{boundsY}
}
\end{table}

\subsection{Single coupling analysis}

Following the above strategies, we present the bounds on the CPV operators in Table~\ref{boundsY}. We assume here that only a single CPV coupling is turned on at the scale $\MT$. From the first line of the table, it is clear that $d_e$ is mainly sensitive to the Yukawa couplings of the heavy quarks, while it does not constrain the up- and down quark-Yukawa couplings and $\theta'$ at a significant level. Considering the excellent theoretical accuracy in case of $d_e$, we always take the central value of the matrix elements and do not consider the conservative or minimized case.

In contrast, $d_n$ and $\dHg$ obtain large contributions from $\theta'$, the light-quark Yukawa couplings, and the top CEDM. Compared to $d_e$, these EDMs are less sensitive to $\mathrm{Im}\,Y'_t$, even when using central matrix elements. With central values, $\dHg$ gives the strongest constraints on $Y_{u,d}'$, but this does not take into account the considerable theoretical uncertainties. Once these are taken into account, several bounds are changed dramatically. Moving from the central to the conservative strategy, the matrix elements  for the Weinberg operator decrease by a factor five, which is reflected in the constraints on $\tilde d_t$ and $Y_{c,b}'$ which mainly induce EDMs via $d_W$. Similarly, the constraint on $Y_s'$, which mainly induces $d_s$, is softened by a factor ten due to the uncertainty in the nucleon matrix element. Finally, the $\dHg$ constraints on $Y_{u,d}'$ are severely weakened due to the uncertain status of the nuclear matrix elements connecting $\dHg$ and $\bar g_1$.

Moving to the minimized case, we see that the bounds become softer. In most cases, the bounds from $d_n$ are only mildly effected. 
The main exceptions are the bounds on $Y_b'$ and, to lesser extent, $\tilde d_t$  for which several contributions of similar size contribute to $d_n$. These contributions can mutually cancel within the minimization strategy, such that no significant constraint on $Y_b'$ remains. The dramatic change in the constraint on $Y_s$ arises because the allowed range of the matrix element connecting $d_s$ to $d_n$ includes zero. The minimizing strategy has more severe consequences for the bounds from $\dHg$. For all operators, the uncertainties in the matrix elements are large enough to kill the constraints. This clearly reflects the additional uncertainty due to the nuclear many-body problem. Although this might sound as an extremely conservative conclusion, we show in the next section that modest theory improvements could drastically change the impact of diamagnetic measurements.

When combining the constraints from $d_n$, $d_{\mathrm{Hg}}$, and $d_e$, we obtain a significant constraint in all cases, even when using the minimizing strategy. Within the context of constraining non-standard Higgs couplings,  this shows once more the  importance of complementary EDM probes. 

Finally, we briefly discuss the dependence of the constraints on the scale of new physics, $\MT$. This dependence enters in two ways. First of all, the couplings scale as $\MT^{-2}$, such that the constraints on the dimensionless couplings are less stringent for higher values of $\MT$. Second, the value of $\MT$ affects  logarithmically the evolution to lower energies. To illustrate this effect, we show the resulting constraints assuming three values for the scale of new physics $\MT=1,\,10,\,100$ TeV in Table \ref{boundsYHigherScales}. As might be expected, these constraints differ by factors of $\Or(1)$. The constraints on the Yukawa couplings for $\MT= 10\,(100)$ TeV are strengthened by a factor $1.13$ ($1.24$) with respect to those for $\MT=1$ TeV. Similarly, the bounds on $\theta'$ and $\tilde d_t$ are scaled by a factor of $2.0$ ($2.8$) and $0.60$ ($0.72$) for $\MT = 10 \,(100)$ TeV. The evolution weakens the bound on  $\tilde d_t$ at $10$ TeV compared to $1$ TeV, while it strengthens the limits for $100$ TeV compared to $10$ TeV. This nontrivial scaling occurs because the contributions from $\tilde d_t$ to $d_W$ decrease with increasing $\MT$, while those to the quark CEDMs increase with $\MT$. In any case, the $O(1)$ factors are rather mild and from now on we present results for $\MT=1$ TeV. 

\begin{table}[t]
\begin{center}
\footnotesize
\begin{tabular}{||c|cccccccc||}
\hline
Comb. Min.  &$v^2 \mathrm{Im}\, Y'_u$ & $v^2 \mathrm{Im} \, Y'_d $  &$v^2 \mathrm{Im} \, Y'_c$ & $v^2 \mathrm{Im} \, Y'_s$ & $v^2 \mathrm{Im} \, Y'_t$& $v^2 \mathrm{Im} \, Y'_b$ & $v^2\, \theta^\prime$& $v^2 \tilde d_t/m_t$\\
\hline
$\MT=1$ TeV &  $2.8\cdot 10^{-6}$ & $1.5\cdot 10^{-6}$  &$6.3\cdot 10^{-3}$ & $0.42$ &$7.8\cdot 10^{-3}$&$0.041$&$0.23$&$4.3\cdot 10^{-2}$ \\
$\MT=10$ TeV&  $2.5\cdot 10^{-6}$ & $1.3\cdot 10^{-6}$  &$5.6\cdot 10^{-3}$ & $0.37$ &$7.0\cdot 10^{-3}$&$0.037$&$0.12$&$7.1\cdot 10^{-2}$ \\
$\MT=100$ TeV&  $2.2\cdot 10^{-6}$ & $1.2\cdot 10^{-6}$  &$5.1\cdot 10^{-3}$ & $0.33$ &$6.3\cdot 10^{-3}$&$0.033$&$0.083$&$5.9\cdot 10^{-2}$ \\
\hline
\end{tabular}
\end{center}
\caption{\small $90\%$ upper bounds on the CPV operators due to EDM constraints, assuming that a single operator dominates at the high scale. The constraints result from the minimization procedure, assuming three different values for the scale of new physics, $\MT=1,\,10,\,100$ TeV.
\label{boundsYHigherScales}
}
\end{table}
\subsubsection{
Impact of more accurate hadronic and nuclear 
matrix elements}\label{future_matrix}
It is extremely instructive  to study the impact of better theoretical control on the hadronic and nuclear matrix elements appearing in the EDM expressions. There are a number of matrix elements which have the largest uncertainty:
\begin{itemize}
\item The dependence of $d_{n,p}$ on $d_s$ and $d_W$. We investigate what happens if these matrix elements were known with $50\%$ accuracy. That is, we take $d_n=(0.008\pm0.004)d_s$ and $d_n = (50\pm 25)\, \mathrm{MeV}\, d_W$. Similarly for $d_p$, but with a relative sign on the $d_W$ element. 
\item The dependence of $d_{n,p}$ on $\tilde d_{u,d}$ has an uncertainty of $50\%$. We reduce this to $25\%$.
\item The dependence of $\bar g_{0,1}$ on $\tilde d_{u,d}$. We give this $50\%$ uncertainty, that is $\bar g_0 = (2.5\pm 1.25)(\tilde d_u + \tilde d_d)\, \mathrm{fm}^{-1}$ and $\bar g_1 = (10\pm 5)(\tilde d_u - \tilde d_d)\, \mathrm{fm}^{-1}$.
\item The dependence $S_{\mathrm{Hg}}$ on $\bar g_0$ and $\bar g_1$. We assume $50\%$ uncertainty on the central values. That is $a_0 = 0.13\pm0.065$ and $a_1 = 0.25\pm 0.125$. In the next section, we do the same for $S_{\mathrm{Xe}}$ and $S_{\mathrm{Ra}}$.
\end{itemize}
In the bottom row of Table~\ref{boundsY}, we present the bounds on the CPV operators assuming these improved matrix elements.  We see that the bounds on $Y_u^\prime$ and $Y_d^\prime$, are only slightly improved, while $Y_t^\prime$ is unaffected. The consequences for the limits on the other couplings are larger, with improvements of a factor $3$ to $25$ depending on the coupling. The bound on $Y_s^\prime$ would be improved by three orders of magnitude. 

An important observation is that once we include the improved matrix elements, the minimized constraints come close to the central values constraints. That is, a comparison of the rows ``Comb. Cen." and ``Future Min." tells us that almost all constraints only differ by a factor $2$. The exceptions are the bounds on $Y_{u,d}'$ which differ by a factor of $5$ and $3$. This indicates that once the hadronic/nuclear theory is at this level of precision, there is very little room for mutual cancellations between contributions. At this point, we can exploit the full power of the impressive experimental constraints on EDMs.

\subsubsection{Impact of improved experimental bounds and additional probes}\label{newexp}

\begin{table}[t]
\begin{center}
\scriptsize
\begin{tabular}{||c|cccccccc||}
\hline
 &$v^2 \mathrm{Im}\, Y'_u$ & $v^2 \mathrm{Im} \, Y'_d $  &$v^2 \mathrm{Im} \, Y'_c$ & $v^2 \mathrm{Im} \, Y'_s$ & $v^2 \mathrm{Im} \, Y'_t$& $v^2 \mathrm{Im} \, Y'_b$ & $v^2\, \theta^\prime$& $v^2 \tilde d_t/m_t$\\
\hline
\hline
$d_e(5\cdot 10^{-30})$&  $0.11$ & $0.23$  &$1.2 \Exp{3} $ & $2.4\Exp{2}$ &$4.5 \Exp{4} $&$2.4 \Exp{3} $&$0.44$&$6.7 \Exp{3} $ \\
$d_n(10^{-28})$&  $5.4 \Exp{9}$ & $2.8 \Exp{9}$  &$4.1 \Exp{6} $ & $1.8 \Exp{6}$ &$1.4 \Exp{4} $&$3.2 \Exp{5} $&$1.9 \Exp{4}$&$3.7 \Exp{6} $ \\
$d_p(10^{-29})$& $2.2 \Exp{10}$ & $5.5 \Exp{10}$  &$4.1 \Exp{7} $ & $1.8 \Exp{7}$ &$1.6 \Exp{5} $&$3.1 \Exp{6} $&$2.3 \Exp{5}$&$3.7 \Exp{7} $ \\
$d_D(10^{-29})$ & $5.0 \Exp{11}$ & $5.2 \Exp{11}$  &$1.3 \Exp{5}{}^\star $ & $9.3 \Exp{8}$ &$6.7 \Exp{6} $&$7.5 \Exp{6}{}^\star $&$7.3 \Exp{6}$&$1.2 \Exp{6}{}^\star $ \\
$\dHg(10^{-29})$&  $1.5 \Exp{7}$ & $1.3 \Exp{7}$  &$8.8 \Exp{4} $ & $2.9 \Exp{4}$ &$1.3 \Exp{2} $&$9.0 \Exp{3} $&$0.016$&$9.7 \Exp{4} $ \\
$\dXe(10^{-30})$&  $3.5 \Exp{7}$ & $5.1 \Exp{7}$  &$0.21^\dagger$ & x$^\dagger$& $0.11$ &$0.12^\dagger$ &$0.12$&$0.020^\dagger $ \\
$\dRa(10^{-27})$&  $1.8 \Exp{8}$ & $1.7 \Exp{8}$  &$3.8\Exp{3}{}^\dagger $ & $0.026{}^\dagger$ &$2.0\Exp{3}  $&$2.3\Exp{3}{}^\dagger$&$2.2\Exp{3}$&$3.6\Exp{4}{}^\dagger $ \\
\hline
\hline
\end{tabular}
\end{center}
\caption{ \small Sensitivity of future EDM experiments (reach shown in brackets in units of e cm)
to various anomalous couplings (at the scale $\MT=1$ TeV).  Central values are used for the matrix elements, such that the bounds do not take into account the theoretical uncertainties. Stars denote entries that are sensitive to the contribution of the Weinberg operator to the sum $d_n+d_p$, which vanishes for the chosen matrix elements. Daggers denote entries which might not be reliable because the contribution from the nucleon EDMs to $\dXe$ and $\dRa$ are not taken into account.}
\label{boundsY_central_improved}
\end{table}

We now study how the constraints would change with improved measurements of $d_e$, $d_{n}$, and $\dHg$ or with measurements of  $\dXe$ and $\dRa$ that are currently not competitive. 
The impact of improved limits on $d_{n}$, $d_{\rm Hg}$, or $d_{e}$, can be simply obtained by rescaling the bounds in Table \ref{boundsY}. In Table \ref{boundsY_central_improved} we show constraints using expected future experimental sensitivities and central values of the matrix elements. For all couplings a measurement of $d_n$ at $10^{-28}$ e cm would be more constraining than $d_e$ at $5\cdot 10^{-30}$ e cm. This observation, however, does not take into account the hadronic uncertainties which we discuss below. 

Future experiments on light-nuclear EDMs such as $d_p$ and $d_D$ can have a large impact as well. The projected $d_p$ measurement at $10^{-29}$ e cm, would be roughly $10$ times better than the $d_n$ measurement. This factor is not surprising considering that the matrix elements for $d_n$ and $d_p$ are very similar. 
For several couplings, a measurement of $d_D$ at $10^{-29}$ e cm would even be more constraining  than a $d_p$ measurement with the same accuracy. In particular, couplings such as $Y_{u,d}'$ that induce light-quark CEDMs give relatively large contributions to $d_D$. This behavior illustrates the complementarity of a $d_D$ measurement \cite{Pospelov_deuteron,deVries2011a}. On the other hand, $d_D$ is less sensitive than $d_p$ to couplings such as $Y_{c,b}'$ and $\tilde d_t$ which induce relatively large contributions to $d_W$. The central values of the matrix elements linking $d_W$ to $d_n$ and $d_p$ have opposite signs, and therefore the sum of nucleon EDMs, $d_n+d_p$, that enters in $d_D$, see Eq.~\eqref{eq:h2edm}, vanishes. This conclusion strongly depends on the relative sign of the nucleon matrix elements which is highly uncertain. 

Moving on to the diamagnetic atoms, we see that the prospected $\dRa$ measurement would be the most constraining measurement of the three with respect to $Y_{u,d}'$, but less sensitive than planned $d_{n,p,D}$ experiments. Of the diamagnetic atoms, $Y_{c,s}'$ are mostly constrained by $\dHg$. The reason for the lesser sensitivity of $\dXe$ and $\dRa$ is that for these diamagnetic atoms, the dependence on the constituent nucleon EDMs is not known and has not been included. 
The constraints on $Y_{c,s}'$ and, to lesser extent, $Y_{b}'$ and $\tilde d_t$ from $\dXe$ and $\dRa$ are therefore not very trustworthy. Even if these 
missing matrix elements were included, the sensitivities to these couplings would most likely still be reduced with respect to direct measurement of $d_{n}$ and $d_p$ due to the atomic screening factors. 

In Table~\ref{boundsY_improved_XeRa} we perform the same analysis using the minimization strategy. In the first two rows, we repeat the combined constraints with current and future matrix elements. In the next two rows, we include the expected increase in sensitivity of future $d_n$ and $d_e$ experiments. 
In rows $5$ and $6$ we add the prospected $\dXe$ and $\dRa$ measurements with and without improved matrix elements. From row $5$ we see that  $\dXe$ and $\dRa$ mainly improve the constraints on  $\mathrm{Im}\,Y_{u,d}'$, but at most a factor $4$. From the comparison between rows $2$ and $5$, we conclude that theory improvements can have as much impact as additional experimental probes.
 Once improved matrix elements are added, measurements of $\dXe$ and $\dRa$ improve the constraints on $\mathrm{Im}\,Y_{u,d}'$ by roughly an order of magnitude and on $\theta'$ by a factor of $5$ over current constraints with improved matrix elements (\textit{i.e.} row 2 of the table). This observation reflects that improvements in experiments and theory must go hand in hand.

In the last two rows we study the impact of $d_p$ and $d_D$ measurements. Due to the very high accuracy ($10^{-29}$ e cm) the bounds are strongly improved over the current constraints, but it must be said that these experiments have a longer time-scale. Interestingly,  the bounds on $\textrm{Im}\, Y^\prime_{s,b}$ are dramatically improved if more accurate matrix elements are used, once more underscoring the strong impact of 
hadronic and nuclear theory.

\begin{table}[t]
\begin{center}
\tiny
\begin{tabular}{||c|cccccccc||}
\hline
 &$v^2 \mathrm{Im}\,Y'_u$ & $v^2 \mathrm{Im} \,Y'_d $  &$v^2 \mathrm{Im} \,Y'_c$ & $v^2 \mathrm{Im} \,Y'_s$ & $v^2 \mathrm{Im} \,Y'_t$& $v^2 \mathrm{Im} \,Y'_b$ & $v^2\, \theta^\prime$& $v^2 \tilde d_t/m_t$\\
\hline
\hline
Current&  $2.8 \Exp{6}$ & $1.5 \Exp{6} $  &$6.3 \Exp{3} $ & $0.42$ &$7.8 \Exp{3} $&$0.041$&$0.23$&$0.043$ \\
Current+Th.&  $1.9\cdot 10^{-6}$ & $9.7\cdot 10^{-7}$  &$2.3\cdot 10^{-3}$ & $8.7\cdot 10^{-4}$ &$7.8\cdot 10^{-3}$&$0.011$&$0.052$&$1.6\cdot 10^{-3}$\\
\hline
$d_{n}+d_{\rm ThO}$&  $9.5\cdot 10^{-9}$ & $5.1\cdot 10^{-9}$  &$2.3\cdot 10^{-5}$ & $0.024{}$ &$2.9\cdot 10^{-4}$&$2.4 \cdot 10^{-3}$&$8.0 \Exp{4}$&$1.6\cdot 10^{-4}$\\
$d_{n}+d_{\rm ThO}$+Th.&  $7.0\cdot 10^{-9}$ & $3.6\cdot 10^{-9}$ &$8.4\cdot 10^{-6}$ & $3.5\Exp{6}$ &$1.7\cdot 10^{-4}$&$8.9 \cdot 10^{-5}$&$3.3 \Exp{4}$&$9.4\cdot 10^{-6}$\\\hline
$d_{\rm Xe}+d_{\rm Ra}$&  $1.3\cdot 10^{-6}$ & $3.4\cdot 10^{-7}$  &$6.3\cdot 10^{-3}{}^\dagger$ & $0.41{}^\dagger$ &$7.8\cdot 10^{-3}$&$0.040{}^\dagger$&$0.14$&$0.023{}^\dagger$\\
$d_{\rm Xe}+d_{\rm Ra}$+Th.&  $1.6\cdot 10^{-7}$ & $8.8\cdot 10^{-8}$  &$2.2\cdot 10^{-3}{}^\dagger$ & $8.7\cdot 10^{-4}{}^\dagger$ &$6.1\cdot 10^{-3}$&$0.010{}^\dagger$&$0.011$&$1.5\cdot 10^{-3}{}^\dagger$\\
\hline
$d_{p}+d_D$& $1.9\cdot 10^{-10}$ & $2.1\cdot 10^{-10}$  &$2.2\cdot 10^{-6}{}^\star$ & $0.13$ &$2.3 \cdot 10^{-5}{}$&$0.014{}^\star$&$3.1\cdot 10^{-5}$&$7.5\cdot 10^{-6}{}^\star$\\
$d_{p}+d_D$+Th.& $1.5\cdot 10^{-10}$ & $1.8\cdot 10^{-10}$  &$8.4\cdot 10^{-7}{}^\star$ & $1.7 \cdot 10^{-7}$ &$1.8 \cdot 10^{-5}$&$8.2 \cdot 10^{-6}{}^\star$&$2.2\cdot 10^{-5}{}$&$8.9\cdot 10^{-7}{}^\star$\\
\hline
\end{tabular}
\end{center}
\caption{ \small The first two rows denotes combined minimized constraints with current and improved matrix elements. Rows $3$ and $4$ are similar but for future $d_n$ and ThO measurements. Rows $5$ and $6$ do the same but now for future measurements of $\dXe$ and $\dRa$, while Rows $7$ and $8$  include $d_p$ and $d_D$ measurements. For explanation of asterisks and daggers, see caption of Table~\ref{boundsY_central_improved}.}
\label{boundsY_improved_XeRa}
\end{table}

\section{Constraints from colliders}
\label{Sec4}

In this section we discuss  the constraints that collider observables impose 
on the couplings  $\theta^\prime$, the pseudoscalar Yukawa couplings of the light quarks, $\textrm{Im}\, Y^\prime_u$, $\ldots$, $\textrm{Im}\, Y_b^{\prime}$, and 
the top pseudoscalar Yukawa and CEDM, $\textrm{Im}\, Y_t^{\prime}$ and $\tilde d_t$. 
We focus on total production cross sections and on branching ratios, that are sensitive to the square of the coefficients of the CPV operators. Additional information could be obtained by a study of observables that depend linearly on the CPV coefficients.

The operators we study are all constrained by the Higgs signal strengths, which are observed to be compatible with the SM \cite{Khachatryan:2014jba,Aad:2015gba}. 
For a given Higgs production mechanism, $i \rightarrow h$, followed by the decay of the Higgs to the final state $f$,  
the signal strength in the presence of the dimension-6 operator $\mathcal O$ is defined as 
\begin{equation}
\mu^{\mathcal O}_{i \rightarrow h \rightarrow f} = \mu^{\mathcal O}_i\, \mu^{\mathcal O}_f =   \left(1+ \frac{ \sigma^{\mathcal O}_{i \rightarrow h} }{\sigma^{SM}_{i \rightarrow h}} \right) \frac{  1 + \frac{\Gamma^{\mathcal O}_{h \rightarrow f}}{\Gamma^{SM}_{h \rightarrow f} }  }{
1 + \frac{\Gamma^{\mathcal O}_{\textrm{tot}}}{\Gamma^{SM}_{\textrm{tot}}}
},
\end{equation}
where $\sigma^{SM}$ and $\sigma^{\mathcal O}$ are, respectively, the production cross sections in the SM and the correction induced by $\mathcal O$. 
$\Gamma^{SM, \mathcal O}_{h \rightarrow f}$ are the decay widths in the channel $f$ and $\Gamma^{SM, \mathcal O}_{\textrm{tot}}$
the Higgs total width.
In Sections \ref{thetaYuk} and \ref{topCPV} we discuss how the operators we consider affect the production and decay signal strength, $\mu_i$ and $\mu_f$, and extract bounds from the LHC Run 1 \cite{Khachatryan:2014jba,Aad:2015gba}.

In addition, we discuss the bounds on the top CEDM  from the $t \bar t$ cross section, and bounds on $\tilde d_t$ and $\textrm{Im}\, Y^\prime_t$ from the $t\bar t h$ cross section.

Earlier discussions of non-standard CPV top couplings at hadron colliders 
can be found in Refs.~\cite{DeRujula:1990db,Degrande:2012gr,Hayreter:2013kba,Choudhury:2012np,Bramante:2014gda,Buckley:2015vsa} 
(in connection with  modified $t \bar t h$ production) and in   Refs.~\cite{Bernreuther:1992be,Brandenburg:1992be,Atwood:1992vj,Bernreuther:1993hq,Choi:1997ie,Sjolin:2003ah,Martinez:2007qf,Antipin:2008zx,Gupta:2009wu,Gupta:2009eq,Choudhury:2009wd,Hioki:2009hm,HIOKI:2011xx,Kamenik:2011dk,Ibrahim:2011im,Hioki:2012vn,Biswal:2012dr,Baumgart:2012ay,Hioki:2013hva,Bernreuther:2013aga,Kobakhidze:2014gqa,Englert:2014oea,Rindani:2015vya,Gaitan:2015aia,Bernreuther:2015yna,Hayreter:2015ryk} (in connection with $t \bar t $ pair production and decay).

\begin{figure}
\includegraphics[width=16.5cm]{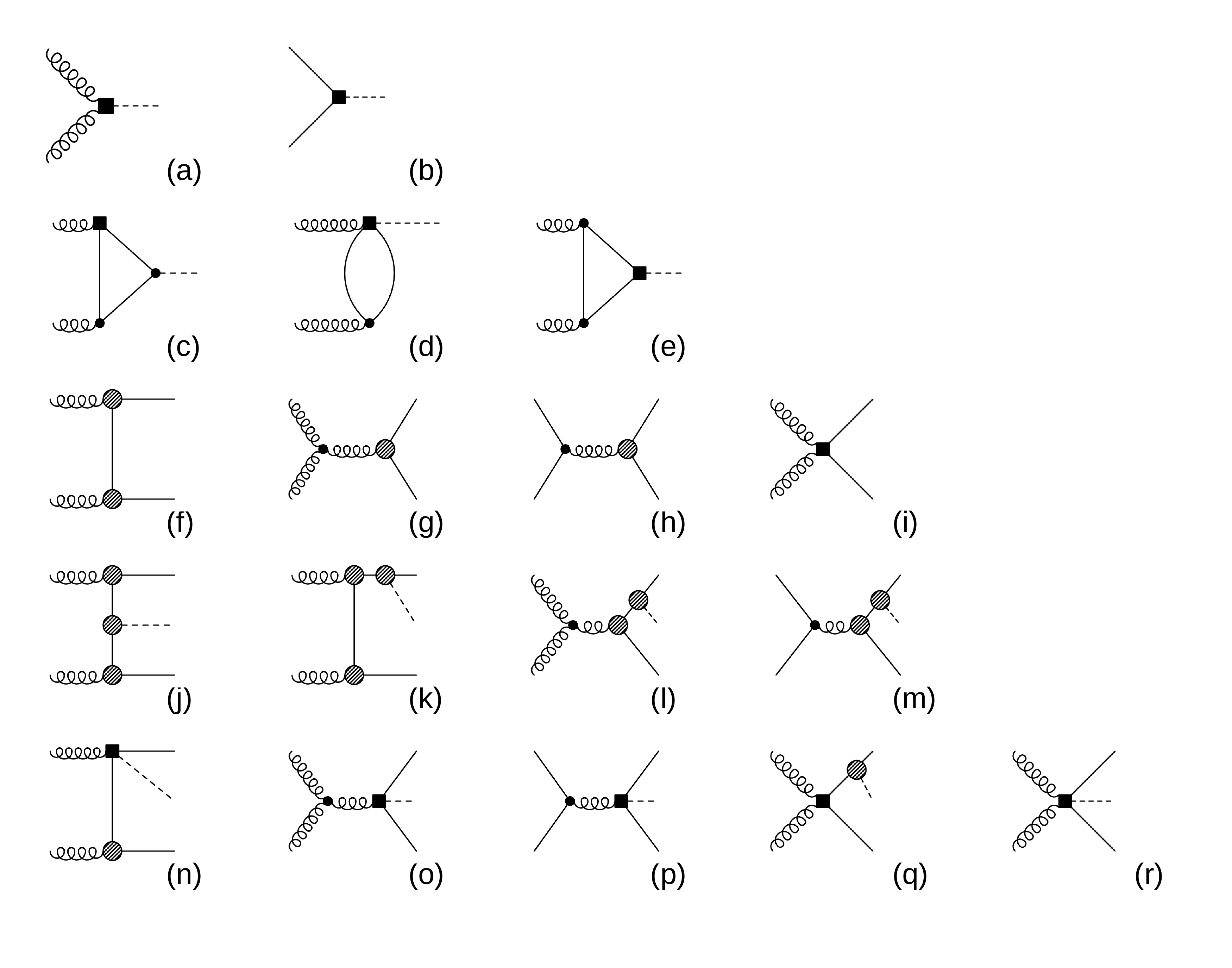}
\caption{Contributions of $\theta^\prime$, 
$\textrm{Im}\, Y_q^\prime$, $\textrm{Im}\, Y_t^\prime$ and $\tilde d_t$ 
to the Higgs production cross section, $t\bar t$ and $t\bar t h$. Solid, dashed and curly lines denote, respectively, quarks, Higgs bosons and gluons. Circles denote SM vertices, and squares insertion of the CPV operators. 
Diagrams (a) and (b) denote the LO contributions of $\theta^\prime$ and $\textrm{Im}\, Y_q^\prime$ to the Higgs production cross section.
Diagrams (c), (d) illustrate the mixing of  $\tilde d_t$ and $\theta^\prime$, while (e) the matching correction to $\theta^\prime$ induced by $\textrm{Im}\, Y_t^\prime$.
Diagrams (f)-(i) and (j)-(r) exemplify the contributions of CPV operators to the $t\bar t$ and $t\bar t h$ cross section. Here the shaded blobs  denote the sum of the SM 
top-gluon and Higgs-top vertices and insertions of $\tilde d_t$ or $\textrm{Im}\, Y^\prime_t$, while squares vertices that are originated only by $\tilde d_t$.}\label{FeynDiagCollider}
\end{figure}

\subsection{Limits on $\theta^\prime$ and $\textrm{Im}\, Y^\prime_q$ from Higgs production and decay}\label{thetaYuk}

The most important manifestations of the operator $\theta^{\prime}$ and of the non-standard Yukawa couplings of the light quarks at colliders are the modification of the Higgs production cross section
and decay width.
In the SM, the dominant mechanism of Higgs production is gluon fusion through a top loop.
The gluon fusion cross section has been computed at N${}^2$LO in $\alpha_s$  \cite{Harlander:2002wh,Anastasiou:2002yz,Ravindran:2003um}, with the inclusion of 
top quark mass effects and of electroweak corrections. Recently, the inclusion of  N${}^3$LO corrections has been completed \cite{Anastasiou:2015ema}. 
Here, for consistency with the calculation of the production cross section induced by $\theta^\prime$ and $\textrm{Im}\, Y^{\prime}_q$, we use the N${}^2$LO expression, and, at this order, for a Higgs mass of $m_h = 125$ GeV, the SM gluon fusion cross section is \cite{Heinemeyer:2013tqa}
\begin{equation}\label{eq:coll.1}
\sigma^{SM}_{ggF} = 19.2 ^{+1.4}_{-1.5} {}^{+2.3}_{-2.1} \; \textrm{pb}.
\end{equation}
The first uncertainty takes into account the effects of missing terms in the perturbative expansion of the cross section, and it is obtained
by varying the renormalization and factorization scales. The second uncertainty is the combination of the uncertainties due to the parton distribution
functions (PDFs) and  $\alpha_s$. Here and below, we always quote PDF and $\alpha_s$ uncertainties at the 90\% c.l.
Higgs production through the SM light-quark Yukawa couplings is negligible, with the exception, to some extent, of the $b$ quark \cite{Harlander:2003ai}.
Since the $b$ quark contribution to the Higgs production cross section is only a few percent of Eq. \eqref{eq:coll.1} \cite{Harlander:2003ai,Wiesemann:2014ioa}, we neglect it
in our discussion. 

The tree-level contributions of $\theta^\prime$ and the pseudoscalar Yukawa couplings
to the Higgs production cross section are shown in Fig. \ref{FeynDiagCollider}(a,b).
The cross section induced by $\theta^{\prime}$ and the pseudoscalar Yukawa couplings
has been computed at N${}^2$LO in Refs.~\cite{Anastasiou:2002wq,Harlander:2002vv} and \cite{Harlander:2003ai}, respectively.
The calculation of Higgs production through the coupling to $G \tilde G$ was performed in the framework of supersymmetric models,
where a neutral pseudoscalar Higgs boson $A$ couples to the top quark with a pseudoscalar Yukawa-type coupling. The coupling  $A G \tilde G$ is then induced by integrating out the top quark, and, in the MSSM, 
its coefficient is $v^2 \theta^{\prime}  = \cot \beta$. The calculation of Refs.~\cite{Anastasiou:2002wq,Harlander:2002vv} can be simply adapted to the CPV coupling of the scalar Higgs boson to $G \tilde G$, 
by not fixing the coupling to $\cot\beta$, and by neglecting 
higher-order corrections to the matching coefficients of the effective operators that are
specific to the model of Refs.~\cite{Anastasiou:2002wq,Harlander:2002vv}. 
The Higgs production cross section induced by scalar and pseudoscalar Yukawa couplings was computed at N${}^2$LO in Ref. \cite{Harlander:2003ai}. While Ref. \cite{Harlander:2003ai} focuses on the $b$ quark Yukawa coupling, the
results can be used for any massless quark.

At $8$ TeV, the gluon fusion cross section induced by $\theta^{\prime}$ is
\begin{equation}\label{eq:coll.2}
\sigma^{\theta^{\prime}}_{ggF} = ( 39.8^{+4.2}_{-3.4} \pm 2.6 )  \; \left( v^2 \theta^{\prime} \right)^2  \; \textrm{pb} ,
\end{equation}
where we neglected electroweak corrections. 
The first  error in Eq. \eqref{eq:coll.2} is the scale uncertainty, obtained by varying the factorization  scale $\mu$ between $m_h/2$ and $2 m_h$.
The second error is the combination of the PDF and $\alpha_s$ uncertainties.
To estimate the central value and the PDF error, we followed the recipe of the PDF4LHC working group \cite{Botje:2011sn}.
We evaluated the cross section using  three N${}^2$LO PDF sets, CT10 \cite{Lai:2010vv}, MSTW08 \cite{Martin:2009iq} and NNPDF2.3 \cite{Ball:2012cx},
following the prescriptions of each collaboration to extract the 90\% c.l.  PDF and $\alpha_s$ uncertainties.   
As our central value and PDF error we quote the midpoint and the width of the envelope provided by the central values and PDF and $\alpha_s$ errors obtained 
with these three different PDF sets.
Eq. \eqref{eq:coll.2} shows that scale variations and PDF errors have approximately the same importance.

The coupling $\theta^{\prime}$ in Eq. \eqref{eq:coll.2} is evaluated at the renormalization scale $\mu$, which we set to $\mu = 125$ GeV.
The RGE of $\theta^\prime$ is discussed in Eq. \eqref{RGEs}. 
If  only  $\theta^{\prime}$ is turned on at the scale $\MT$, 
to a very good approximation we can neglect the running of $\theta^\prime$ and interpret the coupling in Eq. \eqref{eq:coll.2} as the coupling at the  scale $M_{\slashT}$.

We summarize the Higgs production cross section induced by $\textrm{Im}\, Y_q^{\prime}$ in Table \ref{YukErrors}.
The couplings $\textrm{Im}\,Y^{\prime}_q$ are scale dependent. Since the calculation of the cross section neglects all mass effects, we also neglected the mixing of the light quark Yukawas to $\theta^\prime$ and to the light quark
CEDM. In this approximation, the RGE of $\textrm{Im}\, Y_q^{\prime}$ is diagonal, and, by an appropriate choice of scheme, can be made identical to that of the quark masses \cite{Larin:1993tq}. 
For consistency with the calculation of Ref. \cite{Harlander:2003ai}, we used the three-loop anomalous dimension  to run the couplings 
from the reference scale $\mu_0 = 1$ TeV to  the renormalization scale $\mu = m_h$.
For the $u$, $d$, and $s$ quark, once N${}^2$LO corrections are included, the uncertainty is dominated by PDF errors. For the $c$ and $b$ quarks, PDF errors and scale variations are comparable.
Notice that the PDF error is particularly large for $\textrm{Im} \, Y^\prime_s$, reflecting some issues in the determination of the strange quark PDF.

\begin{table}[t]
\center \small
\begin{tabular}{||c |  c| c c c c c||}
\hline 
& $\sigma$ $  ( 10^3 \, \textrm{pb} )  $ & $\left(v^2 \textrm{Im}\,Y^{\prime}_u\right)^2$ & $ \left(v^2 \textrm{Im} \,Y^{\prime}_d\right)^2$ & $ \left(v^2 \textrm{Im} \,Y^{\prime}_s\right)^2$ & $ \left(v^2 \textrm{Im} \,Y^{\prime}_c\right)^2$ & $ \left(v^2 \textrm{Im} \,Y^{\prime}_b\right)^2$ \\
\hline
&central                  &  100.8    & 68.2      & 20.3  &  8.3  &  2.4 \\
8 TeV &scale    	 	 &   $^{+0.9}_{-1.7}$ & $^{+0.5}_{-1.1}$       & $^{+0.3}_{-0.4}$          & $^{+0.3}_{-0.3}$      &  $^{+0.2}_{-0.1}$           \\
&pdf  \& $\alpha_s$       &   6.4    &  4.8      &  7.0  &  0.9  &  0.2 \\
\hline
& $\sigma$ $  ( 10^3 \, \textrm{pb} )  $ & $\left(v^2 \textrm{Im}\,Y^{\prime}_u\right)^2$ & $ \left(v^2 \textrm{Im}\, Y^{\prime}_d\right)^2$ & $ \left(v^2 \textrm{Im}\, Y^{\prime}_s\right)^2$ & $ \left(v^2 \textrm{Im}\, Y^{\prime}_c\right)^2$ & $ \left(v^2 \textrm{Im}\, Y^{\prime}_b\right)^2$ \\
\hline
&central                  &  175.1    & 124.5      & 46.1  &  20.7  &  6.5 \\
14 TeV &scale    	 	 &   $^{+2.6}_{-4.1}$ & $^{+1.8}_{-2.6}$       & $^{+1.0}_{-0.9}$          & $^{+0.9}_{-0.7}$      &  $^{+0.5}_{-0.3}$           \\
&pdf  \& $\alpha_s$       &   9.3    &  8.7      &  14.9  &  1.9  &  0.6 \\
\hline 
\end{tabular}
\caption{\small Higgs production cross sections induced by non-standard Yukawa couplings at 8 TeV and 14 TeV, with theoretical uncertainties.} \label{YukErrors}
\end{table}

\begin{table}[t]
\center\small
\begin{tabular}{|| c | c| c c c c c||}
\hline 
\hline 
&$\mu_{ggF} -1$   & $\left(v^2 \textrm{Im}\, Y^{\prime}_u\right)^2$ & $ \left(v^2 \textrm{Im}\, Y^{\prime}_d\right)^2$ & $ \left(v^2 \textrm{Im}\, Y^{\prime}_s\right)^2$ & $ \left(v^2 \textrm{Im} \,Y^{\prime}_c\right)^2$ & $ \left(v^2 \textrm{Im}\, Y^{\prime}_b\right)^2$ \\
\hline
&central                  &   5804    		& 3985               & 1160  &  483  &  136 \\
8 TeV &scale    	 	 &   $^{+636}_{-665}$   & $^{+425}_{-452}$   & $^{+126}_{-127}$      &  $^{+68}_{-63}$      &  $^{+27}_{-21}$    \\
&pdf  \& $\alpha_s$       &   491     		&  324               &  454  &  30  &   5 \\
\hline
& $\mu_{ggF} -1$   & $\left(v^2 \textrm{Im}\, Y^{\prime}_u\right)^2$ & $ \left(v^2 \textrm{Im} \,Y^{\prime}_d\right)^2$ & $ \left(v^2 \textrm{Im}\, Y^{\prime}_s\right)^2$ & $ \left(v^2 \textrm{Im}\, Y^{\prime}_c\right)^2$ & $ \left(v^2 \textrm{Im}\, Y^{\prime}_b\right)^2$ \\
\hline
&central                  &   3925    		& 2809               & 1015  &  463  &  145 \\
14 TeV &scale    	 	 &   $^{+402}_{-441}$   & $^{+286}_{-308}$   & $\pm 106  $    &  $^{+61}_{-55}$      &  $^{+25}_{-19}$    \\
&pdf  \& $\alpha_s$       &   261     		&  196               &  371  &  24  &   4 \\
\hline
\end{tabular}\caption{\small Production signal strength at 8 TeV and 14 TeV, with theoretical uncertainties.} \label{YukErrors14}
\end{table}

We can use the cross sections in Eq.~\eqref{eq:coll.2} and Table~\ref{YukErrors} to construct the production signal strength
$\mu_{ggF}$ appropriate for our scenario, given by  
the ratio of the single Higgs production cross section in the presence of the CPV operators in Eq.~\eqref{Dim6}
and the SM cross section.
For $\theta^{\prime}$, the signal strength is
\begin{eqnarray}\label{eq:coll3}
\mu^{\theta^\prime}_{ggF} &=& 
1 + (2.28 \pm 0.01)  \,  \left( v^2 \theta^{\prime} \right)^2,
\end{eqnarray}
where we neglected electroweak corrections to the SM and $\theta^\prime$ cross sections, and worked in the $m_t \rightarrow \infty$ limit. 
The Higgs production cross section induced by the $h G \tilde{G}$ operator is very similar to the SM cross section, that proceeds via the $h G G$ operator \cite{Anastasiou:2002wq,Harlander:2002vv}. As a consequence,
the signal strength is very close to the tree level value of $9/4$, and the PDF, $\alpha_s$
and scale errors  cancel almost completely in the ratio, leaving a negligible error on $\mu_{ggF}$. 
The production signal strengths for the pseudoscalar Yukawa couplings are summarized in Table \ref{YukErrors14}. In this case, the scale variations are almost completely determined by the error on the SM cross section, 
while PDF errors of the same size appear in both SM and $\textrm{Im} Y^\prime_q$ cross sections.

\begin{table}
\center\small
\begin{tabular}{ ||c|c c c c c ||}
\hline
 	& $\left| v^2 \textrm{Im}\, Y^{\prime}_u \right |$ & $\left| v^2 \textrm{Im}\, Y^{\prime}_d\right|$ & $\left| v^2 \textrm{Im} \,Y^{\prime}_c \right|$ & $\left| v^2 \textrm{Im}\, Y^{\prime}_s \right|$ & $\left| v^2 \textrm{Im}\, Y^{\prime}_b \right|$ \\
	\hline
ATLAS 	&  $1.2 \cdot 10^{-2}$ & $1.0 \cdot 10^{-2}$ & $0.6 \cdot 10^{-2}$  & $0.7 \cdot 10^{-2}$ & $0.5 \cdot 10^{-2}$ \\
CMS     & $1.4 \cdot 10^{-2}$ & $1.3 \cdot 10^{-2}$ & $1.0 \cdot 10^{-2}$ & $1.1 \cdot 10^{-2}$ & $1.0 \cdot 10^{-2}$ \\
\hline
future &  $0.6 \cdot 10^{-2}$ & $0.5 \cdot 10^{-2}$ & $0.4 \cdot 10^{-2}$ & $0.4 \cdot 10^{-2}$ & $0.4 \cdot 10^{-2}$ \\
\hline
\end{tabular}
\caption{\small 90$\%$   bounds on pseudoscalar Yukawa couplings (at $\mu_0 = \MT = 1$~TeV). 
The last line assumes that the 14 TeV LHC run will observe the SM value of the Higgs signal strengths. 
We assume 10\% uncertainty on the gluon fusion signal strength, with Higgs decaying in $\gamma \gamma$, $Z Z^*$ and $W W^*$,
20\%  uncertainty on the vector boson fusion signal strength, with Higgs decaying in $W W^*$, and 30\% uncertainty on the $H \rightarrow b\bar b$ signal strength \cite{CMS:2013xfa,ATL-PHYS-PUB-2014-016}.  }\label{TabY}
\end{table}

The operator $\theta^\prime$ does not significantly affect the decay channels that are relevant at the LHC, $\gamma \gamma$, $ZZ^*$, $W W^*$ and $b\bar b$.
It contributes, however, to the Higgs decay into gluons, thus affecting the total width. 
The contribution of $\theta^\prime$ to the width can be extracted from the review  \cite{Spira:1997dg}, to which we refer for references to the original calculations.
In Ref. \cite{Spira:1997dg} the decay width of a pseudoscalar Higgs boson $A$ into gluons via the $A G \tilde G$ operator was considered. As for the production cross section, 
the contribution  of the CPV coupling  $h G\tilde G$ to the width  can be obtained by replacing $\cot \beta$ with $v^2 \theta^\prime$, and sending the top mass to infinity. 
We find 
\begin{equation}
\mu^{\theta^\prime}_{\gamma \gamma, Z Z^*, W W^*} = \frac{1}{1 + 0.17 (v^2 \theta^\prime)^2}.
\end{equation}
The contribution of $\theta^\prime$ to the width is thus less than 10\% of the contribution to the production cross section. 
We can approximate $\mu^{\theta^\prime}$ with $\mu^{\theta^\prime}_{ggF}$ and use 
the production signal strengths extracted by the  ATLAS and CMS collaborations \cite{Khachatryan:2014jba,Aad:2015gba},
\begin{equation}\label{muggF}
\mu^{ATLAS}_{ggF} = 1.23^{+0.23}_{-0.20}, \qquad  \mu^{CMS}_{ggF} = 0.85^{+0.19}_{-0.16},
\end{equation}
to derive bounds.
Here, and in the rest of the section, when citing experimental results we quote the uncertainty reported in the original publication, usually at the 68\% c.l.,
and we rescale it to obtain the 90\% bounds.
From Eq. \eqref{muggF} we extract the $90\%$ bounds on $\theta^{\prime}$ to be
\begin{equation}
\left| v^2 \theta^{\prime} \right|_{ATLAS} < 0.52, \qquad  \left| v^2 \theta^{\prime} \right|_{CMS} < 0.27. 
\end{equation}

The corrections to the width are more important in the case of the pseudoscalar Yukawa couplings. 
For the $u$, $d$, $s$ and $c$ quark, the pseudoscalar Yukawa contribute mainly to the total width, while 
$\textrm{Im}\, Y_b^\prime$ also affects the $b \bar b$ decay channel.
$\textrm{Im}\, Y_b^\prime$ and $\textrm{Im}\, Y_c^\prime$ contribute to the $\gamma \gamma$ width, but the effect is negligible with respect to the correction to the total width.
The decay width is related to the decay width of a pseudoscalar $A$ boson into quarks, photons, and gluons, with appropriate replacement of the couplings.
Using the expressions of Ref. \cite{Spira:1997dg}, we obtain
\begin{eqnarray}
\mu^{\textrm{Im}\, Y^{\prime}_q}_{\gamma \gamma,\, W W^*,\, Z Z^*}  &=& \frac{1}{1 + 6068 (v^2 \textrm{Im}\, Y_q^{\prime} )^2},  \\
\mu^{\textrm{Im}\, Y^{\prime}_b}_{b \bar b}  &=& \frac{1 + 10438 (v^2 \textrm{Im}\, Y_b^{\prime} )^2 }{1 + 6068 (v^2 \textrm{Im}\, Y_b^{\prime} )^2}. 
\end{eqnarray}

We performed a fit to the signal strengths in the various production and decay channels observed by the ATLAS and CMS collaboration \cite{Khachatryan:2014jba,Aad:2015gba,Aad:2014eha,Aad:2014eva,ATLAS:2014aga,Khachatryan:2014ira,Chatrchyan:2013mxa,Chatrchyan:2013iaa,Chatrchyan:2014nva}.
In Table \ref{TabY} we show  the $90\%$ bounds we obtain by turning on one CPV coupling at a time. We find that current LHC data exclude  pseudoscalar Yukawa couplings greater than about  1$\%$. 
For the lightest quarks, $u$ and $d$, the correction to the production cross section is compensated by the dilution of the $\gamma \gamma$,
$W W^*$, and $Z Z^*$ decay channels, resulting in weaker bounds. For heavier quarks, $s$, $c$, and $b$, the smaller PDFs suppress the production cross section, and the 
most important effect is the correction to the total width.

In Table~\ref{boundsYHigherScalesColl} we show the dependence of the bound on the scale $M_{\slashT}$. In the case of the pseudoscalar Yukawa couplings, the running of 
$\textrm{Im}\, Y^{\prime}_q$ is such that the bounds get slightly stronger as the new physics scale is increased. The effect is mild, about 20\%-25\% in going from $M_{\slashT} = 1$ TeV to 
$M_{\slashT} = 100$ TeV. For $\theta^\prime$, the effect of the running is negligible.

\begin{table}[t]
\begin{center}
\footnotesize
\begin{tabular}{||c|cccccccc||}
\hline
  &$v^2 \mathrm{Im}\, Y^\prime_u$ & $v^2 \mathrm{Im}\, Y^\prime_d $  &$v^2 \mathrm{Im}\, Y^\prime_c$ & $v^2 \mathrm{Im}\, Y^\prime_s$ & $v^2 \mathrm{Im}\, Y^\prime_t$& $v^2 \mathrm{Im}\, Y^\prime_b$ & $v^2\, \theta^\prime$& $v^2 \tilde d_t/m_t$\\
\hline
$\MT=1$ TeV &  $1.2   \cdot 10^{-2}$ & $1.0\cdot 10^{-2}$  & $0.6 \cdot 10^{-2}$ & $0.7\cdot 10^{-2}$  &  $15	 \cdot 10^{-2} $   & $0.5\cdot 10^{-2}$ & 0.27  &$5.2 \cdot 10^{-2}$ \\
$\MT=10$ TeV&  $1.1   \cdot 10^{-2}$ & $0.8\cdot 10^{-2}$  & $0.5 \cdot 10^{-2}$ & $0.6\cdot 10^{-2}$ &  $14 \cdot 10^{-2}$   & $0.4\cdot 10^{-2}$  & 0.27  &$2.8\cdot 10^{-2}$ \\
$\MT=100$ TeV&  $1.0 \cdot 10^{-2}$ & $0.8\cdot 10^{-2}$  & $0.5 \cdot 10^{-2}$ & $0.5\cdot 10^{-2}$ &$13 \cdot 10^{-2}$  & $0.4\cdot 10^{-2}$  & 0.27  &$2.1\cdot 10^{-2}$ \\
\hline
\end{tabular}
\end{center}
\caption{\small $90\%$ upper bounds on the CPV operators due to single Higgs production, $t\bar t$ and $t \bar t h$ production, assuming that a single operator dominates at the high scale. The shown constraints assume three different values for the scale of new physics, $\MT=1,\,10,\,100$ TeV.
\label{boundsYHigherScalesColl}
}
\end{table}

The pseudoscalar Yukawa couplings modify another important Higgs production mechanism namely that of associated production with a $W$ or $Z$ boson.
However, we find that the associated production cross section is a factor of $10^3$ smaller than the single Higgs production cross section, yielding significantly weaker bounds on $\textrm{Im}\, Y^{\prime}_q$.

At 14 TeV, the ratio $\sigma_{ggF}^{\theta^{\prime}}/\sigma_{ggF}^{SM}$ remains substantially identical to Eq.~\eqref{eq:coll3}, due to the fact that the SM cross section, induced by the $h G G$ effective operator, and the $\theta^\prime$ cross section, induced by $h G \tilde G$, have the same scaling  
with the center-of-mass energy.
In the case of the pseudoscalar couplings to the $u$, $d$ and $s$ quarks, the  cross section grows more slowly than the gluon fusion cross section, leading to smaller corrections to the production signal strength at 14 TeV. 
For $c$ and $b$ quarks, the PDFs are obtained perturbatively, from the splitting of gluons in heavy quark pairs. Thus, the ratio of the Yukawa and gluon fusion cross section remains approximately constant at higher center-of-mass energy. 

The ATLAS and CMS collaborations have released projections on the fractional error on the signal strength \cite{CMS:2013xfa,ATL-PHYS-PUB-2014-016}. With the integrated luminosity of 300 fb$^{-1}$ that
will be reached at the LHC Run 2, the error on $\mu_{ggF}$ will be dominated  by the theoretical uncertainties on the gluon fusion cross section. After the inclusion of the recently completed N$^3$LO corrections  \cite{Anastasiou:2015ema}, these are in turn dominated by PDF and $\alpha_s$ uncertainties.
As far as the experimental errors are concerned, ATLAS projects a reduction of the error to $6\%$ at 300 fb$^{-1}$,  and $4\%$ at 3000 fb$^{-1}$. 
Assuming a central value $\mu_{ggF}=1$ and a combined theoretical and experimental  error of $10\%$, the bound on $\theta^{\prime}$ improves only slightly,  
\begin{equation}
\left| v^2 \theta^{\prime} \right| < 0.21.
\end{equation}
We obtain the projected limits on the Yukawa couplings by assuming  
a 10\% error on the gluon fusion signal strength with Higgs decaying in $\gamma \gamma$, $Z Z^*$ and $W W^*$,
a 20\% error on the vector boson fusion signal strength with  Higgs decaying in $W W^*$, and a 30\% error on
the Higgs to $b \bar b$  signal strength \cite{CMS:2013xfa,ATL-PHYS-PUB-2014-016}.
The projected limits on the Yukawa are given in Table \ref{TabY}, which shows a possible
improvement to the 0.5\% level for all the pseudoscalar Yukawa couplings.

\subsection{Limits on top CPV couplings}\label{topCPV}

The top pseudoscalar Yukawa coupling and CEDM can be probed in two ways. First of all, these couplings contribute 
to the Higgs gluon fusion production cross section at one loop. Because the dominant Higgs production mechanism in the SM also proceeds via top loops, 
we can expect it to be extremely sensitive to anomalous top-Higgs and top-gluon couplings. 
Secondly, these couplings affect processes with top quark pairs in the final state. In particular, we focus on the $t\bar t$ 
total cross section, for which the SM prediction is known very precisely, at the N${}^2$LO accuracy, and on the associated production of the Higgs and a $t\bar t$ pair ($t\bar t h$ production).

As discussed in Sec. \ref{Sec2}, $\theta^\prime$ receives a threshold correction from  $\textrm{Im}\, {Y}^\prime_t$, and mixes  with the top CEDM.
The relevant diagrams are shown in Fig. \ref{FeynDiagCollider} (c)-(e).
The bound on $\theta^\prime$ can therefore be used to constrain $\textrm{Im}\, {Y}^\prime_t$ and $\tilde{d}_t$.
The gluon fusion cross section induced by $\textrm{Im}\, {Y}^\prime_t$ is very similar to Eq. \eqref{eq:coll3}, with the only difference that $\mathcal O(\alpha_s^2)$ corrections to the 
matching coefficient of $\textrm{Im}\, {Y}^\prime_t$ to $\theta^\prime$ need to be considered, as done in Refs. \cite{Anastasiou:2002wq,Harlander:2002vv}. Their effect is to shift the signal strength in Eq. \eqref{eq:coll3} by 0.005, which has no consequence on the constraints. 
$\textrm{Im}\, {Y}^\prime_t$  modifies the $\gamma \gamma$  and $g g$ decay widths. 
In the case of the $\gamma \gamma$ branching ratio, we find that the corrections to $\gamma \gamma$ and to the total width are very similar,
and $\mu_{\gamma \gamma}$ is, accidentally, very close to one
\begin{equation}
\mu_{\gamma \gamma}^{\textrm{Im}\, Y^{\prime}_t} = \frac{1 + 0.57 (v^2  \textrm{Im}\, {Y}^\prime_t)^2}{1 + 0.58  (v^2 \textrm{Im}\, {Y}^\prime_t)^2 } \sim  1.
\end{equation}
The $WW^*$ and $Z Z^*$ branching ratios are affected by the contribution to the total width. As for $\theta^\prime$, the corrections to the decay signal strength are about  10\% of the correction
to production, and, with the current experimental accuracy, can be neglected. Thus we can extract a bound on $\textrm{Im}\, Y^{\prime}_t$ from the total gluon fusion signal strength, Eq. \eqref{muggF}, obtaining
\begin{equation}
| v^2 \textrm{Im}\, Y^{\prime}_t | < 0.27 \frac{m_t}{v} = 0.15,
\end{equation}\textsl{•}
where $\textrm{Im}\, {Y}^\prime_t$ and the top mass are evaluated at $\mu=1$ TeV. Notice that the ratio $\textrm{Im}\, Y^{\prime}_t/m_t$ is RG invariant, since scalar and pseudoscalar currents have the same anomalous dimension up to  three loops \cite{Larin:1993tq}.

The top CEDM mixes with $\theta^\prime$, with anomalous dimension given in Eq. \eqref{RGEs}. We solve the RGE and 
run $\theta^\prime$ and $\tilde d_t$ to the top threshold, where we integrate out the top quark and stop the running.
In this case, the RGE is only known at LO, and $\mathcal O(\alpha_s)$ and $\mathcal O(\alpha_s^2)$ corrections to the evolution and to the matching coefficient of $\tilde d_t$ onto $\theta^\prime$
are not known. Therefore, to put bounds on $\tilde d_t$ we use the tree-level value of the signal strength. We obtain
\begin{equation}
  \frac{v^2}{m_t} | \tilde{d}_t  |_{ATLAS} < 0.10, \quad   \frac{v^2}{m_t}  |\tilde{d}_t  |_{CMS} < 0.052.
\end{equation}
The strong limit is a consequence of the large mixing of the top CEDM and $\theta^\prime$.  
In similar fashion, the gluon fusion cross section can be used to constrain the top chromo-magnetic dipole moment through its mixing onto $h G G$. We discuss this in  more detail in the next section. Looking to the future, a $10\%$ accuracy in the measurement of $\mu_{ggF}$ would allow to slightly improve the constraint to $4\%$. 
In light of the strength of the constraints, it will be interesting to include higher-order corrections to the mixing of $\tilde d_t$ and $\theta^\prime$.

In addition,  $\tilde d_t$ and $\textrm{Im}\, {Y}^\prime_t$ contribute to processes involving the production of  top quarks.
The top CEDM affects the $t\bar t$ and $t\bar t h $ cross sections, while  $\textrm{Im}\, Y_t^{\prime}$
only  contributes to  $t  \bar t h$.

The top CEDM contributions to the $t \bar t$ cross section are shown in Fig. \ref{FeynDiagCollider} (f)-(i).
The cross section induced by the top chromomagnetic and chromoelectric dipole moments was computed in Refs. \cite{Atwood:1994vm,Haberl:1995ek},  and we consider here terms that are at most quadratic in $\tilde d_t$.
The SM $t\bar t$ cross section is known at N${}^2$LO accuracy \cite{Czakon:2013goa}. 
Combining these results, the $t \bar t$ cross section in the presence of a top CEDM at $\sqrt{S} = 8$ TeV is 
\begin{equation}
\sigma_{t\bar t} =\left(252.9^{+6.4}_{-6.0}\pm 19.2 \right)  + (1878 \pm 183) (m_t \tilde d_t)^2  \;  \textrm{pb},
\end{equation}
where the SM cross section was computed using the program TOP++ \cite{Czakon:2011xx}, and includes N${}^2$LO corrections 
and soft gluon resummation. 
The contribution of $\tilde d_t$ was computed at LO, and we included only PDF errors.
The cross section, and $\tilde d_t$, are evaluated at the renormalization scale $\mu = m_t$.
In the SM, NLO and N${}^2$LO corrections to the $t \bar t$ cross section are large \cite{Czakon:2013goa}, suggesting the need to include NLO corrections for the dipole operators as well
\cite{Franzosi:2015osa}.

Some of the Feynman diagrams showing the contribution  of the top CEDM, pseudoscalar Yukawa
and their interference to the associated production of a Higgs boson and a $t\bar t$ pair are shown in Fig. \ref{FeynDiagCollider} (j)-(r). 
We computed the  cross section at LO, retaining terms at most quadratic in the coefficients of the CPV operators.
Taking the ratio with the SM cross section, we obtain the production signal strength
\begin{eqnarray}
\mu_{t \bar t h}( 8\, \textrm{TeV}) &=& 1 + (248 \pm 24) (m_t \tilde d_t)^2 + ( 0.67 \pm 0.04 )  \left( v^2 \textrm{Im}\, Y^{\prime}_t \right)^2 
\nonumber \\
&+&   (0.41 \pm 0.54) (v^2 m_t \, \tilde d_t \textrm{Im}\, Y^{\prime}_t ),
\\
\mu_{t \bar t h}( 13\, \textrm{TeV}) &=& 1 + (379 \pm 27) (m_t \tilde d_t)^2 + ( 0.84 \pm 0.03 )  \left( v^2 \textrm{Im}\, Y^{\prime}_t \right)^2 
\nonumber \\
&+&  (2.65 \pm 0.37) (v^2 m_t \, \tilde d_t \textrm{Im}\, Y^{\prime}_t ), 
\\
\mu_{t \bar t h}( 14\, \textrm{TeV}) &=& 1 + (401 \pm 28) (m_t \tilde d_t)^2 + ( 0.86 \pm 0.03 )  \left( v^2 \textrm{Im}\, Y^{\prime}_t \right)^2 
\nonumber \\
&+&  (2.65 \pm 0.21) (v^2 m_t \, \tilde d_t \textrm{Im}\, Y^{\prime}_t ), 
\label{eq:coll4}\end{eqnarray}
where we evaluated the SM and the cross section induced by $\tilde d_t$ and $\textrm{Im} \, Y_t^\prime$ at LO in $\alpha_s$.
We  chose $m_t$ as factorization scale and Eq. \eqref{eq:coll4} is expressed in terms of couplings at $\mu = m_t$. 
We  performed the calculation with the  CT10, NNPDF2.3 and MSTW08 NLO PDF sets, and included only PDF and $\alpha_s$ error.
Corrections to the Higgs branching ratios are small and can be neglected.

The $t \bar t$ production cross section has been measured both at CMS and ATLAS. At 8 TeV \cite{Chatrchyan:2013faa,Aad:2014kva}
\begin{equation}
\sigma_{t\bar t}^{ATLAS} = 242.4 \pm 1.7 \pm 9.3 \pm 4.2  \qquad \sigma_{t\bar t}^{CMS} = 239 \pm 2 \pm 11 \pm 6,
\end{equation}
where the first uncertainty is due to statistics, the second  to systematics, and the third  to the limited knowledge of the integrated luminosity.
Current Higgs measurements can be used to infer the $t\bar t h$ signal strength, although with large uncertainties. ATLAS and CMS reported 
\begin{equation}
\mu^{ATLAS}_{t\bar t h} = 1.81 \pm 0.80, \qquad \mu^{CMS}_{t\bar t h} = 2.90^{+1.08}_{-0.94}. 
\end{equation}

We can get a  bound on $\tilde d_t$ by demanding that the BSM cross section is less than the difference between the observations and SM predictions. At the 90\% c.l., we find
\begin{equation}
\frac{v^2}{m_t} |\tilde d_t  | <  0.23,    
\end{equation}
both from ATLAS and CMS. 
This bound is in agreement with the analysis of Ref. \cite{Aguilar-Saavedra:2014iga}. 
Similar bounds can be extracted from the first 13 TeV data.

From the $t \bar t h$ signal strength, we obtain
\begin{equation}
 \frac{v^2}{m_t} |\tilde d_t  |_{ATLAS} <  0.21, \qquad   0.07 <  \frac{v^2}{m_t}  | \tilde d_t  |_{CMS} <  0.27.   
\end{equation}
Interestingly, already with current data, and notwithstanding the large uncertainties on the $t \bar t h$ cross section, the $t \bar t$ and $t \bar t h$ processes show comparable sensitivities to a top CEDM. 
The limit on the Yukawa is a factor of 10 weaker,
\begin{equation}
 \left | v^2 \textrm{Im} \,Y^{\prime}_t \right|_{ATLAS} <  1.6, \qquad   0.6 < \left| v^2 \textrm{Im}\, Y^{\prime}_t \right|_{CMS} <  2.1.  
\end{equation}

In  Table \ref{boundsYHigherScalesColl} we show the dependence of the bounds on $\textrm{Im} \, Y_t^\prime$ and $\tilde d_t$ on $M_{\slashT}$.
While the bound on $\textrm{Im} \, Y_t^\prime$ depends mildly on the new physics scale, the strong mixing of $\tilde d_t$ and $\theta^\prime$ causes the bound on $\tilde d_t$ to get stronger by a factor of 2.5 as $M_{\slashT}$
is increased to 100 TeV.

\begin{figure}
\center
\includegraphics[width=11cm]{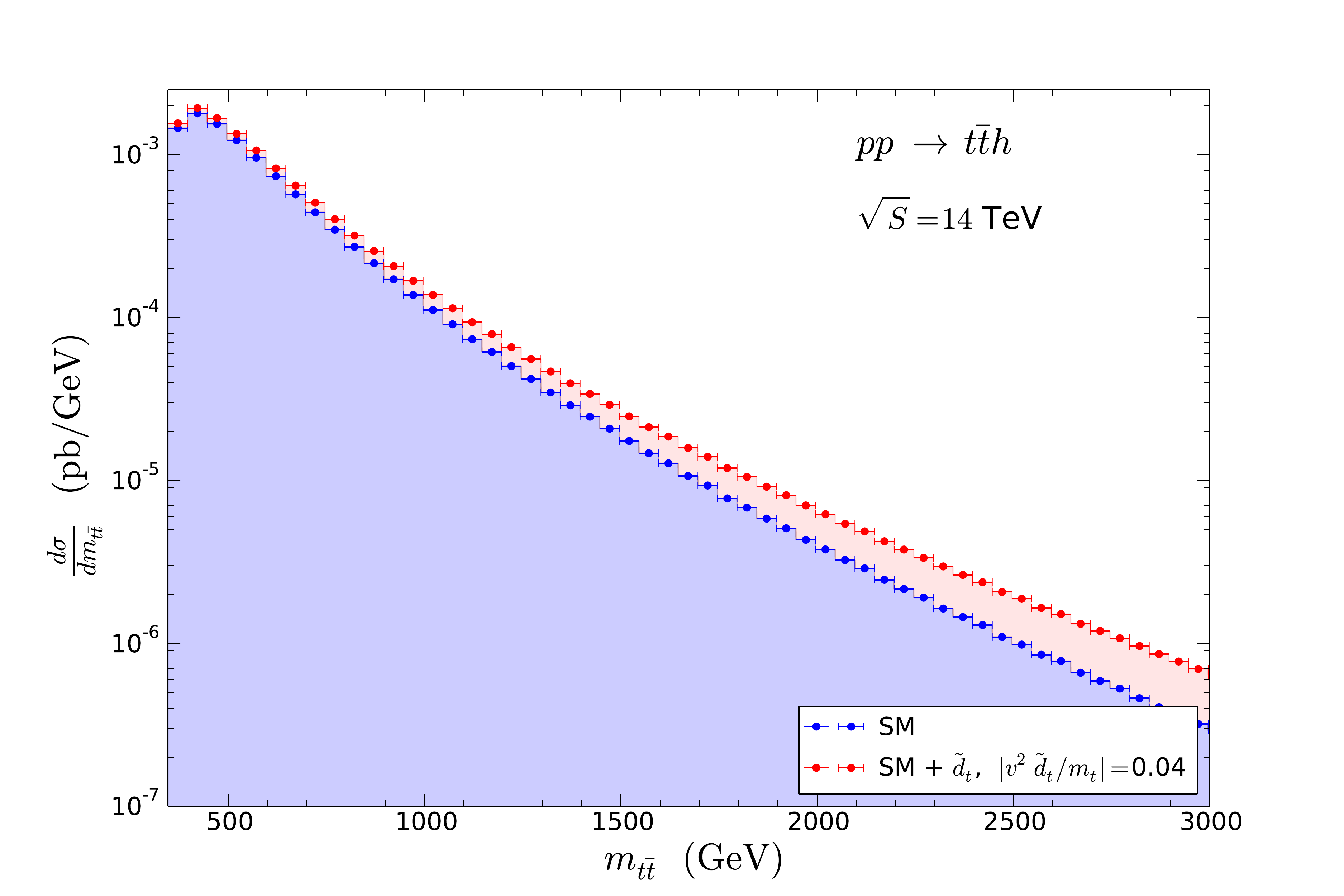}
\caption{ \small The $t\bar t h$ cross section at 14 TeV. We set the coefficient of the top CEDM to  $ v^2 \tilde{d}_t/m_t = 0.04$.}
\label{ttH}
\end{figure}

The bounds will be improved by the LHC Run 2. 
Ref.~\cite{Aguilar-Saavedra:2014iga} discusses how the bounds on the top CEDM from the $t \bar t$ cross section could be reduced to about 8\%,
in particular by studying differential distributions, and focusing on events with large $t \bar t$ invariant mass.

ATLAS and CMS project to reach a $30\%$ ($15\%$) uncertainty on the $t \bar t h$ signal strength with 300 fb$^{-1}$ ( $3000$ fb$^{-1}$) data.
Assuming an observed signal strength compatible with the SM, a $30\%$ accuracy allows to improve the bounds to better than $10\%$ level
\begin{equation}
 \frac{v^2}{m_t} |\tilde d_t | <  0.06.
\end{equation}
At 14 TeV, more information could be gained by looking at differential distributions.   
In Fig. \ref{ttH} we show the differential cross section with respect to the invariant mass of the $t\bar t$ pair, $m_{t\bar t}$, induced by the SM and the top CEDM. We work at LO, and set $\tilde d_t$ to the projected maximum value allowed by the gluon fusion cross section at 14 TeV,
$|v^2 \tilde d_t/m_t | < 0.04$.
While the total $t\bar t h$ cross section will not be able to improve this limit, at large invariant mass the contribution of the top CEDM increases, 
being $60\%$ of the SM at 1.5 TeV, and overtaking the SM for $m_{t\bar t} > 2.5$ TeV.
Thus, the study of events at large $m_{t\bar t}$ could provide a route to further improve the bound on $\tilde d_t$.  
These considerations, of course,  are valid only under the assumption that  new degrees of freedom 
generating the non-standard chromo-electric top couplings are sufficiently heavy 
so that a local operator analysis provides a good description of  the process $pp \to t \bar{t} h$.
Assuming that the top CEDM operator is generated at loop level by new particles 
with a common mass $m_*$ this criterion roughly speaking implies that $m_{t \bar{t}} + m_h \ll m_*$.
This should be kept in mind when analyzing the range of applicability of Fig. \ref{ttH}.
Since current and prospective EDM bounds on $v^2 \tilde{d}_t/m_t$ are consistent with the mass scale $m_*$ 
being in the multi-TeV range (depending on coupling strengths),  there are classes of models in which 
Fig. \ref{ttH} remains valid all the way to $m_{t \bar t} = 3$~TeV.

\begin{figure}
\center
\includegraphics[width=8cm]{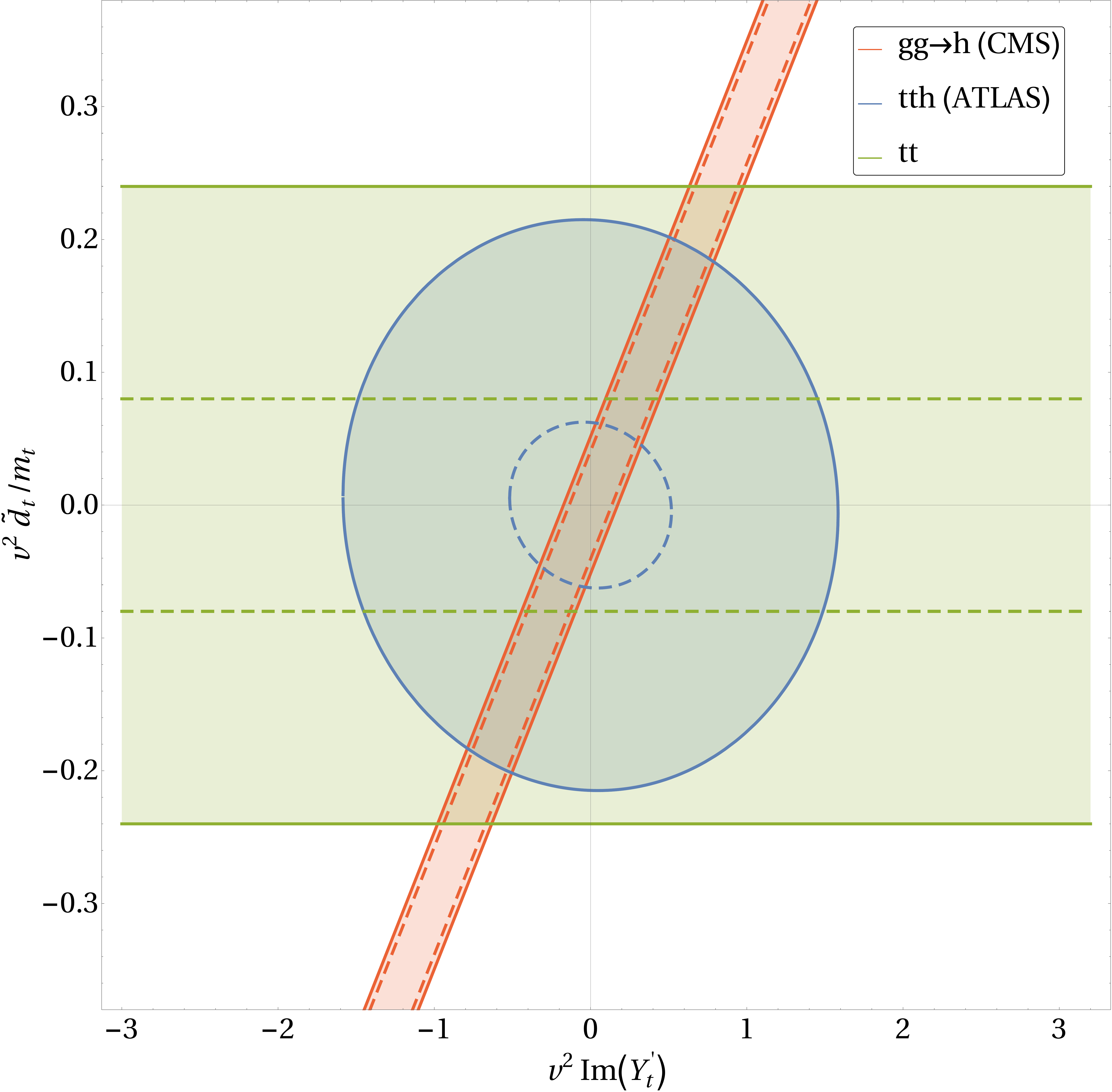}
\caption{ \small Bounds on $\tilde d_t$ and $\textrm{Im} \,Y'_t$ from gluon fusion, $t \bar t$ production, and $t \bar t h$ production.  Solid lines denote current bounds. 
For each observables we show the most stringent limit, namely the CMS limit for gluon fusion, and the ATLAS limit for $t \bar t h$. The $t \bar t$ bound is approximately the same for the two experiments.
Dashed lines denote the projected bounds from the LHC Run 2, assuming that $\mu_{ggF}=\mu_{t\bar t h}=1$ with 10\% and 30\% uncertainty, respectively. The projected $t\bar t$
bound relies on the analysis of Ref. \cite{Aguilar-Saavedra:2014iga}}.
\label{ttH_Bounds}
\end{figure}

Fig. \ref{ttH_Bounds} summarizes the limits on $\textrm{Im}\, Y^\prime_t$ and $\tilde d_t$ set by gluon fusion, $t\bar t$ and $t\bar t h$. Solid lines denote current limits, 
and we show the most stringent bounds, that is the CMS bound for gluon fusion, and the ATLAS bound for $t \bar t h$. 
Dashed lines denote the projected bounds from LHC Run 2.
An interesting feature of Fig. \ref{ttH_Bounds} is that the three observables we considered show comparable sensitivities to $\tilde d_t$ and, to a lesser extent, $\textrm{Im}\, Y^\prime_t$.
Thus, in the presence of a significant deviation from the SM in any of these three observables, it will be possible to look for signals in the remaining two, and gain more insight on the origin of the deviation. EDM constraints on this combination of couplings are discussed in Sect.~\ref{Ytdt}.

\subsection{The top chromo-magnetic dipole moment}

In Sect. \ref{topCPV} we found that the strongest collider limits on $\tilde d_t$ arise from the Higgs gluon fusion production cross section. 
Here we briefly step aside from the main focus of the paper on CPV operators, to remark that a similar observation 
applies to the CP-conserving top chromo-magnetic dipole moment (CMDM) \cite{Degrande:2012gr}.

We consider the effective Lagrangian involving the top quark and the Higgs.
\begin{equation}\label{CMDM.1}
\mathcal L_{t H} =  - \left( \frac{m_t}{v} - v^2 \textrm{Re} Y^\prime_t  \right) h \bar t t +  \frac{\alpha_s}{12 \pi} c_H v h G^a_{\mu \nu} G^{a \mu \nu} 
- \ \frac{1}{2} \tilde c_t\, g_s    \   \bar t     \sigma \cdot G    t \   \left(1+\frac{h}{v}\right).
\end{equation}
Eq. \eqref{CMDM.1} contains the SM Yukawa coupling, and three dimension-6 operators, closely related to those discussed in Eq. \eqref{Dim6}.
The first is a correction to the top Yukawa, the second is a coupling of the Higgs to the gluon field strength, and $\tilde c_t$ is the top CMDM.
In the SM, $\textrm{Re}\, Y^\prime_t$, $\tilde c_t$ and $c_H$ vanish, and the $h G G$ operator is generated below the top threshold, with coefficient $1/v^2$ in the normalization of Eq. \eqref{CMDM.1}.
Thus, the effective Lagrangian  \eqref{CMDM.1} leads to a signal strength
\begin{equation}
\mu_{ggF} = \left(1 - \frac{v}{m_t} \left(v^2 \textrm{Re} Y^\prime_t  \right)  + c_{H} v^2 \right)^2, 
\end{equation}
where this expression is valid at LO in $\alpha_s$.

The dimension-6 operators in Eq. \eqref{CMDM.1} mix, in exactly the same way as their CPV analogs. 
Defining $\vec{C}_t = (\tilde c_t/m_t, \textrm{Re} \, Y^{\prime}_t, c_H)^T$, the one-loop RGE reads
\bea
\frac{d\vec{C_t}(\mu)}{d\ln\mu}=\frac{\al_s}{4\pi}
\bma 
16C_F-4N & 0 & - 1/6\pi\sq\\
 -42 C_F\big(\frac{m_t}{v}\big)^3& -6C_F& 2 C_F\frac{\al_s}{\pi}\frac{m_t}{v}\\
 -12 \frac{4\pi}{\al_s}\big(\frac{m_t}{v}\big)\sq &  0&  0 
\ema \cdot \vec C_t(\mu) ,
\label{RGEeven}\eea
where the differences with Eq. \eqref{RGEs} stem mainly from the different normalization of $c_H$ and $\theta^\prime$.
The most important feature of Eq. \eqref{RGEeven} is the strong mixing of $c_H$ and the top CMDM.
Assuming $M_{\slashT} = 1$ TeV, and taking as boundary condition $c_H = \textrm{Re}\, Y_t^{\prime}  = 0$, we find
\begin{equation}
c_H(m_t) =  7.75  \frac{\tilde c_t (M_{\slashT})}{m_t}, \qquad \textrm{Re}\, Y_t^{\prime}(m_t) = 0.18  \frac{\tilde c_t (M_{\slashT})}{m_t}, \qquad
\frac{\tilde c_t(m_t)}{m_t}  = 0.87 \frac{\tilde c_t(M_{\slashT})}{m_t}.
\end{equation}

At LO, the signal strength induced by the top CMDM is 
\begin{equation}
\mu_{ggF} = \left( 1 + 7.6 \frac{v^2 \tilde c_t}{m_t}   \right)^2.
\end{equation}
As was the case for the top CEDM, the corrections to the Higgs width induced by the top CMDM are less important, and can be neglected.
The requirement that the signal strength is in agreement with the observed $\mu_{ggF}$
can be satisfied in two ways. The CMDM can be negative, and large enough to cancel the SM contribution to 
gluon fusion. This is achieved for a small interval around $v^2 \tilde c_t/m_t = -0.25$, a value already excluded by the LHC and Tevatron measurements of the $t\bar t$ inclusive cross section \cite{Aguilar-Saavedra:2014iga}.
The other solution is
\begin{equation}\label{CMDM.2}
 -0.007 <   \left( \frac{v^2}{m_t}  \tilde{c}_t  \right)_{ATLAS} < 0.035, \qquad  -0.031 < \left(\frac{v^2}{m_t}  \tilde{c}_t \right)_{CMS} < 0.010,
\end{equation}
where the limit is on the coefficient at the scale $M_{\slashT} = 1$ TeV.
The limits are stronger than those on $\tilde d_t$, because $\mu_{ggF}$ has a linear dependence on $\tilde c_t$, while $\tilde d_t$ only contributes quadratically.

The bounds in Eq. \eqref{CMDM.2} cut significantly into the region allowed by the $t\bar t$ cross section, $ v^2 \tilde{c}_t/m_t \in (-0.10,0.05)$,
and are already competitive with the projected bounds from the LHC Run 2, studied in Ref. \cite{Aguilar-Saavedra:2014iga}.
Furthermore, the reduction of the uncertainty on $\mu_{ggF}$ to the $10\%$ level at Run 2 would improve the limits in Eq. \eqref{CMDM.2} to  $|v^2 \tilde c_t/m_t | <  0.006$.
The Higgs gluon fusion production cross section thus appears to be the ideal place to look for anomalous top couplings.

This observation deserves two specifications. First of all, the contribution of $\tilde c_t$ to $c_H$ is generated by running between the scale $M_{\slashT}$, where we assume any BSM degrees of freedom that generate the top CMDM to be integrated out, and $m_t$. If the scale separation between $M_{\slashT}$ and $m_t$ is small, we cannot rely only on logarithmically enhanced terms, and need to consider also the finite contributions 
of $\tilde c_t$ to $c_H$
\begin{equation}\label{CMDM.3}
\delta c_H =   \frac{\tilde c_t m_t}{v^2}  \left(  3 - 3 \beta \log \left(- \frac{1 + \beta}{1-\beta}\right) + 3 \frac{m_t^2}{m_h^2} \log^2 \left(- \frac{1+\beta}{1-\beta}\right)  \right),  
\end{equation}
with $\beta^2 = 1- \frac{4 m_t^2}{m_h^2}$.
In the extreme case $M_{\slashT} = m_t$, where the RGE contribution to $c_H$ vanishes, the finite terms in Eq. \eqref{CMDM.3}
still constrain $v^2 \tilde c_t/m_t$ to be in the range $(-0.10,0.03)$, in the case of ATLAS, or $(-0.03,0.09)$ in the case of CMS.
We note, however, that if $M_{\slashT} = m_t$, our EFT approach is no longer valid and new degrees of freedom should be explicitly accounted for.

Secondly, the strong bounds in Eq. \eqref{CMDM.2} assume that only one coupling, $\tilde c_t$, is turned on at $M_{\slashT}$, so that the  contribution of $\tilde c_t$ to $\mu_{ggF}$ is not influenced by other terms. It therefore remains important to look for direct effects of $\tilde c_t$. In particular, since the sensitivity of the $t \bar t$ and $t\bar t h$ cross sections on $\tilde c_t$ is only moderately weaker than $\mu_{ggF}$, these three observables constitute ideal orthogonal probes to pin down a top CMDM.

\subsection{Summary of collider bounds}

\begin{table}[t]
\begin{center}
\footnotesize
\begin{tabular}{||c|cccccccc||}
\hline
  &$v^2 \mathrm{Im}\, Y^\prime_u$ & $v^2 \mathrm{Im}\, Y^\prime_d $  &$v^2 \mathrm{Im}\, Y^\prime_c$ & $v^2 \mathrm{Im}\, Y^\prime_s$ & $v^2 \mathrm{Im}\, Y^\prime_t$& $v^2 \mathrm{Im}\, Y^\prime_b$ & $v^2\, \theta^\prime$& $v^2 \tilde d_t/m_t$\\
\hline
Current     &  $1.2 \cdot 10^{-2}$ & $1.0 \cdot 10^{-2}$ & $0.6 \cdot 10^{-2}$  & $0.7 \cdot 10^{-2}$ &  $15 \cdot 10^{-2}$ & $0.5 \cdot 10^{-2}$ & 0.27  &$5.2 \cdot 10^{-2}$ \\
LHC Run 2  
& $0.6 \cdot 10^{-2}$ & $0.5 \cdot 10^{-2}$ & $0.4 \cdot 10^{-2}$ & $0.4 \cdot 10^{-2}$ &   $ 12\cdot 10^{-2} $   & $0.4 \cdot 10^{-2}$ & 0.21  &$4.0 \cdot 10^{-2}$ \\
\hline
\end{tabular}
\end{center}
\caption{\small Current bounds from LHC Run 1 and projected bounds from the LHC Run 2 on  
the anomalous couplings defined at $\MT=1$~TeV.
\label{boundsSummaryColl}
}
\end{table}

In Table~\ref{boundsSummaryColl}, we summarize current bounds on $\theta^\prime$, $\textrm{Im}\, Y^\prime_q$, $\textrm{Im}\, Y^{\prime}_t$ and $\tilde d_t$ 
that can be extracted from measurements of the Higgs production and decay processes, of the $t \bar t$ cross section, and of the $t \bar t h$ signal strength performed by the ATLAS and CMS collaborations 
during the LHC Run 1. For each coupling, we listed the strongest bound.  
In the second row, we summarize the projected bound from the LHC Run 2, at 14 TeV center-of-mass energy and with integrated luminosity of 300 fb$^{-1}$.  
The projected bounds are obtained by assuming $\mu_{ggF}$ and $\mu_{t\bar t h}$ to be in agreement with the SM, with uncertainties of 
 $10\%$ and $30\%$ respectively   \cite{CMS:2013xfa,ATL-PHYS-PUB-2014-016}.

The current measurements of the Higgs production cross section and branching ratios allow one to bound the pseudoscalar Yukawa couplings of the light quarks at the level of $0.5$ to $1\%$, that is, they exclude 
pseudoscalar couplings much bigger than the SM bottom Yukawa. 
The higher luminosity of the LHC Run 2, and the consequent reduction of the uncertainties on the signal strength to the 10\%-20\% level, will allow to improve these bounds, especially for lighter quarks.

A comparison with the EDM constraints in Table \ref{boundsY} shows that collider cannot compete with EDM constraints on the pseudoscalar Yukawa of the first generation. Indeed, EDM bounds
already forbid $\textrm{Im}\, Y^{\prime}_{u,d}$ larger than the SM $u$ and $d$ Yukawas. 
The EDM bound on $\textrm{Im}\, Y_c^{\prime}$ is very close to the collider bound. While the LHC Run 2 will improve the bound on $\textrm{Im}\, Y_c^{\prime}$ by at most a factor of two, 
the next generation of EDM experiments will probe this coupling at the $10^{-5}$ level, out of the reach of collider experiments. 
It is nonetheless important to pursue direct probes of this coupling, for example by studying decays of the Higgs to $c \bar c$ \cite{Delaunay:2013pja}. 
Current collider and EDM bounds on $\textrm{Im}\, Y_s^{\prime}$ and $\textrm{Im}\, Y_b^{\prime}$ are comparable, with the LHC having a slight edge. In the case of $\textrm{Im}\, Y_s^{\prime}$, EDMs are not very constraining  because of the poor knowledge of the nucleon matrix element of the strange EDM (and CEDM). A modest improvement on the theory, coupled with the next generation of EDM experiments, will put $\textrm{Im}\, Y_s^{\prime}$ out of the reach of collider experiments.
$\textrm{Im}\, Y_b^{\prime}$ is more interesting because even with improved theory, the collider and EDM bounds are of approximately the same size.
It is therefore important to get as many handles on  $\textrm{Im}\, Y^\prime_b$ 
as possible, by studying   inclusive Higgs production, the decay $h \rightarrow b \bar b$  \cite{Brod:2013cka},
and the associated production of $h$ and a $b \bar b$ pair, with tagged $b$ jets \cite{Wiesemann:2014ioa,Forte:2015hba,Bonvini:2015pxa}. 

The coupling of the Higgs to $G \tilde G$ is constrained by $\mu_{ggF}$ at the 30\% level, with possibility to improve to 20\% in Run 2. Also in this case, collider experiments are  competitive with EDM bounds.

Finally we discuss the CPV  couplings of the top quark, $\textrm{Im}\, Y^\prime_t$ and $\tilde d_t$. 
It is interesting that the  current collider bounds on $\textrm{Im}\, Y^\prime_t$ and $\tilde d_t$
are dominated by the contribution of top loops to $h G \tilde G$. In the case of  $\textrm{Im}\, Y^\prime_t$, the bound from gluon fusion is a factor of ten stronger than the direct bound via $\mu_{t\bar t h}$.
In the case of the top CEDM $t \bar t$ and $t \bar t h$ probe $\tilde d_t$ at the same level, a factor of 3 weaker than the bound from $\mu_{ggF}$. 
However, especially in the presence of a CEDM, $t \bar t$ and $t \bar t h$ have a greater chance of improvement at Run 2, getting much closer to the gluon fusion bound. 
Furthermore, were significant deviations from the SM to be observed in $\mu_{ggF}$, $\mu_{t\bar t h}$ or in the $t \bar t$ cross section, 
the fact that these three observables have roughly the same sensitivity to $\tilde d_t$ would offer the exciting possibility to  prove or exclude 
that the origin of the signal is a top CEDM. 
Comparing to EDM bounds, Table \ref{boundsY}, we see that $\textrm{Im}\, Y_t^\prime$ is strongly constrained by the electron EDM (although this constraint strongly depends on the SM prediction of the electron Yukawa coupling, see the discussion in Ref.~\cite{Brod:2013cka}). On the other hand, the bounds on $\tilde d_t$ are very close to the LHC bounds, which makes the study of the top CEDM (and CMDM) even more interesting.

We conclude by noting that our analysis of collider observables has focused on  CP-even observables that are sensitive to the square of CPV couplings. 
More information could be gained by studying differential observables, see Fig. \ref{ttH}, or observables such as spin correlations 
that are linear in the CPV couplings (see for example  \cite{Demartin:2014fia,Buckley:2015vsa,Hayreter:2015ryk}).

\section{Direct vs indirect constraints:  interplay of couplings}
\label{Sec-interplay}

Although it is interesting to study constraints on the individual CPV dimension-6 operators, in most BSM realizations several will be generated at the same time. Clearly a single EDM experiment can only constrain or identify a single combination of operators and several measurements are needed to isolate the individual couplings. In this section we study how EDM and collider experiments can constrain or identify combinations of CPV couplings and 
to what extent various experiments are complementary. We also focus on the role of the uncertainties in matrix elements which have a strong impact on the EDM analysis. As there are many combinations of CPV operators that can be studied, we focus here on a subset of cases which, in our opinion, are most interesting. Other combinations or specific BSM scenarios can be studied in similar fashion.

\subsection{$\textrm{Im}\, Y_u^\prime$-$\textrm{Im}\, Y_d^\prime$}

\begin{figure}[t]
\centering
\includegraphics[width=0.48\textwidth]{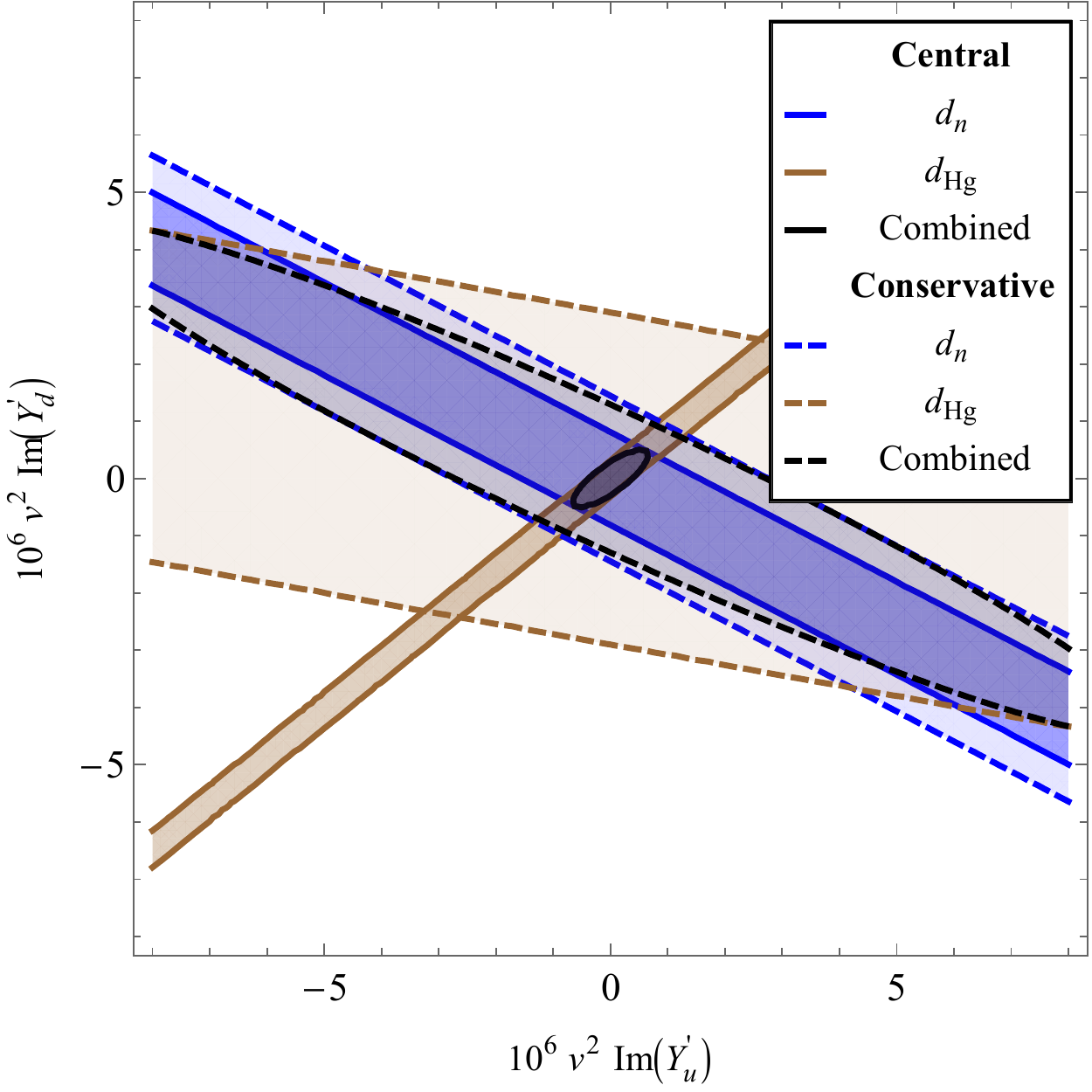} 
\includegraphics[width=0.48\textwidth]{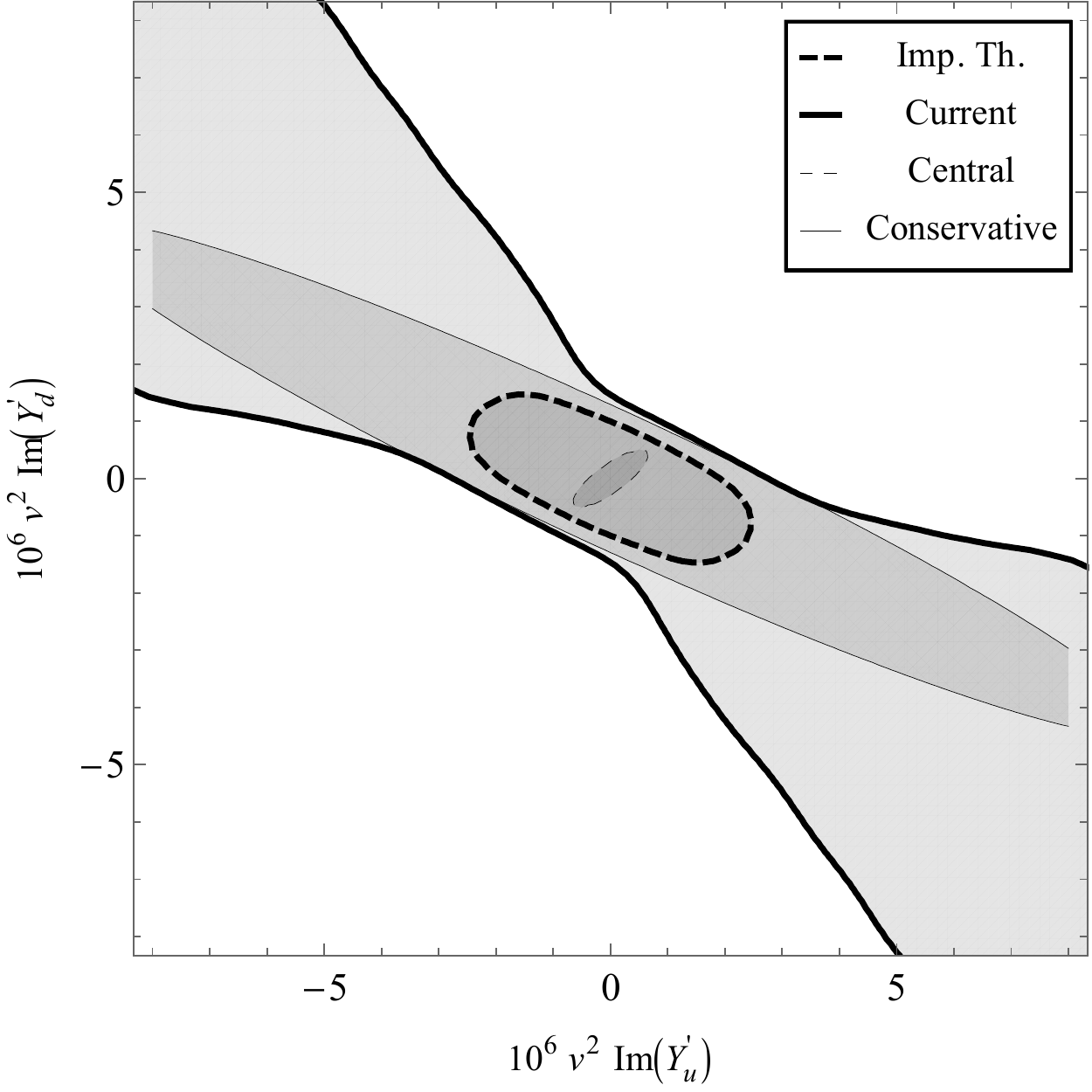} 
\caption{\small Figures showing $90\%$ c.l. contours coming from $d_n$ (blue) and $d_{\mathrm{Hg}}$ (brown), and their combination (black). The left panel uses the central and conservative values for the matrix elements. In the right panel, $90\%$ c.l. combined contours are shown for the three different matrix elements strategies explained in the text: central (thin dashed), conservative (thin), minimized (thick). The thick dashed line denotes  the minimized contour that can be achieved with improved matrix elements.}
\label{Yu_vs_Yd}
\end{figure}

\begin{figure}[t]
\centering
\includegraphics[width=0.48\textwidth]{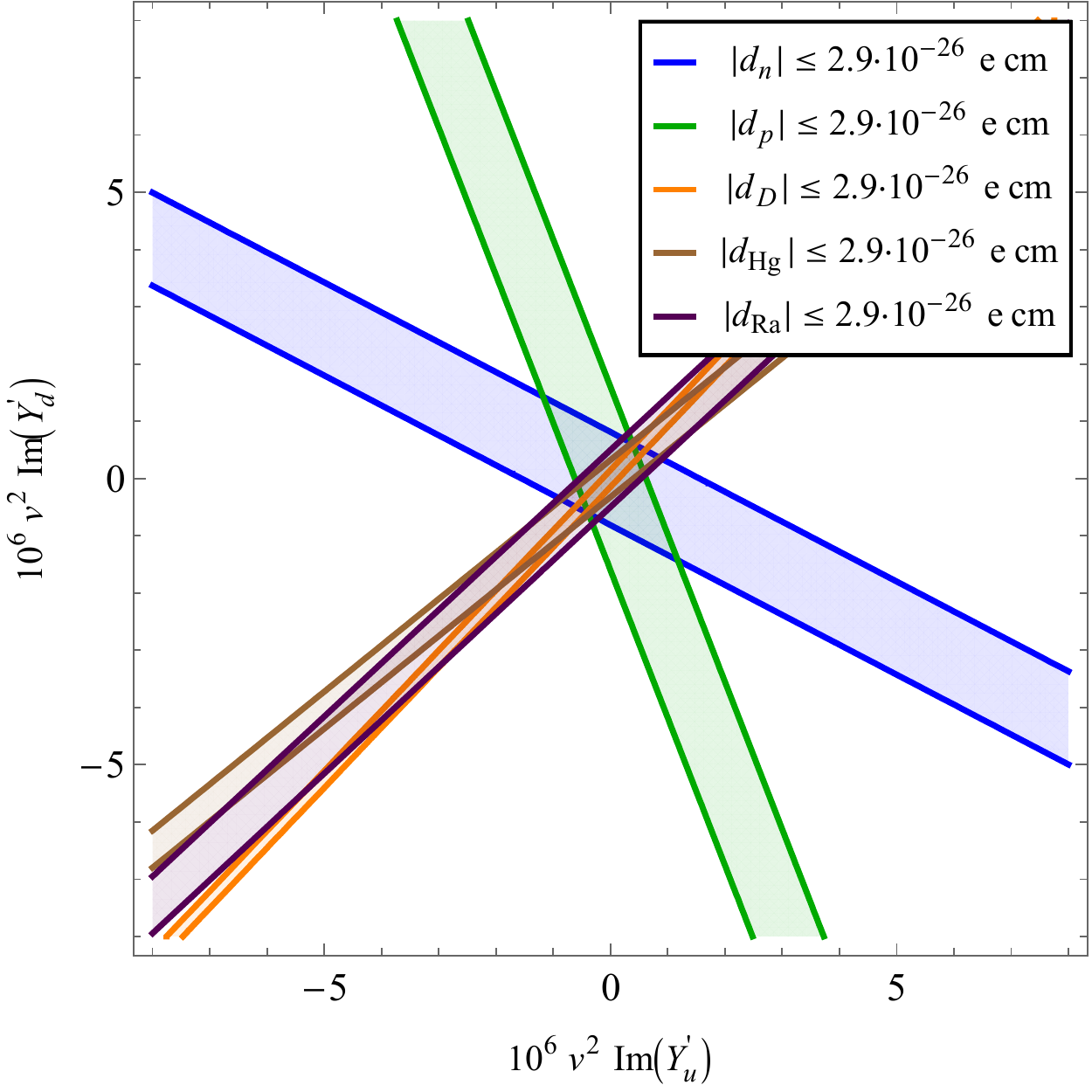} 
\includegraphics[width=0.48\textwidth]{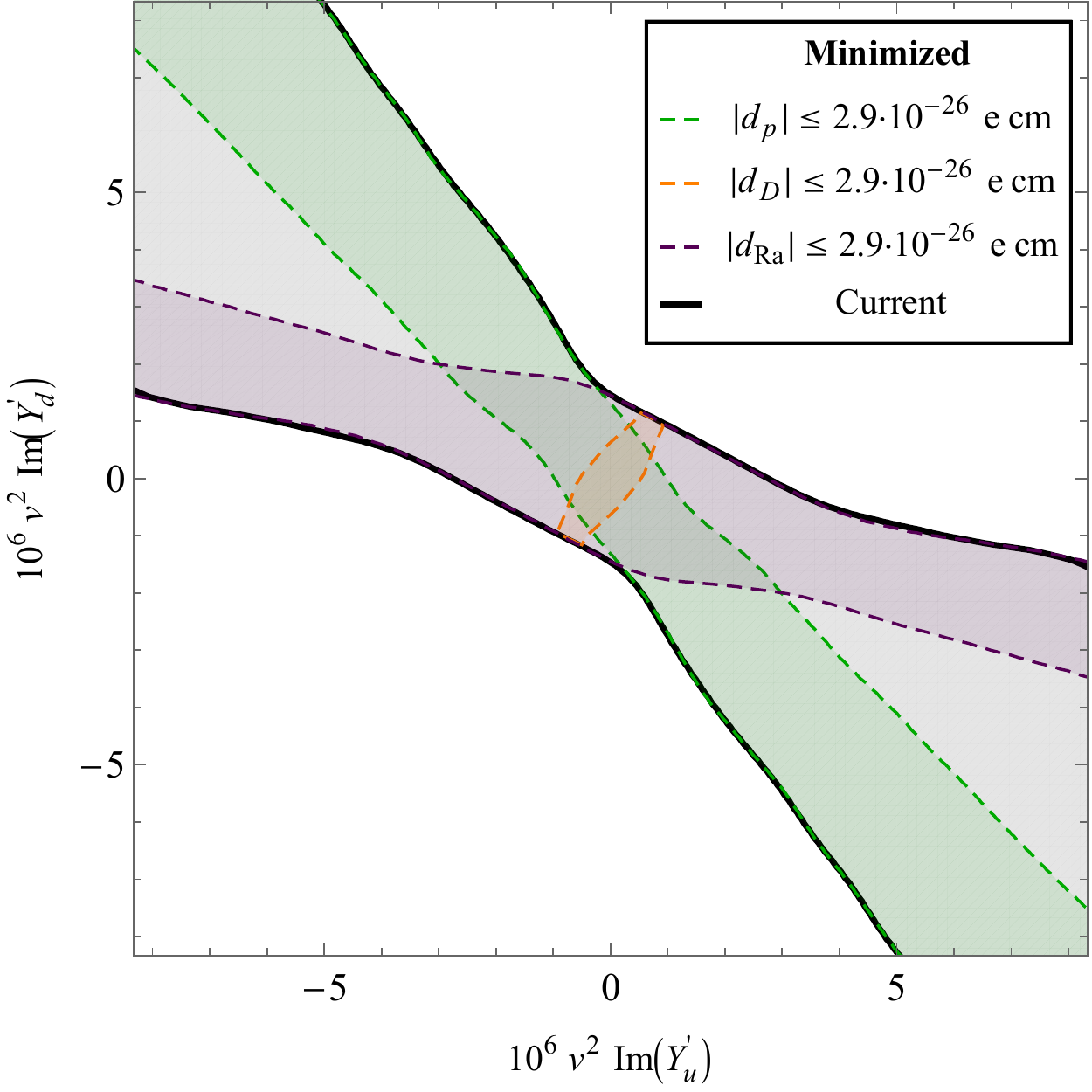} 
\caption{\small The left panel shows the $90\%$ c.l. contours coming from $d_n$ (blue),  $d_{\mathrm{Hg}}$ (brown), $d_{\mathrm{Ra}}$ (purple), $d_p$ (green), and $d_D$ (orange) using the central matrix elements. The right panel shows minimized constraints with current EDM measurements (black), including a future $d_p,\,d_D$, or $\dRa$ measurement (green, orange, and purple, respectively).
} 
\label{Yu_vs_Yd_dDdRa}
\end{figure}

We begin the analysis by studying the case where CPV predominantly occurs in the interactions between the lightest two quarks and the Higgs field. As shown in the previous sections, there are no significant constraint from $d_e$ or collider experiment and we therefore focus on hadronic and nuclear EDMs. In the left panel of Fig.~\ref{Yu_vs_Yd}, we show $90\%$ c.l. contours in the $\textrm{Im}\, Y_u^\prime$-$\textrm{Im}\, Y_d^\prime$ plane, arising from the current $d_n$ and $\dHg$ bounds. The solid lines correspond to the central values of the matrix elements. In this case, the two EDM experiments are very complementary because $\dHg$ is dominated by CPV pion exchange proportional to $\bar g_1$.
 In the conservative case (dashed lines) the complementarity is reduced leading to a significantly larger contour. The loss of complementarity is amplified once we apply the minimization strategy as can be seen in the right panel of Fig.~\ref{Yu_vs_Yd}. A free direction emerges indicating that large values of $\textrm{Im}\, Y_u^\prime$ and $\textrm{Im}\, Y_d^\prime$ cannot be excluded. Note that, as expected,  the central and conservative contours always lie inside the minimized contour. 

Additional information is needed to eliminate the free direction in the $\mathrm{Im}\,Y_u^\prime$-$\mathrm{Im}\,Y_d^\prime$ plane. On the theoretical side,  improved matrix elements definitely help. Using the benchmark matrix elements given in Sec.~\ref{future_matrix}, we obtain the thick dashed contour in the right panel. We see that a modest improvement on the matrix elements (in most cases $50\%$ uncertainty is sufficient) would already greatly improve the bounds and remove the free direction.

Alternatively, we can study additional EDM measurements. This could be achieved by improving $\dHg$ by orders of magnitude, but this is unlikely to happen. Improving $d_n$ alone would still allow for a free direction. Instead we study the impact of EDM measurements on different systems. In the left panel of Fig.~\ref{Yu_vs_Yd_dDdRa}, we show constraints from $d_p$, $d_D$, and $\dRa$ using central matrix elements and assuming the EDMs are measured with the same precision as the current $d_n$ bound. The prospective sensitivities are actually more precise than this. Here we mostly study the complementarity of the experiments which is easier if the experimental bounds are similar.
We see that $d_D$, $\dHg$, and $\dRa$  probe the same combination of couplings as they are all dominated by $\bar g_1$ contributions. The main advantage of the $d_D$ measurement is the status of the nuclear theory. This can be seen in the right panel, where we show the minimized constraints. A future $\dRa$ measurement would not eliminate the free direction, whereas a $d_D$ measurement would. A $d_p$ measurement would also be complementary (left panel), but would not remove the free direction (right panel).

\subsection{$ \textrm{Im}\, Y_b^\prime$-$\textrm{Im}\, Y_s^\prime$}

\begin{figure}[t]
\centering
\includegraphics[width=0.49\textwidth]{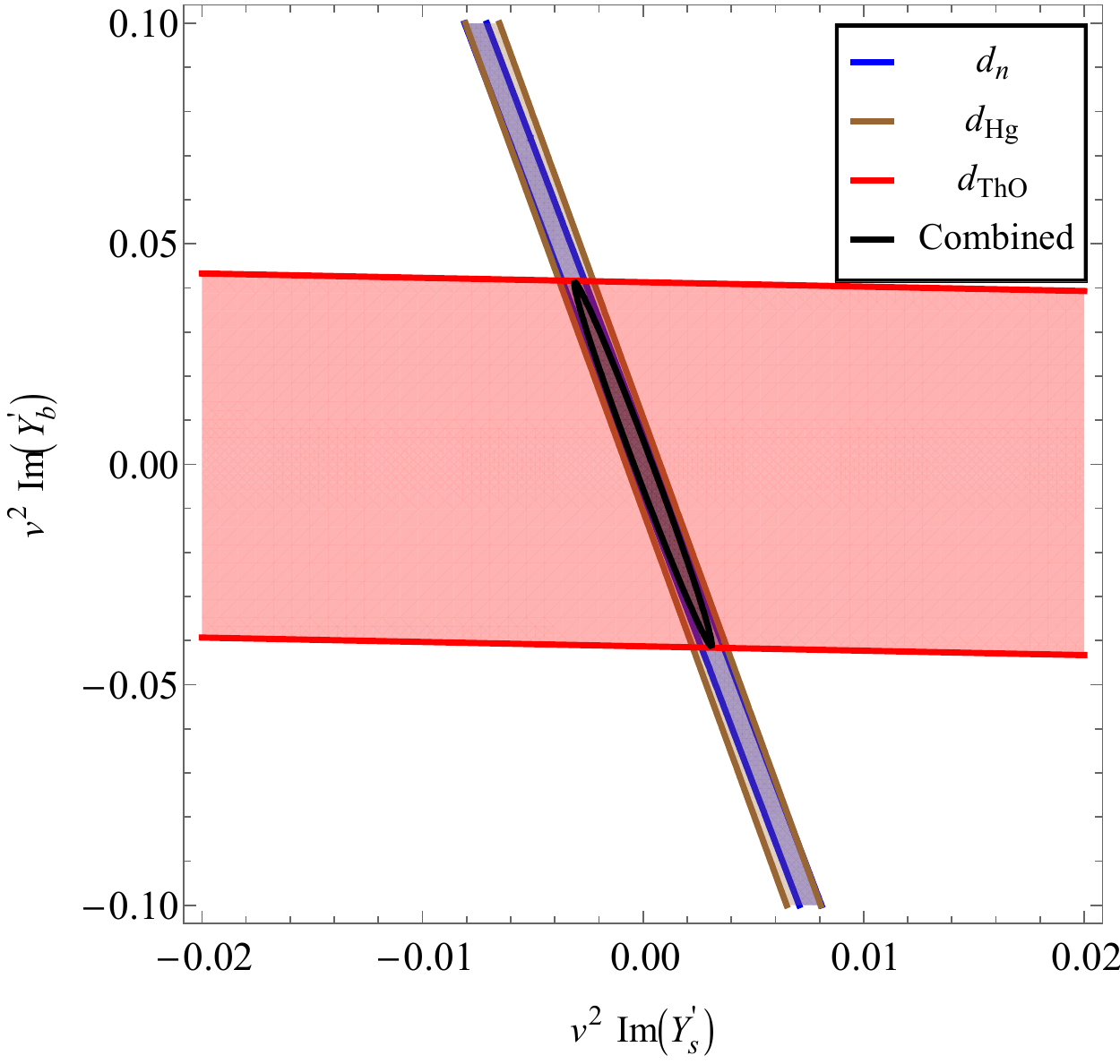} 
\includegraphics[width=0.49\textwidth]{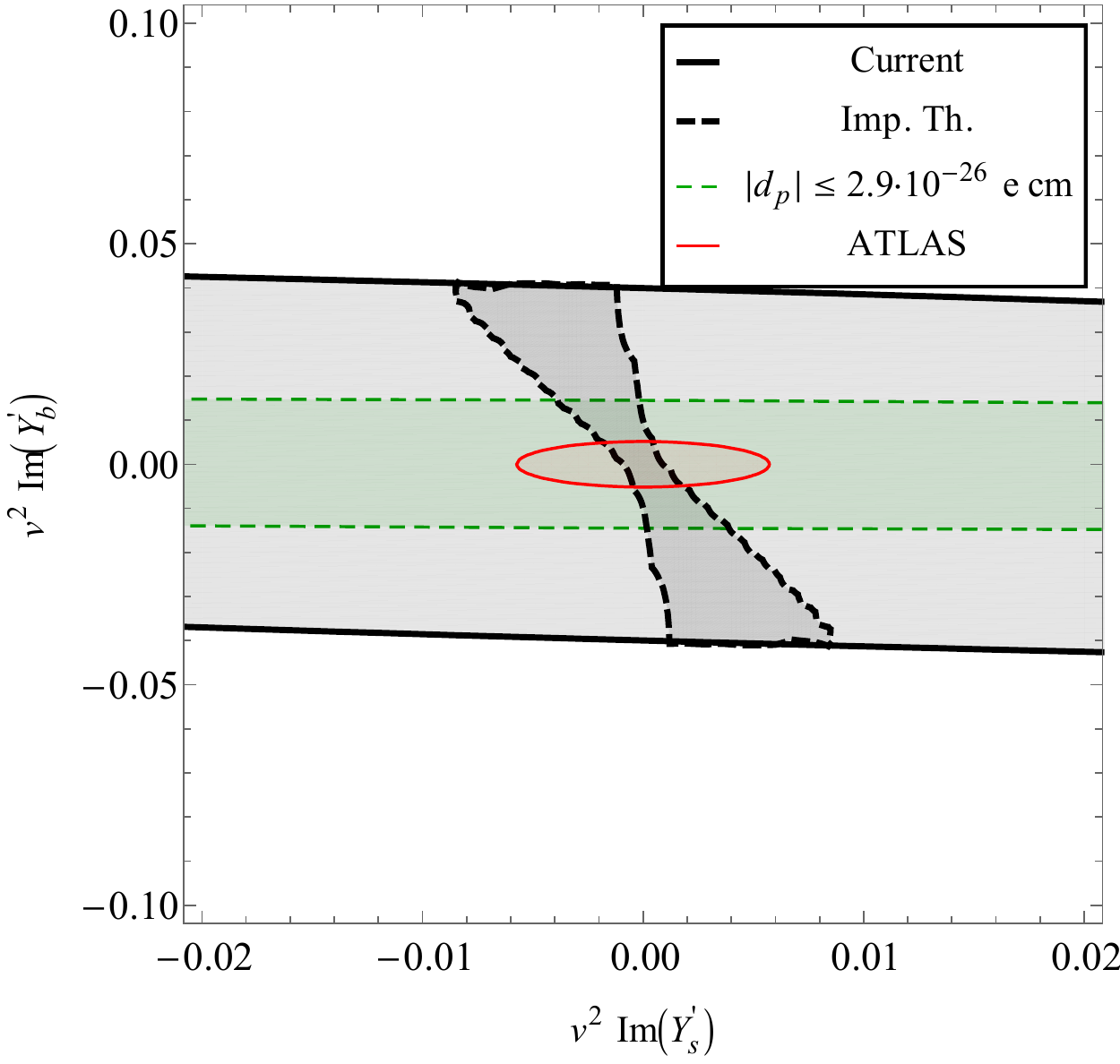} 
\caption{\small The left panel shows the $90\%$ c.l. contours using central matrix elements and current EDM measurements (notation as in Fig.~\ref{Yu_vs_Yd}). The right panel shows in black the current combined minimized EDM constraints. The thick-dashed contour shows the combined limit with improved matrix elements, while a future $d_p$ measurement is shown in green. The red ellipse represents the collider constraints.
} 
\label{Ys_vs_Yb}
\end{figure}

We now focus on the CPV Yukawa couplings of down-type $s$ and $b$ quarks. We show the central constraints in the left panel of  Fig.\ \ref{Ys_vs_Yb}, for the case with a positive Weinberg matrix element. It is clear that the electron EDM mainly constrains one of the couplings, $\mathrm{Im}\,Y_b'$, while the neutron and mercury EDMs probe nearly the same combination of couplings. In case of a negative $d_W$ matrix element, the neutron and mercury constraints are somewhat more complementary, leading to slightly stronger constraints.

We show constraints resulting from the minimized strategy in the right panel of Fig.\ \ref{Ys_vs_Yb}. In this case the $\mathrm{Im}\,Y'_s$ direction becomes unconstrained due to the large uncertainties related to the matrix element of the strange EDM, while the constraint in the $\mathrm{Im}\,Y_b'$ direction is hardly affected. We find that the free direction would not be eliminated by a measurement of $d_p$ (or $d_D$, $d_{\rm Ra}$, $\dXe$) at the current $d_n$ sensitivity, although this would improve the constraint on $\mathrm{Im}\,Y_b'$. In contrast, better knowledge of the strange EDM matrix element does eliminate the free direction.

In this case, current collider bounds, denoted by the red ellipse on the right panel of Fig. \ref{Ys_vs_Yb}, are stronger than EDM bounds in the minimized strategy. In particular,
they constrain the free $\mathrm{Im}\,Y'_s$ direction. Even in the presence of better theoretical handling of the strange matrix elements, the bounds from collider still play an important, complementary role.
The study of the Higgs signal strengths at the LHC Run 2 will improve the bounds on $\textrm{Im}\, Y^\prime_b$ and $\textrm{Im}\, Y^\prime_s$
by 20\% and 40\%, respectively. Further improvement could come from the study of exclusive decays of the Higgs into $b \bar b$ or $s \bar s$ mesons \cite{Bodwin:2013gca,Bodwin:2014bpa,Kagan:2014ila}.

\subsection{$\textrm{Im}\, Y_t^\prime$-$\textrm{Im}\, Y_b^\prime$}

\begin{figure}[t]
\centering
\includegraphics[width=0.49\textwidth]{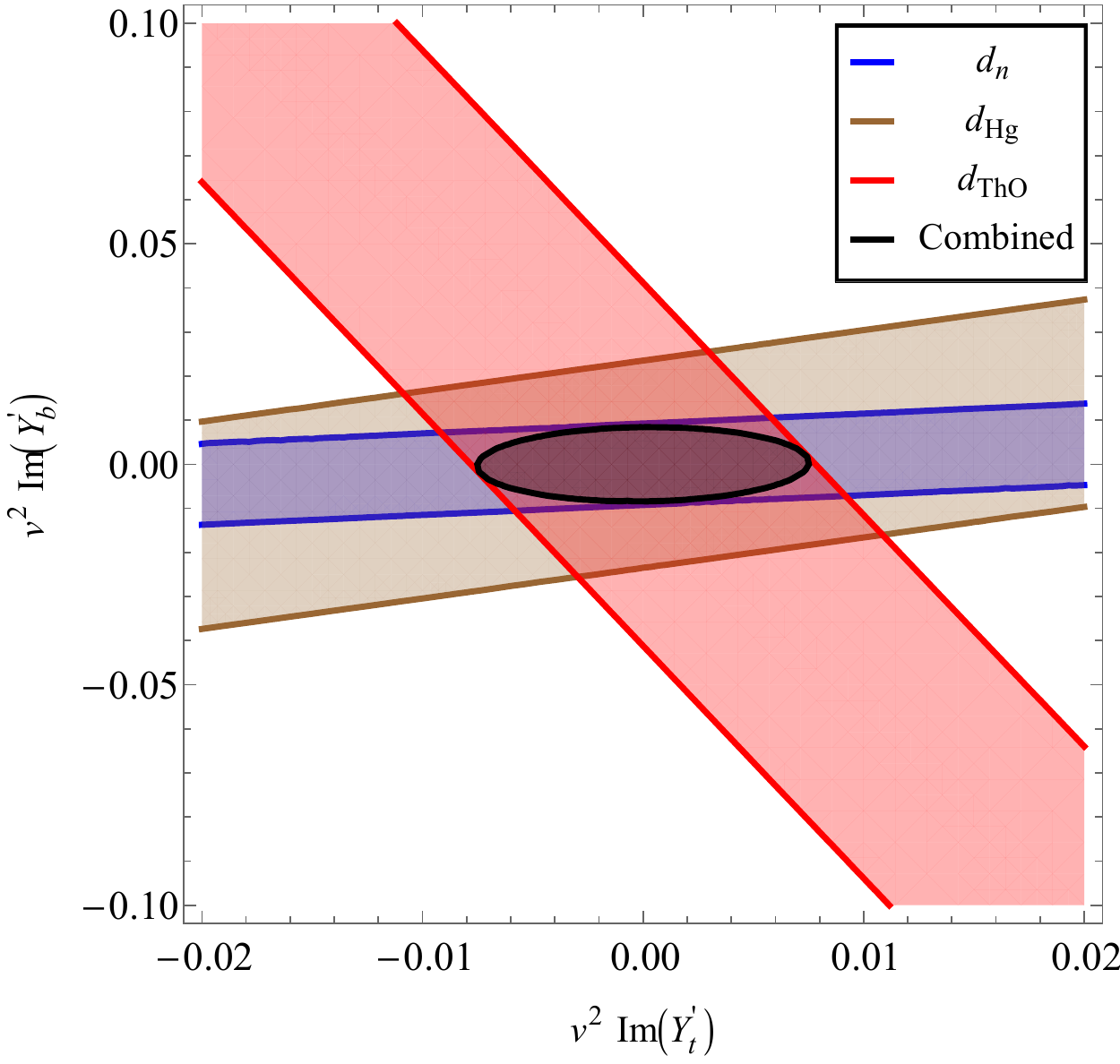} 
\includegraphics[width=0.49\textwidth]{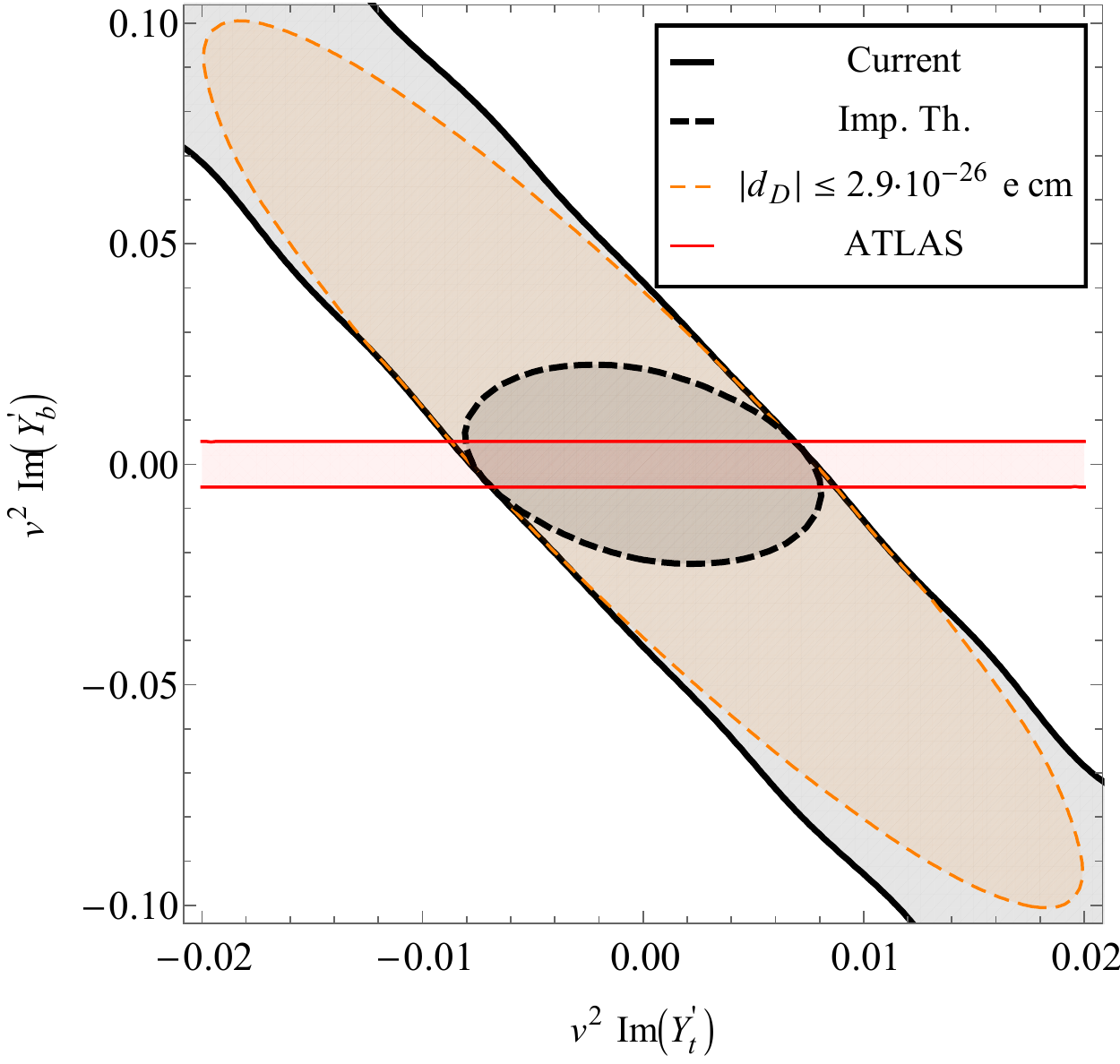} 
\caption{\small The left panel shows the $90\%$ c.l. contours using central matrix elements and current EDM measurements (notation as in Fig.~\ref{Yu_vs_Yd}). The right panel shows in black the combined minimized EDM constraints with current matrix elements 
(solid) and improved matrix elements  (dashed). The impact of a 
 future $d_D$ measurement is shown in orange. The horizontal band  denotes the collider constraints.
} 
\label{Yt_vs_Yb_negative}
\end{figure}

We now turn to the anomalous Yukawa couplings of the third generation. In this case, the constraints depend strongly on the sign of the nucleon matrix element for the Weinberg operator $d_W$. 
In Fig.\ \ref{Yt_vs_Yb_negative} we show the results for the least constrained case, the case of a negative neutron matrix element.
We present constraints from $d_e$, $d_n$, and $\dHg$ using central matrix elements in the left panel of the figure. The constraints originate mainly from the interplay between $d_n$ and $d_e$: the anomalous top-Yukawa couplings is strongly constrained by $d_e$, while $\textrm{Im}\, Y_b'$ is mainly constrained by $d_n$. 

In the right panel of Fig.~\ref{Yt_vs_Yb_negative} we show constraints using the minimization strategy. In this case the constraints are significantly weaker leading to an almost free direction. More precise matrix elements could significantly improve the constraints as can be seen from the black dashed contour. A $d_D$ measurement would give a similar, but weaker, constraint (orange dashed ellipse). 

Also in this case, the collider bound on $\textrm{Im}\, Y_b^\prime$ plays an important role, by eliminating  the free direction in the $\textrm{Im}\, Y_b^\prime$ - $\textrm{Im}\, Y_t^\prime$ plane.
The combined current LHC-EDM bound is indeed better
than  the projected EDM bound with improved matrix elements. 
The LHC Run 2 is likely to probe the pseudoscalar top Yukawa at the 10\% level, still too far from EDM bounds to be relevant. 
As discussed in the $\textrm{Im}\, Y_b^\prime$-$\textrm{Im}\, Y_s^\prime$ case, it will be important to get as many handles as possible on the bottom quark Yukawa.

\subsection{$\textrm{Im}\, Y_t^\prime$-$\tilde d_t$}\label{Ytdt}
\begin{figure}[t]
\centering
\includegraphics[width=0.49\textwidth]{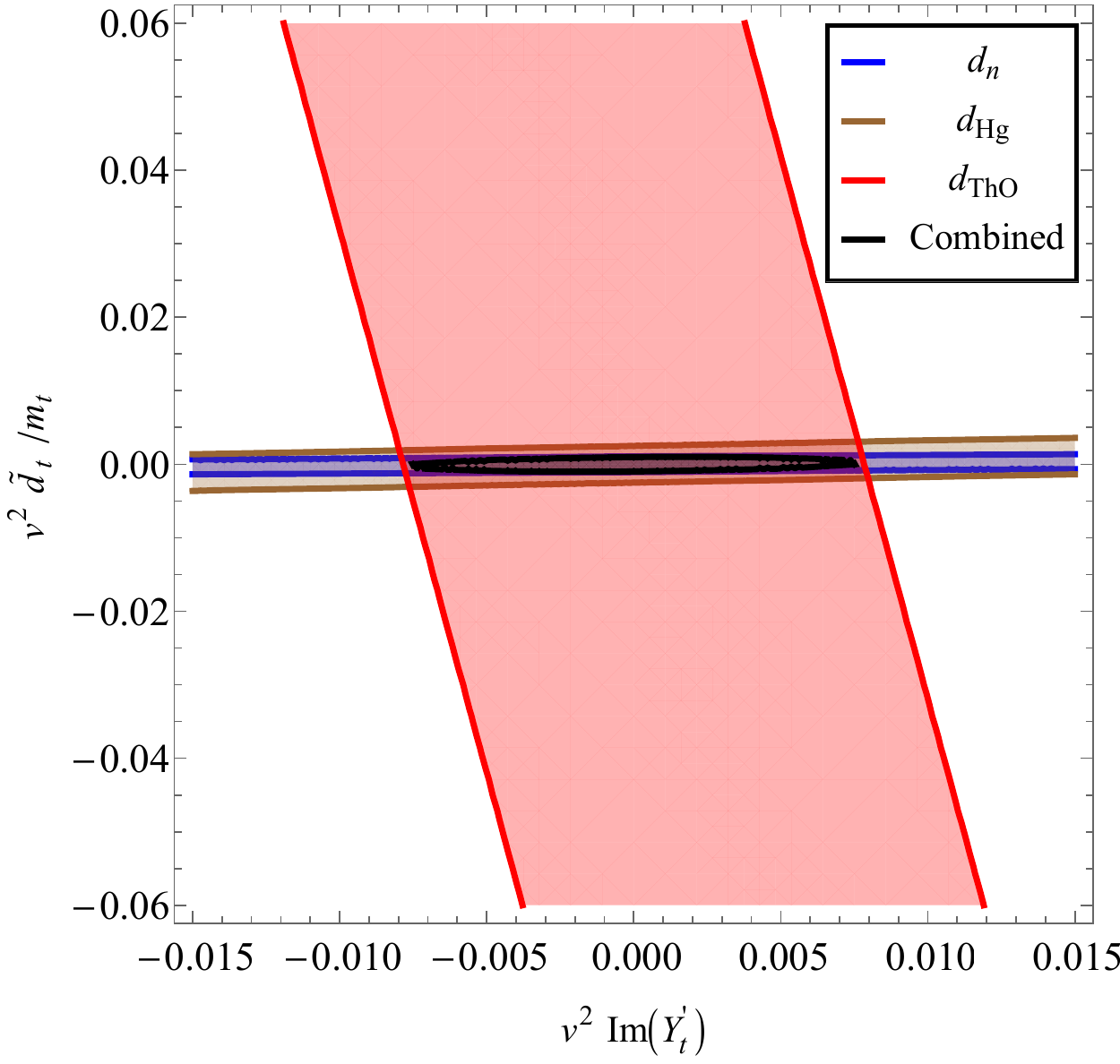} 
\includegraphics[width=0.49\textwidth]{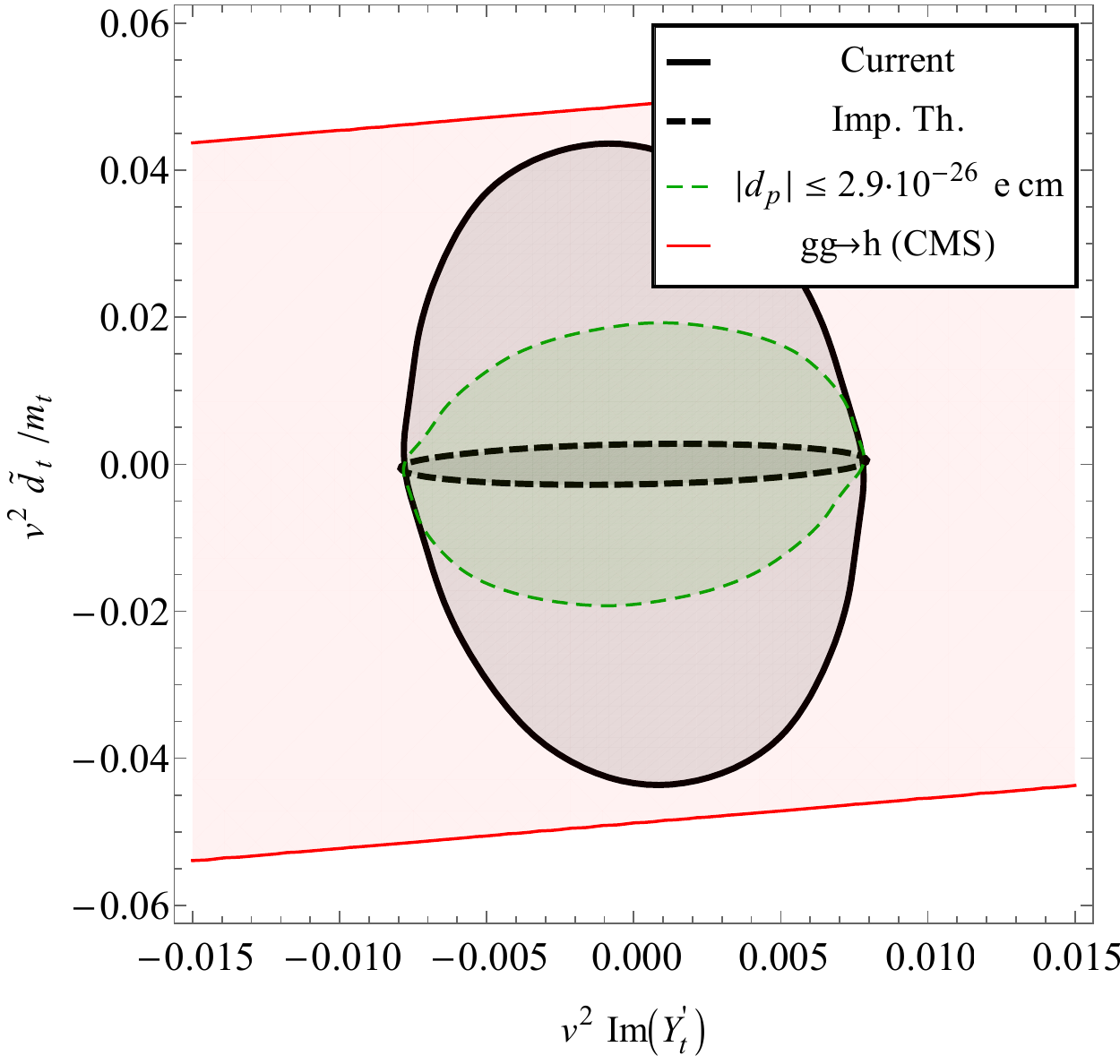} 
\caption{\small The left panel shows the $90\%$ c.l. contours using central matrix elements and current EDM measurements (notation as in Fig.~\ref{Yu_vs_Yd}). The right panel shows in black the combined minimized EDM constraints for current (solid) and improved (dashed) matrix elements. A future $d_p$ measurement is shown in green. 
}
\label{Yt_vs_CEDM_negative}
\end{figure}

Finally, we consider the case that BSM physics contributes mainly to the top CEDM, $\tilde d_t$, and its anomalous Yukawa coupling, $\mathrm{Im}\,Y_t^\prime$. In this case, the sign of the $d_W$ matrix element does not affect the constraints too much. We use a negative value which gives the weakest constraints. We show EDM constraints in the left panel of Fig.\ \ref{Yt_vs_CEDM_negative} using central matrix elements. The plot looks  similar to the $\textrm{Im}\, Y_t^\prime$-$\textrm{Im}\, Y_b^\prime$ plot, with the strongest constraint on $\tilde d_t$ ($\mathrm{Im}\,Y_t^\prime$ ) arising from $d_n$ ($d_e$).

After minimizing over the matrix elements, see the right panel in Fig.\ \ref{Yt_vs_CEDM_negative}, the constraints become weaker by roughly an order of magnitude in the $\tilde d_t/m_t$ direction. The constraint in the $\mathrm{Im}\,Y_t^\prime$ direction is much less affected because it mostly arises from $d_e$ where the uncertainties are smaller. The reduced sensitivity could be almost  completely overcome with improved matrix elements, as can seen by the dotted contour. Finally, we show the impact of a $d_p$ measurement at the current $d_n$ level. A $d_D$ measurement would be complementary as well, but is sensitive to cancellations in the sum of nucleon EDMs, $d_n+d_p$, in  case of the Weinberg operator that is induced by $\tilde d_t$ (see also the discussion in Sect.~\ref{newexp}). Collider constraints from the gluon fusion process, depicted in red, 
are very close to the minimized contour, but  slightly too weak to have an impact. Constraints from $ t \bar t$ and $t \bar t h$ 
are at the moment not competitive with gluon fusion or with EDM constraints. However, they have the largest margin of improvement at the LHC Run 2, and are likely to become relevant, especially in the absence of theoretical improvement on the hadronic matrix elements.

\section{Discussion}
\label{Sec-discussion}

In this paper we have presented a detailed study of  both direct and indirect  constraints on non-standard CP-violating Higgs couplings to quarks and gluons. 
Working within a linear EFT framework, we have focused on the leading flavor-conserving dimension-6 operators coupling the Higgs doublet 
to quarks and gluons, namely the  CPV Yukawa couplings, the chromo-electric dipole operators and the $\varphi^\dagger \varphi G \tilde G$ operator
(see Section \ref{Sec2}).  We  have first obtained bounds on the effective couplings  by assuming that at the high-scale (where we match the EFT to the underlying  new physics model)  only one coupling at the time dominates (Sections \ref{Sec3} and \ref{Sec4}). 
A  summary of current and prospective bounds from EDMs is provided in Table~\ref{boundsY_improved_XeRa}.  
Similarly,  in Table~\ref{boundsSummaryColl} we present   a summary of current and prospective bounds from  single Higgs production and decay, and from  $t \bar t$ and $t \bar t h$ production at the LHC.
In Section \ref{Sec-interplay} we have then studied a few selected cases in which two couplings dominate at the matching scale, 
summarized in Figs.~\ref{Yu_vs_Yd}-\ref{Yt_vs_CEDM_negative}. 

Throughout our analysis, we have payed special attention to the theoretical uncertainties. 
For the extraction of direct bounds from LHC production cross sections, 
uncertainties arise from $\alpha_s$,  parton distribution functions,  and the residual dependence on 
the renormalization and factorization scales.  The uncertainty estimate is straightforward, and typically leads to 
effects of $O(10\%)$ (see Section \ref{Sec4}).~\footnote{The strange quark PDFs induce larger uncertainties, of the order of 30\%.} 
On the other hand,  the non-perturbative matrix elements at the hadronic and nuclear level pose a greater theoretical challenge. 
The uncertainties corresponding to model calculations are quite large, and in some cases  not even the sign of 
a matrix element is  determined reliably. 
We have obtained bounds with different treatments of theoretical input. 
These treatments include two extreme cases: 
(a)  Taking the central value of the matrix elements, ignoring the uncertainty. 
The resulting bounds reflect the maximal physics reach of EDM experiments. 
(b)   Assuming that the matrix elements have a flat distribution in a certain range 
corresponding to existing calculations:   this is  the range-fit method used in  Ref.~\cite{Charles:2004jd} in the context of 
fits to the elements of the  Cabibbo-Kobayashi-Maskawa matrix. 
The resulting bounds account for the theoretical uncertainties in the safest possible way (perhaps over-conservative,  but realistic).  
We have also explored a third option (c):  using the range-fit method with reduced theoretical errors, 
at the 25-50\% level  (see discussion in  Section~\ref{future_matrix}), 
anticipating  progress in the next few years from both lattice QCD and nuclear many-body theory.

We would like to highlight the following  points of our analysis: 

\begin{itemize}

\item Concerning the EDMs,  a key result  of our study  is that 
hadronic and nuclear uncertainties greatly  dilute the nominal  constraining power (i.e. the 
one obtained by using central values for all matrix elements, ignoring their uncertainty). 
The  dilution effect comes about  because a given high-energy coupling  generates via RGE and threshold corrections   
a number of  operators at low-energy, whose contribution can cancel each other due to the poorly known matrix elements. 
From Table~\ref{boundsY} one sees that  when going from central  values of the matrix 
elements to range-fit method, the $^{199}$Hg bounds essentially disappear, while the neutron bounds 
are  weakened by up to an order of magnitude,  depending on the coupling,   and they are eliminated in the case of $\textrm{Im}\, Y_{s,b}'$. 
Nonetheless,  when considering all existing EDM constraints ($d_e, d_n,  d_{\rm Hg}$), it is still possible to obtain bounds 
on non-standard Higgs couplings. Using  current theoretical  uncertainties these bounds are summarized in the next-to-last row of 
Table~\ref{boundsY}.    The bounds on  $\textrm{Im}\, Y_{s,b,t}'$ are currently determined by the ThO  EDM limit, 
while the bounds on $\textrm{Im}\, Y_{u,d,c}'$, $\theta'$, and $\tilde{d}_t$ are set  by the  neutron EDM  limit. 
This complementarity is also quite evident in the two-couplings analysis of Section~\ref{Sec-interplay}.

\item   Another noteworthy result is that  with the improved matrix element precision  advocated in  Section~\ref{future_matrix}
the  bounds obtained with the range-fit method  come very close to the ones obtained  with central value matrix elements. 
That is, comparing the row ``Comb. Cen." and  ``Future Min." in  Table~\ref{boundsY},  most numbers only differ by a factor $2$. 
The exceptions are the bounds on $\textrm{Im}\, Y_{u}'$ and $\textrm{Im}\, Y_{d}'$ which are different by a factor $3$ and $5$, respectively.
This follows from the fact that once matrix elements are known at the 25-50\% level,  there is very little room for cancellations 
and one essentially exploits the full power of experimental constraints. 
We reiterate here that the desirable target uncertainties for hadronic and nuclear matrix elements are:
(i) $d_{n,p} [d_s] $ and  $d_{n,p} [d_W] $    at 50\% level; 
(ii) $d_{n,p} [\tilde{d}_{u,d}] $  at 25\% level; 
(iii)  $\bar g_{0,1} [\tilde d_{u,d}] $ at the  $50\%$ level; 
(iv)  $S_{\mathrm{Hg}} [\bar g_{0,1}]$   at the  $50\%$ level.\footnote{In addition to these targets, we stress that determining the dependence of $S_{\mathrm{Ra}}$ and $S_{\mathrm{Xe}}$ on the 
constituent nucleon EDMs is important. The same can be said about the role of the strange CEDM on the nucleon EDMs which has been under recent debate \cite{Hisano2}.}
These targets  do not seem unrealistic and further motivate 
systematic studies of hadronic and nuclear matrix elements 
with lattice QCD and modern nuclear many-body methods.

\item In Table~\ref{boundsY_improved_XeRa} we summarize possible future scenarios, 
by taking into account
(i)  improved matrix elements according to the benchmarks of  Section~\ref{future_matrix};   
(ii)  improved sensitivities on existing systems ($d_e, d_n$) and EDM measurements in additional systems  
($d_{\textrm{Xe}}$, $d_{\textrm{Ra}}$, $d_p$, $d_D$);  and combinations of both (i) and (ii). 
This exercise shows that improving the theory can have as much, or even more, impact as additional measurements. 
Note that this  point is also evident from the plots in Section~\ref{Sec-interplay}. 
For the couplings under consideration here,  the anticipated improvements in the neutron and ThO EDMs 
(as well as the addition of proton and deuteron EDMs at the level of $10^{-29} \ e$~cm) 
will have the largest impact. 
In any case, regardless of which new experimental probe becomes available,   the constraints on couplings  
dramatically improve by using   more accurate matrix elements as discussed above.

\item  Our analysis of collider observables has focused on (total) production and decay processes, that are sensitive to the square  of CP-odd couplings. 
Additional information could be gained by studying more differential observables, as briefly showed in Fig. \ref{ttH}, or 
observables,  such as  spin correlations, 
that depend  linearly  on the new physics couplings. 

\item   We noticed that the Higgs gluon fusion production cross section provides a very strong bound on the top CMDM, better than the direct constraint from the $t \bar t$ total cross section. With the integrated luminosity of the LHC Run 2, in the absence of deviations from the SM, the limit from gluon fusion will significantly improve, to better than 1\%. 
If multiple couplings are generated at the high scale  $\MT$, the gluon fusion, $t \bar t$ and $t \bar t h$ cross section provide complementary observables, ideal to pin down a top CMDM (and CEDM).

\item Complementarity of EDMs and LHC constraints: 

Currently, our best  knowledge of the non-standard CPV Higgs couplings comes from a combination 
of EDMs and LHC constraints, summarized in Table~\ref{bounds-complementarity}.~\footnote{
For ease of comparison with the existing literature, we also quote here the 
bounds on the non-standard Yukawa couplings in terms of the parameters $\tilde{\kappa}_q$, defined by 
${\cal L} =     (m_q/v) \tilde{\kappa}_q \, \bar{q} i \gamma_5 q   \ h$. 
Multiplying the entries of Table~\ref{bounds-complementarity} by $v/m_q$ we obtain:
$\tilde{\kappa}_u  <  0.45$,
$\tilde{\kappa}_d  <  0.11$,
$\tilde{\kappa}_s  <  37$,
$\tilde{\kappa}_c  <  2.7$,
$\tilde{\kappa}_b  <  0.5$,
$\tilde{\kappa}_t   <  0.01$.}
The strongest  constraints on  $\tilde{d}_{q \neq t}$ and $\textrm{Im}\, Y_{u,d,t}'$ arise  (by far for the light flavors)  from EDMs, while 
for $\textrm{Im}\, Y_{s,c,b}'$,  $\tilde{d}_t$ and $\theta'$ the current bounds from EDM and LHC 
are comparable, once we take into account the uncertainties in hadronic and nuclear matrix elements.
In all cases,  except for $\mathrm{Im}\,Y_b'$, improved matrix elements would strengthen the current EDM constraints and put 
these couplings out of the reach of LHC, at least with the observables we considered. 
Because EDM  and LHC experiments probe different combinations of couplings, they complement each other in cases
where more than one coupling is simultaneously generated at the high scale.  In Sect.~\ref{Sec-interplay}, we have studied several cases where, with the current status of the hadronic and nuclear theory, only by combining LHC and EDM constraints  significant constraints are obtained.

Looking to the future,  the prospects for improving bounds on the non-standard couplings from 
EDMs  are excellent, especially if experimental progress will be accompanied by improved matrix elements.  
On the other hand,  the bounds obtained from the LHC 
will improve little with increased center-of-mass energy and luminosity. Although the constraints on $Y'_{u,d,s}$ are expected to become more stringent by up to a factor of two, the expected improvements for $Y'_{c,b,t}$, $\theta'$ and $\tilde d_t$ are more modest, see Table \ref{boundsSummaryColl}.
The reason for this is that the additional non-standard contributions to Higgs production 
induced by  $\textrm{Im}\, Y_{q}'$ and $\theta'$  grow  with the center-of-mass energy  
more slowly or at the same rate  as the SM gluon fusion cross section.  
So,  significant improvements will be possible 
only with a substantial reduction of the uncertainties on the SM gluon fusion cross section.
Better prospects exists  for the top CEDM  $\tilde{d}_t$, in which case the $t\bar t$ and $t\bar t h$ cross sections
grow faster than the SM, and additional information can be extracted from the shape of differential distributions.
As a result, as  can be seen from Tables~\ref{boundsY_improved_XeRa}  and \ref{boundsSummaryColl}, 
anticipated improvements in the ThO and neutron EDM would put all the couplings considered here  out of reach at the LHC Run 2 
in total cross section measurements.  It would be very interesting, therefore, to  explore  CPV observables 
in Higgs production and decay. 
\end{itemize}

\begin{table}[t]
\begin{center}
\scriptsize
\begin{tabular}{||cccccccc||}
\hline
 $v^2 \mathrm{Im}\, Y^\prime_u$ & $v^2 \mathrm{Im}\, Y^\prime_d $  &$v^2 \mathrm{Im}\, Y^\prime_s$ & $v^2 \mathrm{Im}\, Y^\prime_c$ & $v^2 \mathrm{Im}\, Y^\prime_b$& $v^2 \mathrm{Im}\, Y^\prime_t$ & $v^2\, \theta^\prime$& $v^2 \tilde d_t/m_t$\\
\hline
  $2.8   \cdot 10^{-6} \  (\dagger) $ & $1.5\cdot 10^{-6} \ (\dagger)$  &$0.7\cdot 10^{-2} \ (*)$ & $0.6 \cdot 10^{-2} \ (\dagger\, *)$ &$0.5\cdot 10^{-2} \ (*)$   &$7.8 \cdot 10^{-3} \ (\dagger) $  & 0.23 
  \ $(\dagger)$  &$4.3 \cdot 10^{-2} \ (\dagger)$ \\
\hline
\end{tabular}
\end{center}
\caption{\small 
Summary of current best bounds on  non-standard CPV Higgs couplings (at $\mu=\MT= 1$~TeV) 
coming either from EDMs with minimized matrix elements (denoted by  $\dagger$) or the LHC 
(denoted by  $*$). 
\label{bounds-complementarity}
}
\end{table}

In  this work we have focused on new  CPV couplings of the Higgs to quark and gluons.  
In light of the upcoming Run 2 at the LHC and EDM searches with improved sensitivities,  
we think it will be timely to systematically analyze all possible CPV Higgs couplings. 
In this context,  several new directions are worth exploring.
First, as evident from our discussion, it would be interesting to study observables involving 
the Higgs at the LHC, that  are linearly sensitive to the non-standard couplings. 
Second,  in a framework in which the observed Higgs is part of an EW doublet, 
 additional CPV operators appear at  dimension-6, 
generating CPV Higgs couplings involving electroweak bosons and  fermions~\cite{Buchmuller:1985jz,Grzadkowski:2010es}. 
We plan to study  these in a subsequent work,  focusing again on the best information that can be obtained 
from both direct and indirect probes. 
Finally,  it would be interesting to perform a comparative analysis of the linear EFT versus 
the more general EFT based on the EW chiral Lagrangian with a light Higgs~\cite{Buchalla:2013rka,Buchalla:2013eza}.
In this framework, the non-standard  (possibly CPV) Yukawa couplings rise to the level of leading order couplings, and 
some symmetry relations are lost (e.g. in the dipole operators the coefficients of $O(h^0)$ and $O(h)$ are independent).  
In this context it would be very valuable to identify experimental tests  involving a  combination of 
EDMs and LHC observables  that would discriminate between  the two scenarios, 
and thus shed light on the nature of the Higgs and electroweak symmetry breaking.

\subsection*{Acknowledgements}
This work (JdV) is supported in part by the DFG and the NSFC
through funds provided to the Sino-German CRC 110 ``Symmetries and
the Emergence of Structure in QCD'' (Grant No. 11261130311).
The work of VC and EM is supported by DOE Office of Nuclear Physics and the LDRD program at Los Alamos National Laboratory. 
We acknowledge useful discussions with 
Tanmoy Bhattacharya, 
Dani\"el Boer,
Giuseppe Cerati,
Martin Gonzalez-Alonso, 
Michael Graesser, 
Rajan Gupta, 
Gino Isidori, 
Robert~Harlander, 
Maxim Pospelov,
Maria Ubiali,   
and Andreas Wirzba. We thank the INT at the University of Washington for its hospitality during the completion of this work.

\appendix

\section{The light-quark color EDMs}
\label{sect-appA}

In this Appendix we discuss the quark chromo-EDMs other than the top-quark CEDM. In particular we present results for the running of these operators as well as present and future EDM and collider constraints.

At the LHC the $\tilde d_{q\neq t}$ operators contribute to single-Higgs production and can be bound at the level of $v \tilde{d}_q \sim 4$-$20$\%. 
Much stronger constraints arise from EDMs (four to seven orders of magnitude stronger), and their analysis can be performed in a similar way as for the top CEDM. The evolution of 
 $\tilde d_{q\neq t}$  to low energies is well described by the RGEs of $(d_q,\, \tilde d_q,\, d_W)$ \cite{Degrassi:2005zd}. This means the up, down, and strange CEDMs only give rise to the up, down, and strange EDMs and CEDMs, respectively, at low energies. Instead, the charm (bottom) CEDM induces a threshold correction to the Weinberg operator at $m_c$ ($m_b$), see 
Eq.~\eqref{eq:dWthreshold}. In turn, the induced Weinberg operator generates all the light quark (C)EDMs, $d_{u,d,s}$ and $\tilde d_{u,d,s}$, when evolved to $\Lambda_\chi$ \cite{Sala:2013osa}. This gives rise to the contributions to the operators at $\Lambda_\chi$ shown in Table \ref{Table:CEDMs1GeV}.

\begin{table}\footnotesize
\centering
$
\begin{array}{c||ccccc}
\MT=1\text{ TeV} & \tilde d_u/m_u & \tilde d_d/m_d & \tilde d_c/m_c &
   \tilde d_s/m_s & \tilde d_b/m_b \\\hline\hline
 d_u/ m_u   & 0.26\, e & - & 5.1\cdot 10^{-6}\, e & - & 6.7 \Exp{5}\, e \\
 \tilde{d}_u/ m_u       & 0.44 & -& 2.1\Exp{4} & -& 5.9 \Exp{4} \\
 d_d/m_d  &-& -0.13\, e & -2.5\cdot 10^{-6} \, e &- & -3.4 \Exp{5}\,e \\
 \tilde{d}_d/m_d &-& 0.44 & 2.1\Exp{4} & -& 5.9\Exp{4} \\
 d_s / m_s & - & - & -2.5\cdot 10^{-6}\, e &
   -0.13 \,e & -3.4 \Exp{5}\,e \\
 \tilde{d}_s/m_s  &- &- & 2.1\Exp{4} & 0.44 & 5.9\Exp{4} \\
 d_W\ & -&- & -0.011 & - & -5.8\Exp{3} \\
\end{array}$
\caption{\small The contributions of the quark CEDM operators to the operators which contribute to EDMs (Eq.\ \eqref{ExtendedO}) at low energies, $\MQCD\simeq 1\, {\rm GeV}$. Here we assumed the scale of new physics to be $\MT = 1$ TeV. A dash, $``-"$, indicates no, or a negligible, contribution.  }
\label{Table:CEDMs1GeV}
\end{table}

The operators at $\Lambda_\chi$ can again be related to EDMs as discussed in Sec.\ \ref{Sec:interpretation}. The resulting EDM constraints are presented in Table \ref{boundsCEDMs}. A clear difference with the bounds on the Yukawa couplings and $\tilde d_t$ is that the electron EDM does not constrain any of the operators considered in this appendix. In addition, the constraints from $d_{\rm Hg}$ vanish when applying the minimization procedure, as was the case in Table \ref{boundsY}. 
Nonetheless, there are significant constraints on most quark CEDMs from the neutron EDM. The exception is  $\tilde d_s$, 
which remains unconstrained in the minimized case due to the uncertain $d_s$ matrix element. 
It is important to note that the constraints in the ``Future Min." row differ from those in the ``Comb. Cen." row only by a factor of two for most couplings, and no more than a factor of $5$. Thus, an improvement of the matrix elements, as described in Sec.\ \ref{future_matrix}, would again allow one to exploit the full potential of the experimental limits.

\begin{table}\footnotesize
\centering
$
\begin{array}{||c|ccccc||}
\hline
 \text{} & v^2 \tilde d_u/m_u & v^2 \tilde d_d/m_d & v^2 \tilde d_c/m_c &
   v^2 \tilde d_s/m_s & v^2 \tilde d_b/m_b  \\\hline
 d_e &\text{x} &\text{x} & \text{x} &\text{x} & \text{x} \\\hline
 d_n\text{ Cen.} & 9.0\Exp{5} & 2.3 \Exp{5} &1.6\Exp{4} & 7.0\Exp{4} & 3.1\Exp{4} \\
 d_n\text{ Con.} & 1.6\Exp{4} & 4.0\Exp{5} & 8.1\Exp{4}& 7.0\Exp{3}& 1.6\Exp{3} \\
 d_n\text{ Min.} & 1.6\Exp{4} &4.0\Exp{5} & 8.2\Exp{4} & \text{x} & 1.8\Exp{3} \\\hline
 d_{\text{Hg}}\text{ Cen.} & 2.3\Exp{5} & 9.0\Exp{6} & 3.0\Exp{4} & 1.1\Exp{3} & 6.1\Exp{4} \\
 d_{\text{Hg}}\text{ Con.} & 9.3\Exp{4} & 8.0\Exp{5} & 2.0\Exp{3} & 0.015 & 4.0\Exp{3} \\
 d_{\text{Hg}}\text{ Min.} & \text{x}&\text{x}&\text{x} &\text{x}&\text{x} \\\hline
 \text{Comb. Cen.} & 2.3\Exp{5} & 8.4\Exp{6} & 1.4 \Exp{4}& 5.8\Exp{4} & 2.8\Exp{4} \\
 \text{Comb. Con.} &1.6\Exp{4} & 3.6\Exp{5} & 7.5\Exp{4} & 6.3\Exp{3} & 1.5\Exp{3} \\
 \text{Comb. Min.} &1.6\Exp{4} &4.0\Exp{5} & 8.2\Exp{4} & \text{x}& 1.8\Exp{3} \\\hline
  \text{Future Min.} & 1.1\Exp{4} & 2.7\Exp{5} & 3.1\Exp{4} & 1.2\Exp{3} & 6.2\Exp{4} \\\hline
\end{array}
$
\caption{\small $90\%$ upper bounds on the quark color-EDM operators (for $\MT=1$ TeV) due to current EDM constraints, assuming that a single operator dominates at the high scale. 
Row $1$ is the bound from $d_e$, Rows $2-4$ are bounds from the $d_n$ with the three strategies explained in the text (see section \ref{sec:strategies}), Rows $5-7$ are the same but using $d_{\mathrm{Hg}}$. Rows $8-10$ are bounds due to the combined EDM limits. Row $11$ shows the combined minimized bounds in case of improved matrix elements, see Sect.~\ref{future_matrix} for more details. An `x' indicates that the bound is larger than $1$.
\label{boundsCEDMs}}\end{table}

Finally, in row $3$ and $4$ of Table \ref{boundsCEDMs_improved_XeRa} we show the constraints that would result from the increase in sensitivity of future $d_n$ and $d_e$ experiments ($d_n\leq 10^{-28}\,e$ cm and $d_e\leq 5\cdot 10^{-30}\,e$ cm), with current and future matrix elements. 
Since the $d_n$ measurement would improve by a factor $300$, and the current constraints are dominated by $d_n$, the constraints improve by roughly the same factor.
In rows five and six the same analysis is performed for future measurements of $d_{\rm Xe}$ and  $d_{\rm Ra}$ ($d_{\rm Xe}\leq 10^{-30}\, e$ cm and $d_{\rm Ra}\leq 10^{-27}\, e$ cm). From row five it is again clear that these measurements are mainly sensitive to the up and down quark couplings, as was the case in Table \ref{boundsY_improved_XeRa}. 
In the last two rows we consider the impact of $d_{p}$ and $d_D$ measurements at the level of $10^{-29}\,e$ cm. Experiments at this high level of accuracy would dramatically improve the constraints by up to four orders of magnitude. The most significant effect of improving the matrix elements is an improvement of bound on the strange CEDM by three orders of magnitude in the $d_{\rm Xe}+d_{\rm Ra}$ case, and at least six orders of magnitude in the $d_p+d_D$ and $d_n+d_e$ cases.

\begin{table}\footnotesize
\centering
$
\begin{array}{||c|ccccc||}\hline
 \text{} & v^2 \tilde d_u/m_u & v^2 \tilde d_d/m_d & v^2 \tilde d_c/m_c &
   v^2 \tilde d_s/m_s & v^2 \tilde d_b/m_b  \\\hline\hline
 \text{Current} & 1.6\Exp{4} & 4.0\Exp{5} & 8.2\Exp{4} & \text{x} & 1.8\Exp{3} \\
 \text{Current+Th.} & 1.1\Exp{4} & 2.7\Exp{5} & 3.1\Exp{4} & 1.2\Exp{3} & 6.2\Exp{4} \\\hline
 d_n+d_{\text{ThO}} & 5.5\cdot 10^{-7} & 1.4\cdot 10^{-7} & 2.8\cdot 10^{-6} & \text{x} & 6.2\cdot 10^{-6} \\
 d_n+d_{\text{ThO}}\text{+Th.} & 4.0 \cdot10^{-7} & 1.0\cdot10^{-7} & 1.1\cdot 10^{-6} & 4.8\cdot 10^{-6} & 2.2\cdot 10^{-6} \\\hline
 d_{\text{Xe}}+d_{\text{Ra}} & 7.2\Exp{5} & 9.6\cdot 10^{-6} &8.2\Exp{4} &\text{x}& 1.8\Exp{3} \\
 d_{\text{Xe}}+d_{\text{Ra}}\text{+Th.} & 9.2\cdot 10^{-6} & 2.5\times 10^{-6} & 3.1\Exp{4} & 1.2\Exp{3} & 6.1\Exp{4} \\\hline
 d_p+d_D & 1.1\cdot 10^{-8} & 5.9\cdot 10^{-9} & 2.8\cdot 10^{-7} & 0.75 & 6.1\cdot 10^{-7} \\
 d_p+d_D\text{+Th.} & 8.5\times 10^{-9} & 5.1\cdot10^{-9} & 1.1\cdot 10^{-7} & 2.3\cdot 10^{-7} & 2.2\cdot 10^{-7} \\\hline
\end{array}
$
\caption{ \small The first two rows denote combined minimized constraints with current and improved matrix elements. Rows $3$ and $4$ are similar but for future $d_n$ and ThO measurements. Rows $5$ and $6$ do the same but now for future measurements of $\dXe$ and $\dRa$, while Rows $7$ and $8$  include $d_p$ and $d_D$ measurements. 
}
\label{boundsCEDMs_improved_XeRa}
\end{table}

\bibliographystyle{h-physrev3} 
\bibliography{bibliography,bibVC}

\end{document}